\documentclass[aps, preprint, groupedaddress, floatfix, nofootinbib,longbibliography]{revtex4-1}
\usepackage{epsfig}
\usepackage{graphicx}
\usepackage{amsmath}
\usepackage{bbold}
\usepackage{tabularx, booktabs}
\usepackage[normalem]{ulem}
\usepackage{soul}  
\usepackage{xcolor} 
\usepackage{multirow} 
\usepackage{xr}
\usepackage{subcaption}
\usepackage{ragged2e}
\DeclareCaptionJustification{justified}{\justifying}
\newcommand{\ie}{{\textit i.e.}}

\newcommand{\etal}{{\textit et al.}}
\newcommand{\BFO}{BiFeO$_3$}
\newcommand{\BMO}{BiMnO$_3$}
\newcommand{\PTO}{PbTiO$_3$}

\newcommand{\BPTO}{BPTO}
\newcommand{\SM}{Supplementary Information}

\newcolumntype{Y}{>{\centering\arraybackslash}X}
\newcolumntype{Z}{>{\hsize=1.1\hsize\centering\arraybackslash}X}

\begin{document}

\title{Design of a multifunctional polar metal via
first-principles high-throughput structure screening}

\author{Yue-Wen Fang$^{1,2}$}
\email{fyuewen@gmail.com}
\author{Hanghui Chen$^{2,3}$}
\email{hanghui.chen@nyu.edu}

\affiliation{
 $^1$Department of Materials Science and Engineering, Kyoto University, Kyoto, Japan\\
  $^2$NYU-ECNU Institute of Physics, New York University Shanghai, Shanghai, China\\
  $^3$Department of Physics, New York University, New York, USA\\}
\date{\today}

\begin{abstract}
  Intrinsic polar metals are rare, especially in oxides,
  because free electrons screen electric fields in a metal
  and eliminate the internal dipoles that are needed to
  break inversion symmetry. Here we use first-principles
  high-throughput structure screening to predict a new
  polar metal in bulk and thin film forms.
  After screening more than 1000 different crystal structures,
  we find that ordered BiPbTi$_2$O$_6$ can crystallize in three
  polar and metallic structures, which can be transformed between each other
  via pressure or strain. In a heterostructure of layered BiPbTi$_2$O$_6$
  and PbTiO$_3$, multiple states with different relative orientations of
  BiPbTi$_2$O$_6$ polar displacements, and PbTiO$_3$~polarization, can be stabilized.
  At room temperature the interfacial coupling enables electric fields to first switch
  PbTiO$_3$ polarization and subsequently drive 180$^{\circ}$ change of BiPbTi$_2$O$_6$ polar displacements.
  At low temperature, the heterostructure provides a tunable tunnelling barrier and might be used in multi-state
  memory devices.
\end{abstract}

\maketitle

Polar metals--analogy of ferroelectrics in metals--are characterized
by intrinsic conduction and inversion symmetry breaking.  Polar metals
are rare (especially in oxides) because mobile electrons screen
electric fields in a metal and eliminate internal dipoles that are
needed to break inversion symmetry. The discovery of
LiOsO$_3$~\cite{Shi2013}, a metal that transforms from a
centrosymmetric $R\bar{3}c$ structure to a polar $R3c$ structure at
140 K, has stimulated an active search for new polar metals
in both
theory and experiment ~\cite{xiang2015prb, Puggioni2014NC,
  Filippetti2016NatComm, kim2016polar, Benedek-JMCC-2016,
 PhysRevB.96.235415,PhysRevMaterials.2.125004,Fei2018}.

Density-functional-theory-based first-principles calculations have
proven accurate in describing crystal structures and have been
succesfully applied to predict new functional materials, such as
ferroelectrics, piezoelectrics and multiferroics~\cite{fang2015first}.
Since crystal structure is the essential property of polar metals, we
need to scrutinize the prediction by not presupposing an \textit{a
  priori} favorable crystal structure.  First-principles
high-throughput crystal structure screening method, which is based on the
marriage between first-principles calculations and a multitude of
techniques such as particle-swarm optimization
algorithm~\cite{Wang2012} and evolutionary
algorithm~\cite{USPEX-2006CPC}, has demonstrated its superior power in
effectively searching for the ground state structures and metastable structures
of functional materials with only the given knowledge of chemical
composition~\cite{Wei2018NatMater,He2019-NatComm-PbTe,superconductor-PRL2019}.

In this work, we use \textit{ab initio}
high-throughput structure screening to predict a new
polar metal BiPbTi$_2$O$_6$ (BPTO for short). After screening over
1000 different crystal structures, we find that ordered BPTO can
crystallize in three different polar metallic structures (post-perovskite
$Pmm2$, perovskite $Pmm2$ and perovskite $Pmn2_1$), each of which can
be transformed to another via external pressure or epitaxial
strain. The mechanism is that $6s$ lone-pair electrons of Bi and Pb
ions tend to favor off-center
displacements~\cite{fang2015first}.
On the other hand, in the
perovskite
structures, Bi$^{3+}$ and
Pb$^{2+}$ enforce a fractional valence on Ti, which leads to
conduction; in the post-perovskite structure, strong hybridization
between Bi/Pb $6p$ and O $2p$ states induces a finite density of
states at the Fermi level.

Next we demonstate potential applications of the new polar metal BPTO
by studying a BPTO/PbTiO$_3$ heterostructure. We find that different
states in which BPTO polar displacements are parallel, anti-parallel
and perpendicular to PbTiO$_3$ polarization can be stabilized in the
heterostructure. 180$^{\circ}$ switching of BPTO polar displacements
needs to surmount an energy barrier of about 58 meV per slab.
This implies
that at room temperature where thermal fluctuations can overcome the
switching barrier, the interfacial coupling between the polarization
and polar displacements enables an electric field to first switch
PbTiO$_3$ polarization and subsequently drive BPTO to change its polar
displacements by 180$^{\circ}$; at low temperatures where the
switching barrier dominates over thermal fluctuations, the BPTO polar
displacements can not be switched but the direction of PbTiO$_3$
polarization can be controlled by an electric field. This can
stabilize three distinct states with different tunnelling barriers.

\section*{R\lowercase{esults and discussion}}

\subsection*{Most stable crystal structures of bulk BPTO}
The key question in predicting a new polar metal is to determine its
crystal structure. Since ordered BPTO has not been synthesized in
experiment, we perform a first-principles high-throughput search for
the ground state structure using CALYPSO~\cite{Wang2010, Wang2012}
method, in combination with CrySPY~\cite{Yamashita-CrySPY}.  In the
search, we do not constrain ourselves in any \textit{a priori}
favorable crystal structure.
We screen more than 1000 different
crystal structures among which we consider different Bi/Pb ordering
in perovskite structure:
layered ordering, columnar ordering and
rock-salt ordering; and we also consider many non-perovskite
structures, including post-perovskite structure and hexagonal
structure. The computational details of our first-principles
calculations and high-throughput structure screening method are
provided in \textbf{Methods}.

\begin{figure}[htp]
\includegraphics[angle=0,width=0.9\textwidth]{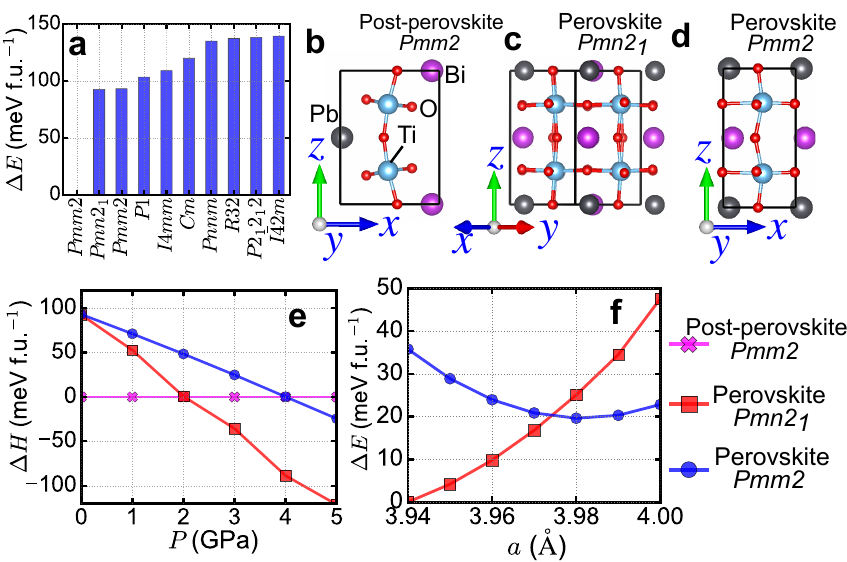}
\caption{\label{fig:calypso} \textbf{The lowest-energy crystal structures and phase transitions.}
\textbf{a} Ten lowest-energy
  crystal structures of BPTO predicted by CALYPSO and DFT
  calculations. \textbf{b} Post-perovskite ${Pmm2}$ structure;
  \textbf{c} Perovskite ${Pmn2_1}$ structure; \textbf{d}
  Perovskite ${Pmm2}$ structure. \textbf{e} Enthalpy of the
  three lowest-energy structures as a function of pressure.  The
  enthalpy of the post-perovskite ${Pmm2}$ structure under each
  pressure is set as the zero point. \textbf{f} Total energy
  of the two lowest-energy perovskite structures as a function of
  epitaxial strain.  The energy of the perovskite ${Pmn2_1}$ structure
  constrained by an in-plane lattice constant of 3.94~\AA~is chosen as
  the zero energy.}
\end{figure}

Fig.~\ref{fig:calypso} shows ten lowest-energy crystal structures of
BPTO from our calculations. The details of these ten crystal
structures are available in Supplementary Table 1.
The lowest energy structure is post-perovskite with a polar symmetry
$Pmm2$ (space group No. 25).  The crystal structure is explicitly
shown in Fig.~\ref{fig:calypso}\textbf{b}. The TiO$_6$ octahedra are
both corner-sharing and edge-sharing. The lack of inversion symmetry
can be appreciated from Ti atoms which have strong polar displacements
with respect to neighboring O atoms towards $x$-axis. The next two
lowest-energy crystal structures are both perovskite with $Pmn2_1$
symmetry (space group No. 31) and $Pmm2$ symmetry (space group
No. 25). Both $Pmn2_1$ and $Pmm2$ symmetries are polar. The two perovskite
structures have almost the same energy.
Fig.~\ref{fig:calypso}\textbf{c} shows the perovskite $Pmn2_1$
crystal structure. Bi and Pb atoms form a rock-salt ordering and their
displacements with respect to O atoms in the $xy$ plane make the
crystal structure acentric.  Fig.~\ref{fig:calypso}\textbf{d} shows
the perovskite $Pmm2$ crystal structure. Bi and Pb atoms have a
layered ordering with a stacking direction along $z$-axis. It is clear
that Bi, Pb and Ti atoms all have strong polar displacements with
respect to O atoms along $x$-axis, which breaks inversion
symmetry. While post-perovskite oxides are interesting by
themselves~\cite{murakami2004post,Science-2018Ohta},
perovskite oxides have been widely studied and are more suitable for
device applications because many perovskite oxide substrates are
available~\cite{BISWAS2017117}, which makes it feasible to grow
perovskite oxide thin films. Therefore, we consider using external
pressure or epitaxial strain to transform BPTO among
different polar structures. Pressure is widely used in bulk
synthesis to isolate metastable phases of matter~\cite{JACS-2001-highpressure,PhysRevMaterials.3.064411}.
Fig.~\ref{fig:calypso}\textbf{e} shows that both perovskite
$Pmn2_1$ and $Pmm2$ structures become more stable than the
post-perovskite $Pmm2$ structure under a few GPa.  The reason is that
the post-perovskite $Pmm2$ structure is very hollow with a very large
volume of 130 $\textrm{\AA}^3$ f.u.$^{-1}$ under ambient conditions, while
the two perovskite structures are more closely packed (122
$\textrm{\AA}^3$ f.u.$^{-1}$ and 126 $\textrm{\AA}^3$ f.u.$^{-1}$ under ambient
conditions, respectively).  Applying pressure favors structures with
smaller volumes.  If we want to grow BPTO thin films on a perovskite
oxide substrate, the post-perovskite structure does not form due to
very large lattice mismatch~(see Supplementary Table 1 for the cell
parameters of post-perovskite structure). The
pseudo-cubic lattice constant of the perovskite $Pmn2_1$ structure
is 3.94~\AA, while that of the perovskite
$Pmm2$ structure is 3.98~\AA.  It
is anticipated that as the substrate lattice constant varies from
3.94~\AA~to 3.98~\AA, the energetically favored structure changes from
the perovskite $Pmn2_1$ structure to the perovskite $Pmm2$
structure. This is indeed what Fig.~\ref{fig:calypso}\textbf{f}
shows. Since the perovskite $Pmm2$ structure has a layered ordering of
Bi/Pb atoms, it is highly suitable for thin film growth methods such
as pulsed layer deposition and molecular beam
epitaxy~\cite{korotcenkov2017metal}.  Substrates such as NdScO$_3$ and
KTaO$_3$ have a proper lattice constant to stabilize the perovskite
$Pmm2$ structure in BPTO thin films. The DFT calculated lattice constants
of KTaO$_3$ and NdScO$_3$ can be found in Supplementary Table 2 with the comparison to the experimental data.
We also enforce the
post-perovskite $Pmm2$ structure to be stabilized on a perovskite
oxide substrate and we expectedly find that its total energy is about
10 eV f.u.$^{-1}$ higher than the two perovskite structures because of the
large lattice mismatch~(Supplementary Figure 1).
In addition to the study of phase transitions under strains and pressures,
the temperature effect on phase transitions can be found in Supplementary Figure 2.

\subsection*{Electronic and magnetic properties of polar metal BPTO}
\begin{figure}[!tp]
  \includegraphics[angle=0,width=0.9\textwidth]{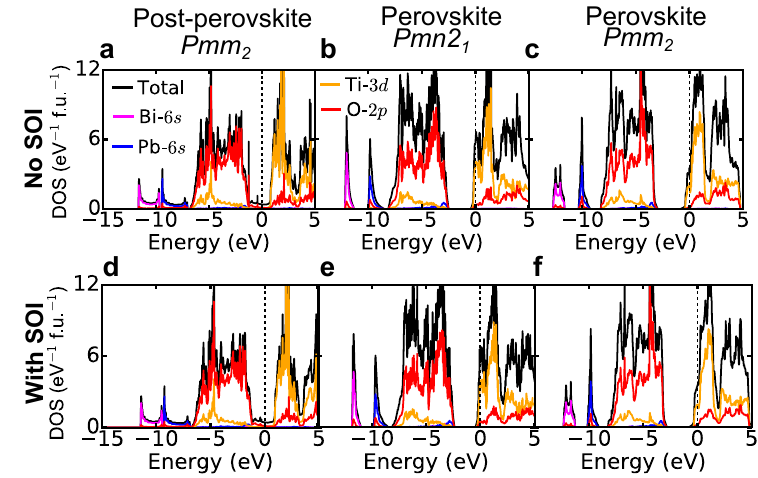}
  \caption{\label{fig:dos} \textbf{Densities of states (DOS) of BPTO
  of the three polar structures.}
  The upper and lower panels show the DOS
  calculated by DFT without spin-orbit interactions (SOI)
  and DFT with SOI, respectively.
  \textbf{a,d} The post-perovskite
  structure with $Pmm2$ symmetry; \textbf{b,e} The perovskite
  structure with $Pmn2_1$ symmetry; \textbf{c,f} The perovskite
  structure with $Pmm2$ symmetry. The black curve is the total DOS.
  The magenta, blue, orange and red curves are Bi-$6s$, Pb-$6s$, Ti-$3d$ and
  O-$2p$ projected DOS, respectively. The dashed line is the Fermi level.}
\end{figure}

The upper panels of Fig.~\ref{fig:dos} show the DFT-calculated
densities of states of the post-perovskite $Pmm2$, perovskite $Pmn2_1$
and perovskite $Pmm2$ structures. We calculate both total density of
states (DOS) and orbital-projected DOS (Ti-$3d$, Bi-$6s$, Pb-$6s$ and
O-$2p$). In three optimized structures, we do not find any
magnetization or charge disproportionation. Therefore spin-up and
spin-down are summed in the DOS. The DOS projected on the two Ti atoms
are identical and hence are also summed. The three DOS share
similarities but also have important differences. All three structures
have a non-zero DOS at the Fermi level; both Bi-$6s$ and Pb-$6s$ are
well below the Fermi level and are fully occupied. The difference is
that in both perovskite structures ($Pmn2_1$ and $Pmm2$), because
nominally Bi$^{3+}$, Pb$^{2+}$ and O$^{2-}$, due to charge neutrality
Ti must have a formal valence of Ti$^{3.5+}$, \ie, every Ti atom has
0.5 electron in the $3d$ conduction bands (no charge
disproportionation is found in the calculations).
Figures.~\ref{fig:dos}\textbf{b} and ~\ref{fig:dos}\textbf{c} show that the Fermi level
crosses Ti-$3d$ states in the DOS of the perovskite $Pmn2_1$ and
$Pmm2$ structures. However, in the post-perovskite structure
(Fig.~\ref{fig:dos}\textbf{a}), Ti-$3d$ states have negligible
contribution around the Fermi level. Instead, Bi-$6p$ and Pb-$6p$ as
well as O-$2p$ states make the largest contribution to the DOS around
the Fermi level, which can also be seen in Supplementary Figure 3 where
the electronic states around the Fermi level are zoomed in.

The Bader (static) charge
analysis in Table~\ref{Tab:charge} shows that Bi and Pb have about 0.25 and
0.19 more electrons in the post-perovskite structure than in the
perovskite structures, which indicates stronger hybridization between
Bi/Pb and O atoms in the post-perovskite structure. Therefore in the
post-perovskite structure, Bi-$6p$ and Pb-$6p$ states are not fully
empty and thus appear around the Fermi level.

\begin{table}[htp]
	\caption{\label{Tab:charge} \textbf{Bader charges for bulk \BPTO.}
          Charges are normalized to per atom.  }
\begin{tabular}{cccccc}
\hline \hline
\multirow{2}{*}{\begin{tabular}[c]{@{}c@{}}Structural\\ type\end{tabular}}
&
\multirow{2}{*}{\begin{tabular}[c]{@{}c@{}}Space\\ group\end{tabular}}
& \multicolumn{4}{c}{Bader charges ($e$)} \\ \cline{3-6} & & Bi & Pb &
Ti & O \\ \hline Post-perovskite & ${Pmm2}$ & +1.48 & +1.13 & +2.19 &
-1.16 \\ Perovskite & ${Pmn2_1}$ & +1.73 & +1.35 & +2.12 & -1.22
\\ Perovskite & ${Pmm2}$ & +1.73 & +1.32 & +2.06 & -1.19
\\ \hline\hline
\end{tabular}
\end{table}


Pb and Bi are heavy elements and their spin-orbit interactions (SOI)
are not negligible.  In the lower panels of Fig.~\ref{fig:dos}, we
take into account SOI and show the corresponding densities of states
of BPTO of the three polar structures.  Similar to the results without
SOI, we do not find any magnization or charge disproportionation in
the fully relaxed structures.  By comparing the densities of states
calculated by DFT without SOI (upper panels of Fig.~\ref{fig:dos}) and
DFT with SOI (lower panels of Fig.~\ref{fig:dos}), SOI almost
unaffects the electronic structure, similar to previous studies on
other polar metals~\cite{Puggioni2014NC,aulesti2018APL}.

\begin{figure}[!tp]
  \includegraphics[angle=0,width=0.7\textwidth]{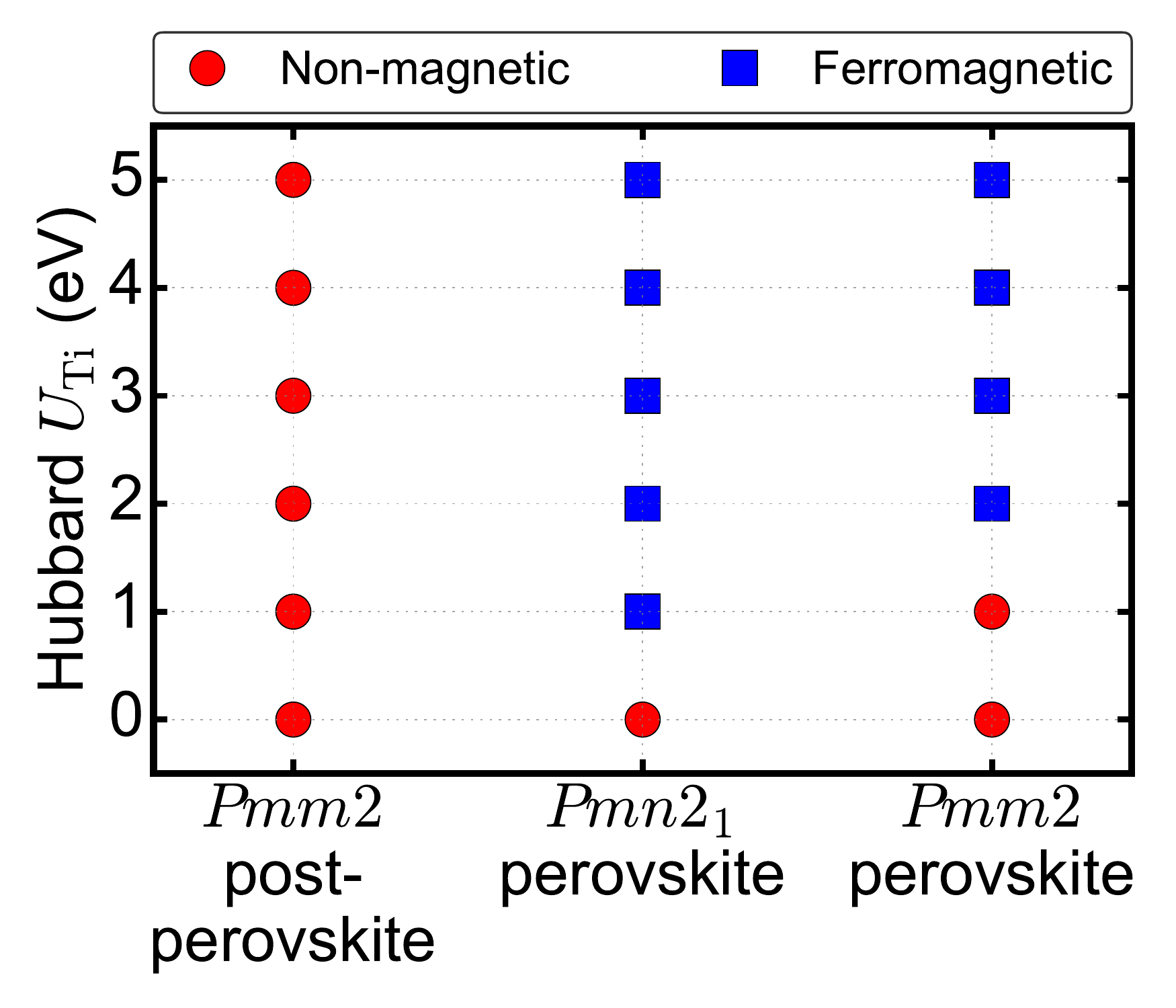}
  \caption{\label{fig:HubbardU} \textbf{Phase diagram as a function of Hubbard $U$.}
  The magnetic phase diagram of the
  post-perovskite ${Pmm2}$, perovskite ${Pmn2_1}$ and perovskite
  ${Pmm2}$ structures as a function of Hubbard $U_{\rm Ti}$.
  Non-magnetic state and ferromagnetic state are shown by
  red circle and blue square, respectively.}
\end{figure}

While DFT with/without SOI calculations do not find any magnetization or
charge disproportionation, correlation effects from Ti-$3d$ orbitals
may favor spin ordering and charge ordering. A long-range magnetic
ordering with a charge disproportionation (Ti$^{3+}$+Ti$^{4+}$) can
result in an insulating ground state~\cite{PhysRevLett.99.016802}.
To test the robustness of our
prediction that BiPbTi$_2$O$_6$ is a polar metal, we apply an
effective Hubbard $U$ correction on the Ti-3$d$ orbitals and calculate
the densities of states for all three low-energy structures. The
accurate value of correlation strength of \BPTO~is not known, but
presumably it should not exceed that of Mott insulator LaTiO$_3$, in
which Hubbard $U_{\textrm{Ti}}$ is about 5 eV~\cite{Pavarini2004}.
Therefore we consider a Hubbard $U_{\textrm{Ti}}$ ranging from 0 to 5
eV.  Within this range of $U_{\textrm{Ti}}$, we do not find charge
disproportionation but find robust metallicity in all the three polar
structures of BPTO.  Furthermore, in this range of
$U_{\textrm{Ti}}$, we find itinerant ferromagnetism in the perovskite
$Pmn2_1$ structure at ${U_{\textrm{Ti}} \geq 1}$ eV and in the
perovskite $Pmm2$ structure at ${U_{\textrm{Ti}} \geq 2}$ eV
(antiferromagnetic ordering is less stable than ferromagnetic
ordering). We do not find any magnetism in the post-perovskite $Pmm2$
structure up to $U_{\textrm{Ti}} = 5$ eV. The magnetic phase diagram
for the three polar structures as a function of Hubbard
$U_{\textrm{Ti}}$ is shown in Fig.~\ref{fig:HubbardU}.  The origin of
itinerant ferromagnetism in BPTO is Stoner
instability~\cite{patrik1999lecture}.  In DFT+$U$ calculations, the
Stoner criterion to induce itinerant ferromagnetism
is~\cite{Janicka2008}:
\begin{equation}
        \label{sEq:stoner-corr}
 U\rho({E_\mathrm{F}}) > 1 ,
\end{equation}
were $U$ and ${\rho({E_\mathrm{F}})}$ are Hubbard $U$ parameter and
density of states at the Fermi level of a non-magnetic state,
respectively. The upper panels of Fig.~\ref{fig:dos} shows that the
perovskite ${Pmn2_1}$ structure has a large density of state
at the Fermi level $\rho(E_\mathrm{F})$ in its non-magnetic state;
the perovskite ${Pmm2}$ structure has a slightly
smaller $\rho(E_\mathrm{F})$. Post-perovskite $Pmm2$ structure, on the other hand,
has a very small $\rho(E_\mathrm{F})$ (9 times smaller than that of the
perovskite ${Pmm2}$ structure and 15 times smaller than that of the
perovskite ${Pmn2_1}$ structure).
This explains that
the critical $U_{\textrm{Ti}}$ to stabilize itinerant
ferromagnetism in the perovskite $Pmn2_1$ structure is the smallest,
while a much larger $U_{\textrm{Ti}}$ (larger than 5 eV) is needed to
induce magnetism in the post-perovskite $Pmm2$ structure.

\subsection*{The role of lone-pair electrons}

A local structural instability arising from lone-pair electrons has been
reported in
ferroelectric insulators and degenerately doped
ferroelectrics~\cite{cohen1992origin,fang2015first,PhysRevB.74.224412,2D-FE-lonepair,ZhaoHJ-PhysRevB.97.054107,gu2017coexistence,PhysRevB.94.224107,2D-FTJ-SnSe}.
However,
lone-pair electrons alone are not sufficient to stabilize a polar
state in metals nor a ferroelectric state in insulators. For example,
\BFO~is ferroelectric~\cite{PhysRevB.74.224412} but \BMO~is
anti-ferroelectric~\cite{baettig2007anti,goian2012absence} although
lone-pair electrons are present in both of them. Therefore,
high-throughput crystal structure prediction is essential in
predicting new polar metals and ferroelectric insulators. In our
study, the crystal structure screening takes into account both polar
and anti-polar states for different cation orderings.

\begin{figure}[!tp]
  \includegraphics[angle=0,width=0.75\textwidth]{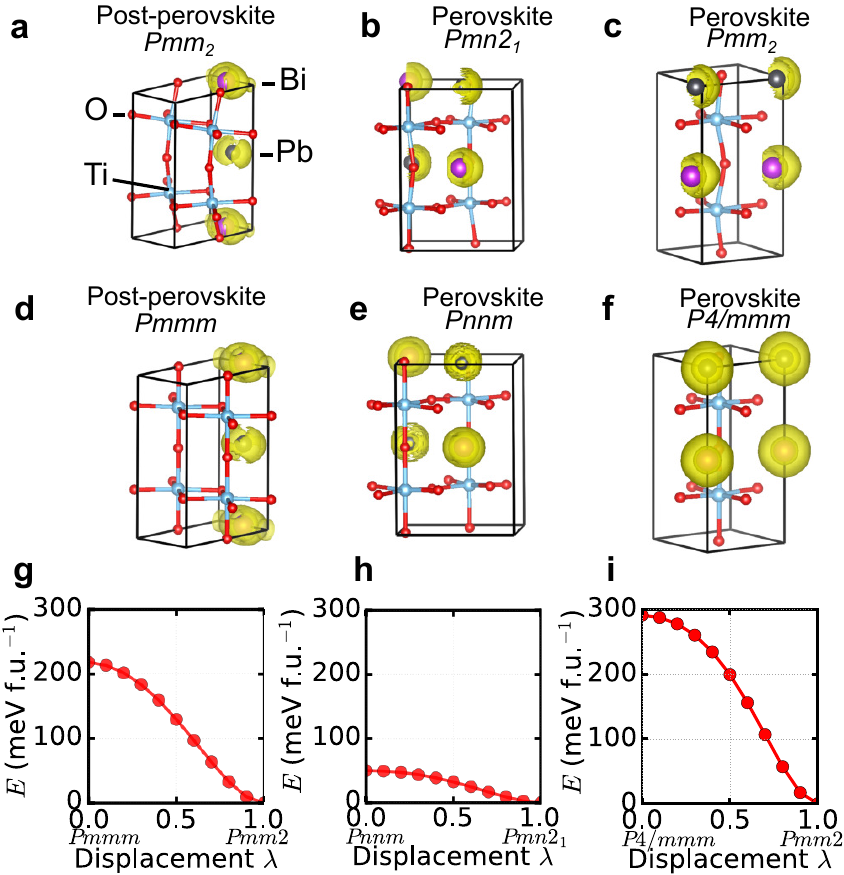}
  \caption{\label{fig:6s} \textbf{Electron localization function (ELF) and the energy curve from the centrosymmetric structure to the corresponding polar structure.}
  \textbf{a} The polar
    structure: post-perovskite ${Pmm2}$ and \textbf{d} the
    centrosymmetric structure: post-perovskite ${Pmmm}$; \textbf{b}
    The polar structure: perovskite ${Pmn2_1}$ and \textbf{e} the
    centrosymmetric structure: perovskite ${Pnnm}$; \textbf{c} The
    polar structure: perovskite ${Pmm2}$ and \textbf{f} the
    centrosymmetric structure: perovskite ${P4/mmm}$.
    The isosurface of ELF is set at a value of $\sim$ 0.5.
   \textbf{g}
    Transition from the post-perovskite ${Pmmm}$ structure to the
    post-perovskite ${Pmm2}$ structure.  \textbf{h} Transition from
    the perovskite ${Pnnm}$ structure to the perovskite ${Pmn2_1}$
    structure.  \textbf{i} Transition from the perovskite ${P4/mmm}$
    structure to the perovskite ${Pmm2}$ structure.  }
\end{figure}

We now show that in the three lowest-energy metallic phases of
  BPTO, the lone-pair $6s$ electrons in Bi and Pb play an important role in
  breaking inversion symmetry. We use electron localization
function (ELF, defined in \textbf{Methods})
to explicitly visualize how lone-pair electrons of Bi and Pb break
inversion symmetry in metallic BPTO.  The panels \textbf{a-c} of
Fig.~\ref{fig:6s} show an iso-surface of ELF of the three polar
structures of BPTO. Only the ELF of Bi and Pb ions are displayed for
clarity. Similar to insulating Bi-based and Pb-based perovskite
oxides~\cite{lonepair-4,PhysRevB.74.224412,cohen1992origin}, the ELF shows that a lobe-like
lone-pair resides on one side of Bi and Pb ions in all three polar
metallic structures, which is the driving force to break inversion
symmetry. On the other hand, ELF in the corresponding centrosymmetric
structures shows spherical-symmetric feature, which is implied in
panels \textbf{d-f} of Fig.~\ref{fig:6s}.  Furthermore, we calculate the
total energy variation as a function of normalized polar displacement
$\lambda$ from the centrosymmetric structures to the polar structures
(see panels \textbf{g-i} of Fig.~\ref{fig:6s}).  In all three cases, the
energy curve monotonically decreases from the centrosymmetric
structure to the polar structure, which indicates a continuous and
spontaneous phase transition below a critical temperature (satisfying
Anderson's and Blount's criterion of a ferroelectric-like
metal~\cite{Anderson1965}).  The energy difference between the polar
structure and the corresponding centrosymmetric structure of \BPTO~in
all three cases is larger than that of LiOsO$_3$ (about 25 meV f.u.$^{-1}$),
implying that the structural transition temperature of \BPTO~is higher
than that of LiOsO$_3$~\cite{Wojde2013arxiv}. We note that the above
second-order structural phase transition is a key property to
distinguish instrinsic polar metals from degenerately doped
ferroelectrics
~\cite{PhysRevLett.109.247601,ZhaoHJ-PhysRevB.97.054107,gu2017coexistence,PhysRevB.94.224107,2D-FTJ-SnSe,PhysRevMaterials.3.054405},
because realistic dopants (cation substitution or oxygen vacancies)
make the crystal symmetry of doped ferroelectrics ill-defined and
correspondingly there is no well-defined continuous structural phase
transition at finite temperatures.

\subsection*{Switching barrier of BPTO thin films in a heterostructure}

Next we study BPTO thin films. The $Pmm2$ perovskite structure of
BPTO, which has a layered Bi/Pb ordering, is highly suitable for thin
film growth and can be stabilized on a perovskite oxide substrate
having a lattice constant of 3.98~\AA~or larger. The Bi/Pb stacking
direction in the $Pmm2$ perovskite structure is chosen as the
$z$-axis, while the polar displacements are in the $xy$-plane.  In
addition, we find that constrained by an in-plane lattice constant of
3.98~\AA~or larger, ferroelectric PbTiO$_3$ is under tensile strain
and favors an in-plane polarization over an out-of-plane
polarization~(Supplementary Figure 4).  Therefore, we study a BPTO/PbTiO$_3$
heterostructure, in which both \BPTO~polar displacements and PbTiO$_3$
polarization are parallel to the interface. We will show that
different from previously studied ferroelectric/polar-metal heterostructures
~\cite{xiang2015prb,PuggioniJAP-2018,Puggioni2015,Filippetti2016NatComm}, multiple states with different relative
orientations of BiPbTi$_2$O$_6$ polar displacements and PbTiO$_3$
polarization can be stabilized. When an electric field is applied
to switch the polarization of PbTiO$_3$, a finite energy barrier
exists for BPTO to change its polar displacements by
180$^{\circ}$. If the temperature is high enough that
thermal fluctuations can overcome the energy barrier, the
polar displacements of BPTO will follow the change of PbTiO$_3$
polarization. Otherwise, the polar displacements of BPTO
stay put and form different configurations when the external electric
field changes the direction of PbTiO$_3$ polarization.

\begin{figure}[!tp]
  \includegraphics[angle=0,width=0.9\textwidth]{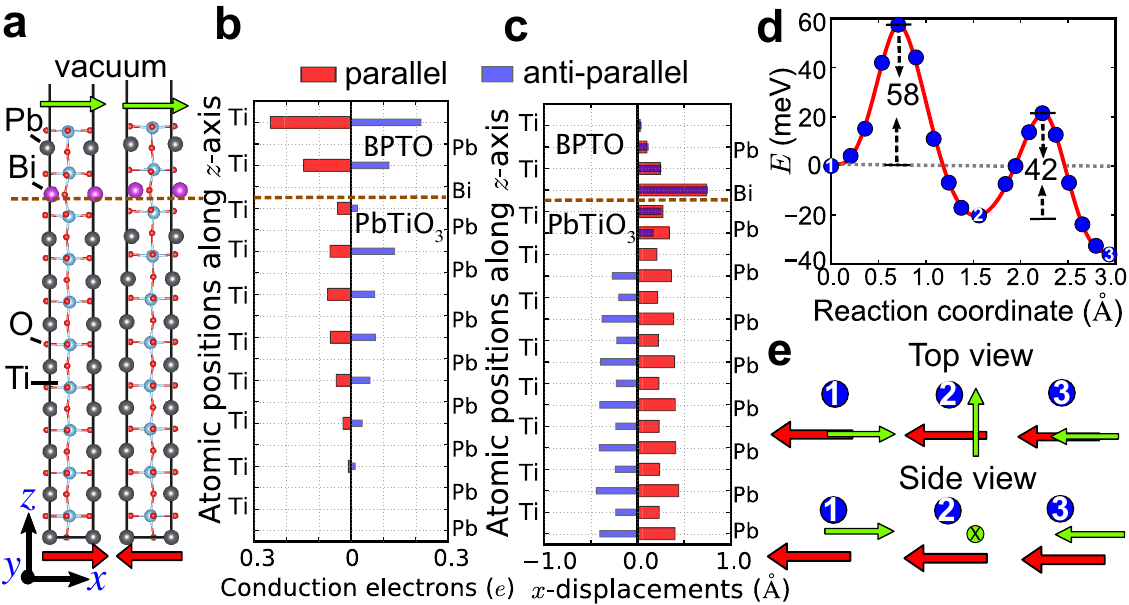}
  \caption{\label{fig:interface}
  \textbf{Switching of polar displacements of BPTO in a BPTO/PbTiO$_3$ heterostructure.}
  \textbf{a} Atomic structure of the \BPTO/\PTO~heterostructure:
  ``parallel state'' (left) and ``anti-parallel state'' (right).
  The red (green) arrow refers to PbTiO$_3$ polarization
  (BPTO polar displacement). \textbf{b}
  Layer-resolved conduction electrons on each Ti atom in parallel
  and anti-parallel states. \textbf{c} Layer-resolved
  polar displacements of metal ions along the $x$-axis in
  parallel and anti-parallel states. In \textbf{a}, \textbf{b} and
  \textbf{c}, the brown dashed lines indicate BPTO/PbTiO$_3$
  interface.  \textbf{d} Calculated energy
  barrier along the transition path from the anti-parallel state to
  the parallel state.  The energy of anti-parallel state is set as the
  zero point.  The black dashed arrows highlight two energy
  barriers. The blue circles represent the images on the transition path
  found in the nudged elastic band calculations. The three energy
  minima along the transition path are schematically shown in
  \textbf{e}: ``1'' is the anti-parallel state, ``3'' is the
  parallel state and ``2'' is a metastable state with BPTO polar
  displacements perpendicular to PbTiO$_3$ polarization in the $xy$ plane.}
\end{figure}

Fig.~\ref{fig:interface}\textbf{a} shows a BPTO/PbTiO$_3$
heterostructure. The in-plane lattice constant is constrained to
4~\AA, which stabilizes both perovskite $Pmm2$ structure of \BPTO~and
an in-plane polarization of PbTiO$_3$.  Experimentally substrates such
as KTaO$_3$ and NdScO$_3$ can provide such a lattice constant.
Fig.~\ref{fig:interface}\textbf{a} shows two different configurations:
on the left (right) is a ``parallel state'' (``anti-parallel state'')
in which PbTiO$_3$ polarization is parallel (anti-parallel) to BPTO
polar displacements. Both configurations are stabilized after
relaxation in our calculations. We first find that one-unit-cell thin
film of \BPTO~is still polar and
metallic. Fig.~\ref{fig:interface}\textbf{b} shows layer-resolved
conduction electrons by integrating the partial density states of
Ti-$d$ orbitals (Supplementary Figure 5).  Conduction electrons are mainly confined
in BPTO with some charge leakage into a few unit cells of
PbTiO$_3$. This charge leakage is due to the proximate effect that
Ti-$d$ states in PbTiO$_3$ are empty while Ti-$d$ states in \BPTO~
nominally have 0.5$e$ per Ti atom. Such a charge leakage can be
effectively prevented by replacing \PTO~with
PbTi$_{1-x}$Zr$_x$O$_3$~\cite{NatCommPZT-hetero}, which is supported by our calculations
in Supplementary Figure 6.
Fig.~\ref{fig:interface}\textbf{c} shows the layer-resolved cation
displacements of Ti and Pb/Bi with respect to O atoms along the
$x$-axis. The polar displacements of Bi and Pb in \BPTO~are almost
bulk-like in both configurations.  We note that the polar property of
\BPTO~is not due to the interfacial coupling with \PTO~(Supplementary Figure 7).  With
\PTO~replaced by a paraelectric SrTiO$_3$ substrate, one-unit-cell
\BPTO~layer still has polar displacements and is metallic~(Supplementary Figure 8).

Next we study thermodynamics and the energy barrier of switching between
``parallel state'' and ``anti-parallel state''.
DFT calculations find that the energy of
``parallel state'' is 37 meV per slab lower than that of ``anti-parallel
state''. This is because in ``anti-parallel state'', a 180$^{\circ}$
domain wall is formed in PbTiO$_3$ close to the interface, which is
clearly seen from Fig.~\ref{fig:interface}\textbf{c}. Forming such a
180$^{\circ}$ domain wall in ferroelectrics increases
energy~\cite{PhysRevB.90.054106,PhysRevB.65.104111}.
Fig.~\ref{fig:interface}\textbf{c} shows that in both configurations,
the interface strongly favors a parallel coupling between \BPTO~polar
displacements and PbTiO$_3$ polarization.

While ``parallel state'' is more stable than ``anti-parallel state'',
``anti-parallel state'' can be stabilized by itself because it is a
local minimum. Therefore a finite energy barrier exists for \BPTO~to
180$^{\circ}$ change its polar displacements from ``anti-parallel
state'' to ``parallel state''. To quantitatively calculate the
  energy barrier and identify a possible switching path for polar
  displacement, we perform the climbing image nudged elastic band
  (NEB) calculations~\cite{climbNEB2000} and use transition state
  theory~\cite{TST-1983}. Transition state theory has been widely used
  in understanding polarization switching in ferroelectric thin
  films~\cite{TADMOR20022989,Wang_20172DMater},
  as well as ferroelectric domain wall motion~\cite{LiXY-JPCC2018}.
We choose ``anti-parallel state'' as the initial state and ``parallel
state'' as the final state. We study a possible switching path in
which \BPTO~ polar displacements are 180$^{\circ}$ ``rotated'' in the
$xy$ plane.  The NEB results are shown in
Fig.~\ref{fig:interface}\textbf{d}.  Along the structural transition
path from ``anti-parallel state'' (labelled as ``1'') to ``parallel
state'' (labelled as ``3''), there is another metastable state
(labelled as ``2'') where \BPTO~polar displacements are perpendicular
to \PTO~polarization in the $xy$ plane
(Fig.~\ref{fig:interface}\textbf{e}).  Between the three stable
states (``1'', ``2'', ``3''), there are two energy barriers. The
larger one, \ie, the energy difference between the anti-parallel state
and the highest saddle point, is 58 meV per slab.

\subsection*{Multifunctions of the BPTO/PTO~heterostructure}

In this section, we discuss potential functions of the BPTO/PTO
heterostructure based on the calculated switching barrier in the
previous section.

\begin{figure}[!t]
  \includegraphics[angle=0,width=0.8\textwidth]{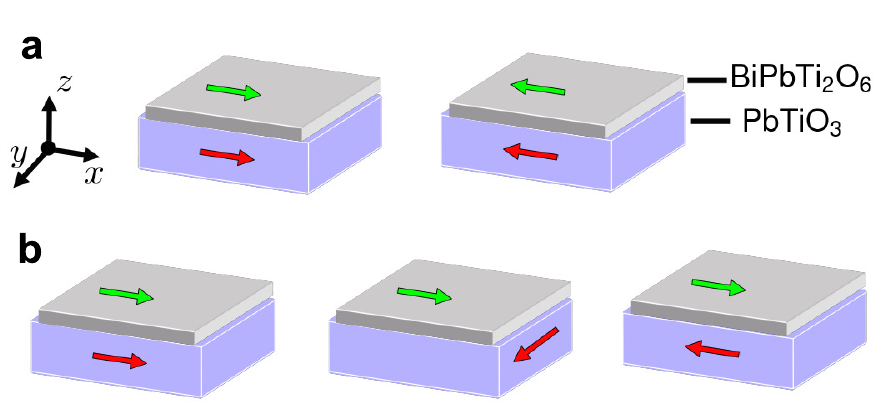}
  \caption{\label{fig:temperature-control}
  \textbf{Multifunctions of the
  BPTO/\PTO~heterostructure.}
  \textbf{a} At high
  temperature, the energy barrier is easily surmounted by thermal fluctuations
  and the polar displacements of \BPTO~can be switched. From the
  left panel to the right panel, as an electric field swiches the
  polarization of \PTO, the \BPTO~thin film follows the change and switches
  its polar displacements via the interfacial coupling. \textbf{b} At low
  temperature, the energy barrier can not be overcome and the
  polar displacements of \BPTO~get ``stuck''. However, an electric
  field can switch the polarization of \PTO~and stabilize multiple
  states with different orientation of \PTO~polarization relative to the
  polar displacements of \BPTO. Each state has different tunnelling barriers.
  From the left panel to the right panel,
  it is ``parallel'', ``perpendicular'',
  and ``anti-parallel'' state.  The red arrow refers to the
  polarization of PbTiO$_3$ thin film. The green arrows refer to the
  polar displacements of BiPbTi$_2$O$_6$ thin film.}
\end{figure}

We first discuss room temperature applications.  The switching barrier
of BPTO is about 58 meV per slab.
From transition state
  theory~\cite{TST-1983}, at a given temperature $T$, an energy
  barrier $\Delta E$ with a magnitude of a few $k_\mathrm{B}T$ can be easily
  surmounted~\cite{surmount} ($k_\mathrm{B}$ is the Boltzmann constant).  Room
  temperature $T = 300$ K is about 26 meV. Our energy barrier is about
  twice room temperature and therefore room temperature is sufficient
  to overcome the barrier.
This implies that the interfacial coupling
at the BPTO/PbTiO$_3$ interface enables an electric field to first
switch PbTiO$_3$ polarization and subsequently drive \BPTO~to
180$^{\circ}$ change its polar displacements. This realizes an
electrically switchable bi-state in the new polar metal \BPTO at room
temperature.  We note that the transition path chosen in the NEB
calculation is only one possiblity. The actual transition path could
be different from the one in our study and the resulting energy
barrier should be even lower, which will make the switching of BPTO
polar displacements more feasible.
Fig.~\ref{fig:temperature-control}\textbf{a} schematically shows how
we can use PbTiO$_3$ polarization to control the polar displacements
of BPTO at room temperature.

\begin{figure}[!t]
    \includegraphics[angle=0,width=0.5\textwidth]{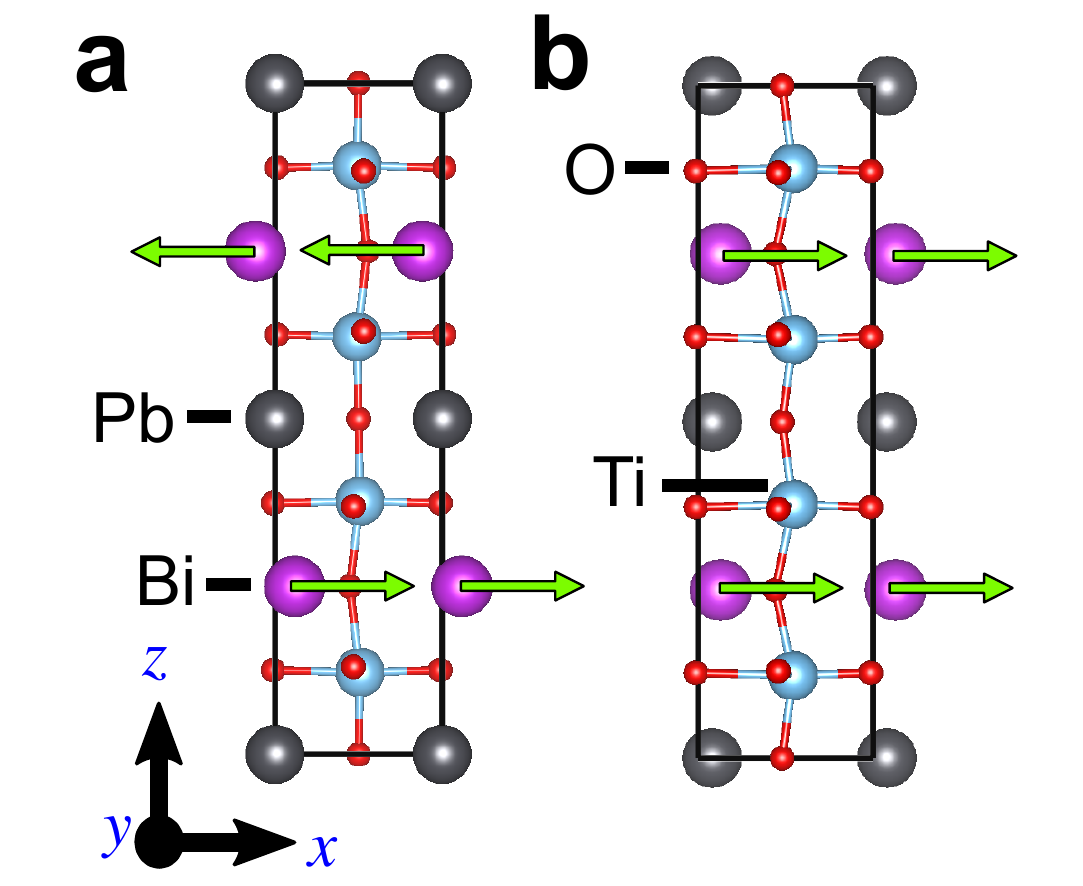}
    \caption{\label{fig:Pmma-anti-polar}
    \textbf{The comparison between anti-polar and polar phases of BPTO.}
    \textbf{a} The unit cell structure of anti-polar phase ${Pmma}$ of BPTO.
    \textbf{b} The two unit cells structure of polar phase ${Pmm2}$ of BPTO.
    The green arrows indicate the polar displacements of the Bi atoms.
    }
\end{figure}

The above switching mechanism is also applicable to multi-layer
BPTO~thin films. The mechanism is as follows. Our calculations find
that bulk BPTO~is more stable in the polar $Pmm2$ perovskite structure
than the anti-polar $Pmma$ perovskite structure by 65 meV f.u.$^{-1}$.  The
anti-polar $Pmma$ perovskite structure is shown in
Fig.~\ref{fig:Pmma-anti-polar}\textbf{a}. The polar ${Pmm2}$
perovskite structure is shown in
Fig.~\ref{fig:Pmma-anti-polar}\textbf{b} for comparison. Therefore,
for multi-layer BPTO~thin films, once the bottom layer of BPTO~is
180$^{\circ}$ switched via the interfacial coupling, the remaining
layers of BPTO~will be driven by thermodynamics to change their polar
displacements in a layer-by-layer manner to avoid an anti-polar state
in the film. The above physical picture is computationally confirmed
in Supplementary Figure 9. However, for device applications, \BPTO~thin films
of  single-unit-cell thick are most desirable, in analogy to two-dimensional
Van der Waals materials~\cite{Fei2018}.

Next we discuss low temperature applications. At sufficiently low temperatures
where the energy barrier is much larger than $k_\mathrm{B}T$, the interfacial
coupling can not drive polar metals to change their polar
displacements when an electric field switches the polarization of
ferroelectrics. However, this has interesting implications: as we use
the electric field to change the direction of PbTiO$_3$ polarization,
we can individually stabilize multiple configurations in which the BPTO
polar displacements are ``parallel'', ``perpendicular'' and
``anti-parallel'' to the PbTiO$_3$ polarization (shown in
Fig.~\ref{fig:temperature-control}\textbf{b}).  Each configuration has
different tunnelling resistance across ferroelectric insulators,
because BPTO polar displacements and PbTiO$_3$ polarization have
different relative orientation. As we use an electric field to change
the direction of ferroelectric polarization (polar displacements do
not follow due to low temperatures), we can tune tunnelling barriers
between different states and therefore the BPTO/PbTiO$_3$
heterostructure can be used in multi-state memory devices.

In conclusion, we demonstrate the power of
  first-principles high-throughput screening in designing new
  functional materials and in particular predict a new polar metal
  BPTO by utilizing the Bi/Pb lone-pair electrons. The three
lowest-energy structures of BPTO are all polar and metallic
(post-perovskite $Pmm2$, perovskite $Pmm2$ and perovskite $Pmn2_1$),
which can be transformed among each other via pressure or strain. In
the perovskite structures, Bi$^{3+}$ and Pb$^{2+}$ enforce a
fractional valence $3.5+$ on Ti, which leads to conduction. In the
post-perovskite structure, strong hybridization between Pb/Bi $6p$ and
O $2p$ states induces a finite density of states at the Fermi level.
In a BPTO/PbTiO$_3$ heterostructures, at room temperature the
interfacial coupling can overcome the switching barrier,
which enables an electric field to first switch
PbTiO$_3$ polarization and subsequently drive BPTO to 180$^{\circ}$
flip its polar displacements. This realizes an electrically switchable
bi-state in the new polar metal BPTO.
The switching method is applicable to other layered polar
metals~\cite{Puggioni2014NC}. At low temperature, an electric
field can control the direction of PbTiO$_3$ polarization and
stabilize multi states in which PbTiO$_3$ polarization and
BPTO polar displacements have different relative orientations,
implying different tunnelling resistance. This property can be used
in tunable multi-state memory devices.
We hope this work will stimulate experimentalists
to synthesize the new polar metal in both bulk and thin-film forms.

\section*{M\lowercase{ethods}}
\label{sec:Methods}
\subsection*{First-principles calculations}

For bulk structures, density functional theory (DFT)
calculations are performed using a plane wave basis set and
projector-augmented wave method~\cite{Blochl1994}, as implemented in
the Vienna Ab-initio Simulation Package
(VASP)~\cite{Kresse1996,Kresse-PRB-1996}.  PBEsol, a revised
Perdew-Burke-Ernzerhof (PBE) generalized gradient approximation for
improving equilibrium properties of densely-packed
solids~\cite{PhysRevLett.100.136406PBEsol}, is used as the exchange
correlation functional and has been applied successfully to
interpreting the experimental observations of polar metal LiOsO$_3$ in
our previous work~\cite{aulesti2018APL}. The Brillouin zone
integration is performed with a Gaussian smearing of 0.05 eV
over a $\Gamma$-centered \textbf{k}-mesh up to 12 $\times$ 12 $\times$
12 and a 600 eV plane-wave cutoff. The threshold of energy
convergence is $10^{-6}$~eV. Hubbard $U$ corrections are also considered
in our calculations to model the effects of strong
correlation on electronic and magnetic properties. The
rotationally-invariant approach of Hubbard $U$ proposed by
Dudarev~\etal~\cite{Dudarev-LDAU-PRB1998}~is used in our DFT+$U$
calculations. Spin-orbit coupling (SOC) is also considered to
study electronic structure in our DFT+$U$+SOC calculations
~\cite{Giovannetti2014}.

For the calculations of
\BPTO/\PTO~structures, a $\Gamma$-centered \textbf{k}-mesh of 10
$\times$ 10 $\times$ 1 is used. The periodic slabs are separated by
vacuum of 20~\AA~thick to diminish the interaction between them.
Since asymmetrical interface modelling is used in our calculations, we
employ dipole correction to eliminate the artificial electric field in
the vacuum~\cite{PhysRevB.51.4014,PhysRevB.46.16067}.  In all the
interface calculations, the in-plane lattice constant is fixed to be
4~\AA~and the bottom layer of PbTiO$_3$ is fixed to simulate the
bulk-like interior that is under tensile strain. All the other atoms
are fully relaxed along the three axes.
We consider two possible terminations of the heterostructure, \ie,
BaO- and BiO-terminations. The former one is less stable
than the latter one by $\sim$ 220 meV per slab. Hence, we only report the
BiO-terminated \BPTO/\PTO~interface in our study.

The energy barriers between
the parallel and anti-parallel states, as well as the saddle points
along the transition path are found by the nudged elastic band (NEB)
calculations through the climbing image NEB
method~\cite{climbNEB2000}. In NEB calculations, a set of
intermediate structures (\ie, images) between the initial state
(anti-parallel state) and the final state (parallel state) are
generated. They are iteratively adjusted so as to minimize the
increase in energy along the transition path.

The electron localization function in our study, which is used
to visualize lone-pair electrons in the real space
is defined as~\cite{silvi1994Nature}:
\begin{equation}
{\rm ELF} = \Big[1+\Big(\frac{D}{D_h}\Big)^2\Big]^{-1}
\end{equation}
where
\begin{equation}
D = \frac{1}{2}\sum_i |\nabla \phi_i|^2 - \frac{1}{8}\frac{|\nabla \rho|^2}{\rho}
\end{equation}
and
\begin{equation}
D_h = \frac{3}{10}(3\pi^2)^{\frac{5}{3}} \rho^{\frac{5}{3}}
\end{equation}
Here $\rho$ is the electron density and ${\phi_i}$ are
the Kohn-Sham wave functions.

\subsection*{Crystal structure search}
The crystal structure search for bulk~\BPTO~is carried out using the
particle swarm optimization algorithm implemented in CALYPSO
code~\cite{Wang2010, Wang2012}, with the assistance of
CrySPY~\cite{Yamashita-CrySPY}.  More than 1000 structures (50$\%$
10-atom BiPbTi$_2$O$_6$ and 50$\%$ 20-atom Bi$_2$Pb$_2$Ti$_4$O$_{12}$)
are created in 20 generations.  The structural optimization and
computation of total energy are performed using VASP.  In the first
step of high-throughput screening of these 1000 crystal structure, we
used non-spin polarized calculations with the exchange-correlation
functional of PBEsol.  The cutoff energy of 450 eV and the
\textbf{k}-mesh grid density is about 2000 per atom.  In the second step,
the lowest 50 structures are re-calculated by the spin-polarized
calculations in which the cutoff energy is increased to 600 eV and the
\textbf{k}-mesh grid density is more than 2500 per atom.  We consider
ferromagnetic ordering and different types of antiferromagnetic
orderings such as $A$-type, $C$-type and $G$-type~\cite{EPL-DING2012}
to examine possible magnetic properties. The global structure
search is performed under 0 GPa.  The five lowest energy structures
after screening are also studied under pressure.  The space groups of
the predicted crystal structures are examined by the FINDSYM
code~\cite{aroyo2011crystallography}.

\subsection*{Visualization}
We use software VESTA to show crystal structures and real-space electron
localized functions~\cite{momma2011vesta}.

\section*{D\lowercase{ATA AVAILABILITY}}
The authors declare that all the data supporting the findings of this study
are available within the paper and its \SM.

\section*{C\lowercase{ode availability}}
The high-throughput crystal structural predictions were carried out
using the proprietary code VASP~\cite{Kresse1996,Kresse-PRB-1996}, with the
combination of CALYPSO~\cite{Wang2010, Wang2012} and
CrySPY~\cite{Yamashita-CrySPY}.  CALYPSO (http://www.calypso.cn/) is
freely distributed on academic use under the license of Copyright
Protection Center of China (registration No. 2010SR028200 and
classification No. 61000-7500).  CrySPY
(https://github.com/Tomoki-YAMASHITA/CrySPY) is released under the
Massachusetts Institute of Technology (MIT) License and is open
source.  The electronic structure calculations were all performed
using VASP.  The thermal properties are calculated by
Phonopy~\cite{Togo-phonopy2015}.  Phonopy
(https://github.com/atztogo/phonopy) is released under the
BSD-3-Clause License and is open source.  The software
VESTA~\cite{momma2011vesta} is distributed free of charge for academic
users under the VESTA License
(https://jp-minerals.org/vesta/jp/download.html).

\section*{A\lowercase{CKNOWLEDGEMENTS}}
We thank Kevin Garrity, Hongjun Xiang and T. Yamashita for valuable
discussions.  We acknowledge support from National Natural Science
Foundation of China (No. 11774236), Pujiang Talents program
(No. 17PJ1407300), the Seed Grants of NYU-ECNU Joint Research
Institutes and the 2019 University Research Challenge Fund.
This research was carried out on the High Performance Computing resources
at New York University New York, Abu Dhabi and Shanghai.


\section*{A\lowercase{UTHOR CONTRIBUTIONS}}
Y.-W.F. and H.C. designed the project,
performed the calculations, analyzed the results, and wrote the manuscript.

\section*{A\lowercase{DDITIONAL INFORMATION}}
\textbf{Supplementary information} accompanies the paper on the \textit{Communications Materials} website (DOI).

\textbf{Competing interests}: We declare that none of the authors have competing financial or non-financial interests.


\begin{thebibliography}{66}%
\makeatletter
\providecommand \@ifxundefined [1]{%
 \@ifx{#1\undefined}
}%
\providecommand \@ifnum [1]{%
 \ifnum #1\expandafter \@firstoftwo
 \else \expandafter \@secondoftwo
 \fi
}%
\providecommand \@ifx [1]{%
 \ifx #1\expandafter \@firstoftwo
 \else \expandafter \@secondoftwo
 \fi
}%
\providecommand \natexlab [1]{#1}%
\providecommand \enquote  [1]{``#1''}%
\providecommand \bibnamefont  [1]{#1}%
\providecommand \bibfnamefont [1]{#1}%
\providecommand \citenamefont [1]{#1}%
\providecommand \href@noop [0]{\@secondoftwo}%
\providecommand \href [0]{\begingroup \@sanitize@url \@href}%
\providecommand \@href[1]{\@@startlink{#1}\@@href}%
\providecommand \@@href[1]{\endgroup#1\@@endlink}%
\providecommand \@sanitize@url [0]{\catcode `\\12\catcode `\$12\catcode
  `\&12\catcode `\#12\catcode `\^12\catcode `\_12\catcode `\%12\relax}%
\providecommand \@@startlink[1]{}%
\providecommand \@@endlink[0]{}%
\providecommand \url  [0]{\begingroup\@sanitize@url \@url }%
\providecommand \@url [1]{\endgroup\@href {#1}{\urlprefix }}%
\providecommand \urlprefix  [0]{URL }%
\providecommand \Eprint [0]{\href }%
\providecommand \doibase [0]{http://dx.doi.org/}%
\providecommand \selectlanguage [0]{\@gobble}%
\providecommand \bibinfo  [0]{\@secondoftwo}%
\providecommand \bibfield  [0]{\@secondoftwo}%
\providecommand \translation [1]{[#1]}%
\providecommand \BibitemOpen [0]{}%
\providecommand \bibitemStop [0]{}%
\providecommand \bibitemNoStop [0]{.\EOS\space}%
\providecommand \EOS [0]{\spacefactor3000\relax}%
\providecommand \BibitemShut  [1]{\csname bibitem#1\endcsname}%
\let\auto@bib@innerbib\@empty
\bibitem [{\citenamefont {Shi}\ \emph {et~al.}(2013)\citenamefont {Shi},
  \citenamefont {Guo}, \citenamefont {Wang}, \citenamefont {Princep},
  \citenamefont {Khalyavin}, \citenamefont {Manuel}, \citenamefont {Michiue},
  \citenamefont {Sato}, \citenamefont {Tsuda}, \citenamefont {Yu},
  \citenamefont {Arai}, \citenamefont {Shirako}, \citenamefont {Akaogi},
  \citenamefont {Wang}, \citenamefont {Yamaura},\ and\ \citenamefont
  {Boothroyd}}]{Shi2013}%
  \BibitemOpen
  \bibfield  {author} {\bibinfo {author} {\bibfnamefont {Youguo}\ \bibnamefont
  {Shi}}, \bibinfo {author} {\bibfnamefont {Yanfeng}\ \bibnamefont {Guo}},
  \bibinfo {author} {\bibfnamefont {Xia}\ \bibnamefont {Wang}}, \bibinfo
  {author} {\bibfnamefont {Andrew~J.}\ \bibnamefont {Princep}}, \bibinfo
  {author} {\bibfnamefont {Dmitry}\ \bibnamefont {Khalyavin}}, \bibinfo
  {author} {\bibfnamefont {Pascal}\ \bibnamefont {Manuel}}, \bibinfo {author}
  {\bibfnamefont {Yuichi}\ \bibnamefont {Michiue}}, \bibinfo {author}
  {\bibfnamefont {Akira}\ \bibnamefont {Sato}}, \bibinfo {author}
  {\bibfnamefont {Kenji}\ \bibnamefont {Tsuda}}, \bibinfo {author}
  {\bibfnamefont {Shan}\ \bibnamefont {Yu}}, \bibinfo {author} {\bibfnamefont
  {Masao}\ \bibnamefont {Arai}}, \bibinfo {author} {\bibfnamefont {Yuichi}\
  \bibnamefont {Shirako}}, \bibinfo {author} {\bibfnamefont {Masaki}\
  \bibnamefont {Akaogi}}, \bibinfo {author} {\bibfnamefont {Nanlin}\
  \bibnamefont {Wang}}, \bibinfo {author} {\bibfnamefont {Kazunari}\
  \bibnamefont {Yamaura}}, \ and\ \bibinfo {author} {\bibfnamefont {Andrew~T.}\
  \bibnamefont {Boothroyd}},\ }\bibfield  {title} {\enquote {\bibinfo {title}
  {A ferroelectric-like structural transition in a metal},}\ }\href@noop {}
  {\bibfield  {journal} {\bibinfo  {journal} {Nat. Mater.}\ }\textbf {\bibinfo
  {volume} {12}},\ \bibinfo {pages} {1024--1027} (\bibinfo {year}
  {2013})}\BibitemShut {NoStop}%
\bibitem [{\citenamefont {Xiang}(2014)}]{xiang2015prb}%
  \BibitemOpen
  \bibfield  {author} {\bibinfo {author} {\bibfnamefont {H.~J.}\ \bibnamefont
  {Xiang}},\ }\bibfield  {title} {\enquote {\bibinfo {title} {Origin of polar
  distortion in {LiNbO$_3$}-type ``ferroelectric'' metals: Role of {$A$}-site
  instability and short-range interactions},}\ }\href@noop {} {\bibfield
  {journal} {\bibinfo  {journal} {Phys. Rev. B}\ }\textbf {\bibinfo {volume}
  {90}},\ \bibinfo {pages} {094108} (\bibinfo {year} {2014})}\BibitemShut
  {NoStop}%
\bibitem [{\citenamefont {Puggioni}\ and\ \citenamefont
  {Rondinelli}(2014)}]{Puggioni2014NC}%
  \BibitemOpen
  \bibfield  {author} {\bibinfo {author} {\bibfnamefont {Danilo}\ \bibnamefont
  {Puggioni}}\ and\ \bibinfo {author} {\bibfnamefont {James~M}\ \bibnamefont
  {Rondinelli}},\ }\bibfield  {title} {\enquote {\bibinfo {title} {Designing a
  robustly metallic noncenstrosymmetric ruthenate oxide with large thermopower
  anisotropy},}\ }\href@noop {} {\bibfield  {journal} {\bibinfo  {journal}
  {Nat. Commun.}\ }\textbf {\bibinfo {volume} {5}},\ \bibinfo {pages} {3432}
  (\bibinfo {year} {2014})}\BibitemShut {NoStop}%
\bibitem [{\citenamefont {Filippetti}\ \emph {et~al.}(2016)\citenamefont
  {Filippetti}, \citenamefont {Fiorentini}, \citenamefont {Ricci},
  \citenamefont {Delugas},\ and\ \citenamefont
  {{\'I}{\~n}iguez}}]{Filippetti2016NatComm}%
  \BibitemOpen
  \bibfield  {author} {\bibinfo {author} {\bibfnamefont {Alessio}\ \bibnamefont
  {Filippetti}}, \bibinfo {author} {\bibfnamefont {Vincenzo}\ \bibnamefont
  {Fiorentini}}, \bibinfo {author} {\bibfnamefont {Francesco}\ \bibnamefont
  {Ricci}}, \bibinfo {author} {\bibfnamefont {Pietro}\ \bibnamefont {Delugas}},
  \ and\ \bibinfo {author} {\bibfnamefont {Jorge}\ \bibnamefont
  {{\'I}{\~n}iguez}},\ }\bibfield  {title} {\enquote {\bibinfo {title}
  {Prediction of a native ferroelectric metal},}\ }\href@noop {} {\bibfield
  {journal} {\bibinfo  {journal} {Nat. Commun.}\ }\textbf {\bibinfo {volume}
  {7}},\ \bibinfo {pages} {11211} (\bibinfo {year} {2016})}\BibitemShut
  {NoStop}%
\bibitem [{\citenamefont {Kim}\ \emph {et~al.}(2016)\citenamefont {Kim},
  \citenamefont {Puggioni}, \citenamefont {Yuan}, \citenamefont {Xie},
  \citenamefont {Zhou}, \citenamefont {Campbell}, \citenamefont {Ryan},
  \citenamefont {Choi}, \citenamefont {Kim}, \citenamefont {Patzner},
  \citenamefont {Ryu}, \citenamefont {Podkaminer}, \citenamefont {Irwin},
  \citenamefont {Ma}, \citenamefont {Fennie}, \citenamefont {Rzchowski},
  \citenamefont {Pan}, \citenamefont {Gopalan}, \citenamefont {Rondinelli},\
  and\ \citenamefont {Eom}}]{kim2016polar}%
  \BibitemOpen
  \bibfield  {author} {\bibinfo {author} {\bibfnamefont {TH}~\bibnamefont
  {Kim}}, \bibinfo {author} {\bibfnamefont {D}~\bibnamefont {Puggioni}},
  \bibinfo {author} {\bibfnamefont {Y}~\bibnamefont {Yuan}}, \bibinfo {author}
  {\bibfnamefont {L}~\bibnamefont {Xie}}, \bibinfo {author} {\bibfnamefont
  {H}~\bibnamefont {Zhou}}, \bibinfo {author} {\bibfnamefont {N}~\bibnamefont
  {Campbell}}, \bibinfo {author} {\bibfnamefont {PJ}~\bibnamefont {Ryan}},
  \bibinfo {author} {\bibfnamefont {Y}~\bibnamefont {Choi}}, \bibinfo {author}
  {\bibfnamefont {J-W}\ \bibnamefont {Kim}}, \bibinfo {author} {\bibfnamefont
  {JR}~\bibnamefont {Patzner}}, \bibinfo {author} {\bibfnamefont
  {S}~\bibnamefont {Ryu}}, \bibinfo {author} {\bibfnamefont {JP}~\bibnamefont
  {Podkaminer}}, \bibinfo {author} {\bibfnamefont {J}~\bibnamefont {Irwin}},
  \bibinfo {author} {\bibfnamefont {Y}~\bibnamefont {Ma}}, \bibinfo {author}
  {\bibfnamefont {CJ}~\bibnamefont {Fennie}}, \bibinfo {author} {\bibfnamefont
  {MS}~\bibnamefont {Rzchowski}}, \bibinfo {author} {\bibfnamefont
  {XQ}~\bibnamefont {Pan}}, \bibinfo {author} {\bibfnamefont {V}~\bibnamefont
  {Gopalan}}, \bibinfo {author} {\bibfnamefont {JM}~\bibnamefont {Rondinelli}},
  \ and\ \bibinfo {author} {\bibfnamefont {CB}~\bibnamefont {Eom}},\ }\bibfield
   {title} {\enquote {\bibinfo {title} {Polar metals by geometric design},}\
  }\href@noop {} {\bibfield  {journal} {\bibinfo  {journal} {Nature}\ }\textbf
  {\bibinfo {volume} {533}},\ \bibinfo {pages} {68} (\bibinfo {year}
  {2016})}\BibitemShut {NoStop}%
\bibitem [{\citenamefont {Benedek}\ and\ \citenamefont
  {Birol}(2016)}]{Benedek-JMCC-2016}%
  \BibitemOpen
  \bibfield  {author} {\bibinfo {author} {\bibfnamefont {Nicole~A.}\
  \bibnamefont {Benedek}}\ and\ \bibinfo {author} {\bibfnamefont {Turan}\
  \bibnamefont {Birol}},\ }\bibfield  {title} {\enquote {\bibinfo {title}
  {`ferroelectric' metals reexamined: fundamental mechanisms and design
  considerations for new materials},}\ }\href@noop {} {\bibfield  {journal}
  {\bibinfo  {journal} {J. Mater. Chem. C}\ }\textbf {\bibinfo {volume} {4}},\
  \bibinfo {pages} {4000--4015} (\bibinfo {year} {2016})}\BibitemShut {NoStop}%
\bibitem [{\citenamefont {Luo}\ \emph {et~al.}(2017)\citenamefont {Luo},
  \citenamefont {Xu},\ and\ \citenamefont {Xiang}}]{PhysRevB.96.235415}%
  \BibitemOpen
  \bibfield  {author} {\bibinfo {author} {\bibfnamefont {Wei}\ \bibnamefont
  {Luo}}, \bibinfo {author} {\bibfnamefont {Ke}~\bibnamefont {Xu}}, \ and\
  \bibinfo {author} {\bibfnamefont {Hongjun}\ \bibnamefont {Xiang}},\
  }\bibfield  {title} {\enquote {\bibinfo {title} {Two-dimensional
  hyperferroelectric metals: A different route to ferromagnetic-ferroelectric
  multiferroics},}\ }\href@noop {} {\bibfield  {journal} {\bibinfo  {journal}
  {Phys. Rev. B}\ }\textbf {\bibinfo {volume} {96}},\ \bibinfo {pages} {235415}
  (\bibinfo {year} {2017})}\BibitemShut {NoStop}%
\bibitem [{\citenamefont {Mochizuki}\ \emph {et~al.}(2018)\citenamefont
  {Mochizuki}, \citenamefont {Kumagai}, \citenamefont {Akamatsu},\ and\
  \citenamefont {Oba}}]{PhysRevMaterials.2.125004}%
  \BibitemOpen
  \bibfield  {author} {\bibinfo {author} {\bibfnamefont {Yasuhide}\
  \bibnamefont {Mochizuki}}, \bibinfo {author} {\bibfnamefont {Yu}~\bibnamefont
  {Kumagai}}, \bibinfo {author} {\bibfnamefont {Hirofumi}\ \bibnamefont
  {Akamatsu}}, \ and\ \bibinfo {author} {\bibfnamefont {Fumiyasu}\ \bibnamefont
  {Oba}},\ }\bibfield  {title} {\enquote {\bibinfo {title} {Polar metallic
  behavior of strained antiperovskites {$A$CNi$_{3}$} ({$A$ = Mg, Zn, and Cd})
  from first principles},}\ }\href@noop {} {\bibfield  {journal} {\bibinfo
  {journal} {Phys. Rev. Materials}\ }\textbf {\bibinfo {volume} {2}},\ \bibinfo
  {pages} {125004} (\bibinfo {year} {2018})}\BibitemShut {NoStop}%
\bibitem [{\citenamefont {Fei}\ \emph {et~al.}(2018)\citenamefont {Fei},
  \citenamefont {Zhao}, \citenamefont {Palomaki}, \citenamefont {Sun},
  \citenamefont {Miller}, \citenamefont {Zhao}, \citenamefont {Yan},
  \citenamefont {Xu},\ and\ \citenamefont {Cobden}}]{Fei2018}%
  \BibitemOpen
  \bibfield  {author} {\bibinfo {author} {\bibfnamefont {Zaiyao}\ \bibnamefont
  {Fei}}, \bibinfo {author} {\bibfnamefont {Wenjin}\ \bibnamefont {Zhao}},
  \bibinfo {author} {\bibfnamefont {Tauno~A.}\ \bibnamefont {Palomaki}},
  \bibinfo {author} {\bibfnamefont {Bosong}\ \bibnamefont {Sun}}, \bibinfo
  {author} {\bibfnamefont {Moira~K.}\ \bibnamefont {Miller}}, \bibinfo {author}
  {\bibfnamefont {Zhiying}\ \bibnamefont {Zhao}}, \bibinfo {author}
  {\bibfnamefont {Jiaqiang}\ \bibnamefont {Yan}}, \bibinfo {author}
  {\bibfnamefont {Xiaodong}\ \bibnamefont {Xu}}, \ and\ \bibinfo {author}
  {\bibfnamefont {David~H.}\ \bibnamefont {Cobden}},\ }\bibfield  {title}
  {\enquote {\bibinfo {title} {{Ferroelectric switching of a two-dimensional
  metal}},}\ }\href@noop {} {\bibfield  {journal} {\bibinfo  {journal}
  {Nature}\ }\textbf {\bibinfo {volume} {560}},\ \bibinfo {pages} {336}
  (\bibinfo {year} {2018})}\BibitemShut {NoStop}%
\bibitem [{\citenamefont {Fang}\ \emph {et~al.}(2015)\citenamefont {Fang},
  \citenamefont {Ding}, \citenamefont {Tong}, \citenamefont {Zhu},
  \citenamefont {Shen}, \citenamefont {Gong}, \citenamefont {Wan},\ and\
  \citenamefont {Duan}}]{fang2015first}%
  \BibitemOpen
  \bibfield  {author} {\bibinfo {author} {\bibfnamefont {Yue-Wen}\ \bibnamefont
  {Fang}}, \bibinfo {author} {\bibfnamefont {Hang-Chen}\ \bibnamefont {Ding}},
  \bibinfo {author} {\bibfnamefont {Wen-Yi}\ \bibnamefont {Tong}}, \bibinfo
  {author} {\bibfnamefont {Wan-Jiao}\ \bibnamefont {Zhu}}, \bibinfo {author}
  {\bibfnamefont {Xin}\ \bibnamefont {Shen}}, \bibinfo {author} {\bibfnamefont
  {Shi-Jing}\ \bibnamefont {Gong}}, \bibinfo {author} {\bibfnamefont
  {Xian-Gang}\ \bibnamefont {Wan}}, \ and\ \bibinfo {author} {\bibfnamefont
  {Chun-Gang}\ \bibnamefont {Duan}},\ }\bibfield  {title} {\enquote {\bibinfo
  {title} {First-principles studies of multiferroic and magnetoelectric
  materials},}\ }\href@noop {} {\bibfield  {journal} {\bibinfo  {journal} {Sci.
  Bull.}\ }\textbf {\bibinfo {volume} {60}},\ \bibinfo {pages} {156--181}
  (\bibinfo {year} {2015})}\BibitemShut {NoStop}%
\bibitem [{\citenamefont {Wang}\ \emph
  {et~al.}(2012{\natexlab{a}})\citenamefont {Wang}, \citenamefont {Lv},
  \citenamefont {Zhu},\ and\ \citenamefont {Ma}}]{Wang2012}%
  \BibitemOpen
  \bibfield  {author} {\bibinfo {author} {\bibfnamefont {Yanchao}\ \bibnamefont
  {Wang}}, \bibinfo {author} {\bibfnamefont {Jian}\ \bibnamefont {Lv}},
  \bibinfo {author} {\bibfnamefont {Li}~\bibnamefont {Zhu}}, \ and\ \bibinfo
  {author} {\bibfnamefont {Yanming}\ \bibnamefont {Ma}},\ }\bibfield  {title}
  {\enquote {\bibinfo {title} {Calypso: A method for crystal structure
  prediction},}\ }\href@noop {} {\bibfield  {journal} {\bibinfo  {journal}
  {Comput. Phys. Commun.}\ }\textbf {\bibinfo {volume} {183}},\ \bibinfo
  {pages} {2063--2070} (\bibinfo {year} {2012}{\natexlab{a}})}\BibitemShut
  {NoStop}%
\bibitem [{\citenamefont {Glass}\ \emph {et~al.}(2006)\citenamefont {Glass},
  \citenamefont {Oganov},\ and\ \citenamefont {Hansen}}]{USPEX-2006CPC}%
  \BibitemOpen
  \bibfield  {author} {\bibinfo {author} {\bibfnamefont {Colin~W.}\
  \bibnamefont {Glass}}, \bibinfo {author} {\bibfnamefont {Artem~R.}\
  \bibnamefont {Oganov}}, \ and\ \bibinfo {author} {\bibfnamefont {Nikolaus}\
  \bibnamefont {Hansen}},\ }\bibfield  {title} {\enquote {\bibinfo {title}
  {{USPEX}-—{Evolutionary} crystal structure prediction},}\ }\href@noop {}
  {\bibfield  {journal} {\bibinfo  {journal} {Comput. Phys. Commun.}\ }\textbf
  {\bibinfo {volume} {175}},\ \bibinfo {pages} {713--720} (\bibinfo {year}
  {2006})}\BibitemShut {NoStop}%
\bibitem [{\citenamefont {Wei}\ \emph {et~al.}(2018)\citenamefont {Wei},
  \citenamefont {Nukala}, \citenamefont {Salverda}, \citenamefont {Matzen},
  \citenamefont {Zhao}, \citenamefont {Momand}, \citenamefont {Everhardt},
  \citenamefont {Agnus}, \citenamefont {Blake}, \citenamefont {Lecoeur},
  \citenamefont {Kooi}, \citenamefont {{\'I}{\~n}iguez}, \citenamefont
  {Dkhil},\ and\ \citenamefont {Noheda}}]{Wei2018NatMater}%
  \BibitemOpen
  \bibfield  {author} {\bibinfo {author} {\bibfnamefont {Yingfen}\ \bibnamefont
  {Wei}}, \bibinfo {author} {\bibfnamefont {Pavan}\ \bibnamefont {Nukala}},
  \bibinfo {author} {\bibfnamefont {Mart}\ \bibnamefont {Salverda}}, \bibinfo
  {author} {\bibfnamefont {Sylvia}\ \bibnamefont {Matzen}}, \bibinfo {author}
  {\bibfnamefont {Hong~Jian}\ \bibnamefont {Zhao}}, \bibinfo {author}
  {\bibfnamefont {Jamo}\ \bibnamefont {Momand}}, \bibinfo {author}
  {\bibfnamefont {Arnoud~S.}\ \bibnamefont {Everhardt}}, \bibinfo {author}
  {\bibfnamefont {Guillaume}\ \bibnamefont {Agnus}}, \bibinfo {author}
  {\bibfnamefont {Graeme~R.}\ \bibnamefont {Blake}}, \bibinfo {author}
  {\bibfnamefont {Philippe}\ \bibnamefont {Lecoeur}}, \bibinfo {author}
  {\bibfnamefont {Bart~J.}\ \bibnamefont {Kooi}}, \bibinfo {author}
  {\bibfnamefont {Jorge}\ \bibnamefont {{\'I}{\~n}iguez}}, \bibinfo {author}
  {\bibfnamefont {Brahim}\ \bibnamefont {Dkhil}}, \ and\ \bibinfo {author}
  {\bibfnamefont {Beatriz}\ \bibnamefont {Noheda}},\ }\bibfield  {title}
  {\enquote {\bibinfo {title} {A rhombohedral ferroelectric phase in
  epitaxially strained {Hf$_{0.5}$Zr$_{0.5}$O$_2$} thin films},}\ }\href@noop
  {} {\bibfield  {journal} {\bibinfo  {journal} {Nat. Mater.}\ }\textbf
  {\bibinfo {volume} {17}},\ \bibinfo {pages} {1095--1100} (\bibinfo {year}
  {2018})}\BibitemShut {NoStop}%
\bibitem [{\citenamefont {He}\ \emph {et~al.}(2019)\citenamefont {He},
  \citenamefont {Xia}, \citenamefont {Naghavi}, \citenamefont {Ozoli\ifmmode
  \mbox{\c{n}}\else \c{n}\fi{}\ifmmode~\check{s}\else \v{s}\fi{}},\ and\
  \citenamefont {Wolverton}}]{He2019-NatComm-PbTe}%
  \BibitemOpen
  \bibfield  {author} {\bibinfo {author} {\bibfnamefont {Jiangang}\
  \bibnamefont {He}}, \bibinfo {author} {\bibfnamefont {Yi}~\bibnamefont
  {Xia}}, \bibinfo {author} {\bibfnamefont {S.~Shahab}\ \bibnamefont
  {Naghavi}}, \bibinfo {author} {\bibfnamefont {Vidvuds}\ \bibnamefont
  {Ozoli\ifmmode \mbox{\c{n}}\else \c{n}\fi{}\ifmmode~\check{s}\else
  \v{s}\fi{}}}, \ and\ \bibinfo {author} {\bibfnamefont {Chris}\ \bibnamefont
  {Wolverton}},\ }\bibfield  {title} {\enquote {\bibinfo {title} {Designing
  chemical analogs to {PbTe} with intrinsic high band degeneracy and low
  lattice thermal conductivity},}\ }\href@noop {} {\bibfield  {journal}
  {\bibinfo  {journal} {Nat. Commun.}\ }\textbf {\bibinfo {volume} {10}},\
  \bibinfo {pages} {719} (\bibinfo {year} {2019})}\BibitemShut {NoStop}%
\bibitem [{\citenamefont {Zhao}\ \emph {et~al.}(2019)\citenamefont {Zhao},
  \citenamefont {Zhang}, \citenamefont {Yu}, \citenamefont {Xu}, \citenamefont
  {Bergara},\ and\ \citenamefont {Yang}}]{superconductor-PRL2019}%
  \BibitemOpen
  \bibfield  {author} {\bibinfo {author} {\bibfnamefont {Ziyuan}\ \bibnamefont
  {Zhao}}, \bibinfo {author} {\bibfnamefont {Shoutao}\ \bibnamefont {Zhang}},
  \bibinfo {author} {\bibfnamefont {Tong}\ \bibnamefont {Yu}}, \bibinfo
  {author} {\bibfnamefont {Haiyang}\ \bibnamefont {Xu}}, \bibinfo {author}
  {\bibfnamefont {Aitor}\ \bibnamefont {Bergara}}, \ and\ \bibinfo {author}
  {\bibfnamefont {Guochun}\ \bibnamefont {Yang}},\ }\bibfield  {title}
  {\enquote {\bibinfo {title} {Predicted pressure-induced superconducting
  transition in electride {Li$_6$P}},}\ }\href@noop {} {\bibfield  {journal}
  {\bibinfo  {journal} {Phys. Rev. Lett.}\ }\textbf {\bibinfo {volume} {122}},\
  \bibinfo {pages} {097002} (\bibinfo {year} {2019})}\BibitemShut {NoStop}%
\bibitem [{\citenamefont {Wang}\ \emph {et~al.}(2010)\citenamefont {Wang},
  \citenamefont {Lv}, \citenamefont {Zhu},\ and\ \citenamefont
  {Ma}}]{Wang2010}%
  \BibitemOpen
  \bibfield  {author} {\bibinfo {author} {\bibfnamefont {Yanchao}\ \bibnamefont
  {Wang}}, \bibinfo {author} {\bibfnamefont {Jian}\ \bibnamefont {Lv}},
  \bibinfo {author} {\bibfnamefont {Li}~\bibnamefont {Zhu}}, \ and\ \bibinfo
  {author} {\bibfnamefont {Yanming}\ \bibnamefont {Ma}},\ }\bibfield  {title}
  {\enquote {\bibinfo {title} {Crystal structure prediction via particle-swarm
  optimization},}\ }\href@noop {} {\bibfield  {journal} {\bibinfo  {journal}
  {Phys. Rev. B}\ }\textbf {\bibinfo {volume} {82}},\ \bibinfo {pages} {094116}
  (\bibinfo {year} {2010})}\BibitemShut {NoStop}%
\bibitem [{\citenamefont {Yamashita}\ \emph {et~al.}(2018)\citenamefont
  {Yamashita}, \citenamefont {Sato}, \citenamefont {Kino}, \citenamefont
  {Miyake}, \citenamefont {Tsuda},\ and\ \citenamefont
  {Oguchi}}]{Yamashita-CrySPY}%
  \BibitemOpen
  \bibfield  {author} {\bibinfo {author} {\bibfnamefont {Tomoki}\ \bibnamefont
  {Yamashita}}, \bibinfo {author} {\bibfnamefont {Nobuya}\ \bibnamefont
  {Sato}}, \bibinfo {author} {\bibfnamefont {Hiori}\ \bibnamefont {Kino}},
  \bibinfo {author} {\bibfnamefont {Takashi}\ \bibnamefont {Miyake}}, \bibinfo
  {author} {\bibfnamefont {Koji}\ \bibnamefont {Tsuda}}, \ and\ \bibinfo
  {author} {\bibfnamefont {Tamio}\ \bibnamefont {Oguchi}},\ }\bibfield  {title}
  {\enquote {\bibinfo {title} {Crystal structure prediction accelerated by
  bayesian optimization},}\ }\href@noop {} {\bibfield  {journal} {\bibinfo
  {journal} {Phys. Rev. Materials}\ }\textbf {\bibinfo {volume} {2}},\ \bibinfo
  {pages} {013803} (\bibinfo {year} {2018})}\BibitemShut {NoStop}%
\bibitem [{\citenamefont {Murakami}\ \emph {et~al.}(2004)\citenamefont
  {Murakami}, \citenamefont {Hirose}, \citenamefont {Kawamura}, \citenamefont
  {Sata},\ and\ \citenamefont {Ohishi}}]{murakami2004post}%
  \BibitemOpen
  \bibfield  {author} {\bibinfo {author} {\bibfnamefont {Motohiko}\
  \bibnamefont {Murakami}}, \bibinfo {author} {\bibfnamefont {Kei}\
  \bibnamefont {Hirose}}, \bibinfo {author} {\bibfnamefont {Katsuyuki}\
  \bibnamefont {Kawamura}}, \bibinfo {author} {\bibfnamefont {Nagayoshi}\
  \bibnamefont {Sata}}, \ and\ \bibinfo {author} {\bibfnamefont {Yasuo}\
  \bibnamefont {Ohishi}},\ }\bibfield  {title} {\enquote {\bibinfo {title}
  {Post-perovskite phase transition in mgsio3},}\ }\href@noop {} {\bibfield
  {journal} {\bibinfo  {journal} {Science}\ }\textbf {\bibinfo {volume}
  {304}},\ \bibinfo {pages} {855--858} (\bibinfo {year} {2004})}\BibitemShut
  {NoStop}%
\bibitem [{\citenamefont {Ohta}\ \emph {et~al.}(2008)\citenamefont {Ohta},
  \citenamefont {Onoda}, \citenamefont {Hirose}, \citenamefont {Sinmyo},
  \citenamefont {Shimizu}, \citenamefont {Sata}, \citenamefont {Ohishi},\ and\
  \citenamefont {Yasuhara}}]{Science-2018Ohta}%
  \BibitemOpen
  \bibfield  {author} {\bibinfo {author} {\bibfnamefont {Kenji}\ \bibnamefont
  {Ohta}}, \bibinfo {author} {\bibfnamefont {Suzue}\ \bibnamefont {Onoda}},
  \bibinfo {author} {\bibfnamefont {Kei}\ \bibnamefont {Hirose}}, \bibinfo
  {author} {\bibfnamefont {Ryosuke}\ \bibnamefont {Sinmyo}}, \bibinfo {author}
  {\bibfnamefont {Katsuya}\ \bibnamefont {Shimizu}}, \bibinfo {author}
  {\bibfnamefont {Nagayoshi}\ \bibnamefont {Sata}}, \bibinfo {author}
  {\bibfnamefont {Yasuo}\ \bibnamefont {Ohishi}}, \ and\ \bibinfo {author}
  {\bibfnamefont {Akira}\ \bibnamefont {Yasuhara}},\ }\bibfield  {title}
  {\enquote {\bibinfo {title} {The electrical conductivity of post-perovskite
  in earth{\textquoteright}s {D}{\textquoteright}{\textquoteright} layer},}\
  }\href@noop {} {\bibfield  {journal} {\bibinfo  {journal} {Science}\ }\textbf
  {\bibinfo {volume} {320}},\ \bibinfo {pages} {89--91} (\bibinfo {year}
  {2008})}\BibitemShut {NoStop}%
\bibitem [{\citenamefont {Biswas}\ \emph {et~al.}(2017)\citenamefont {Biswas},
  \citenamefont {Yang}, \citenamefont {Ramesh},\ and\ \citenamefont
  {Jeong}}]{BISWAS2017117}%
  \BibitemOpen
  \bibfield  {author} {\bibinfo {author} {\bibfnamefont {Abhijit}\ \bibnamefont
  {Biswas}}, \bibinfo {author} {\bibfnamefont {Chan-Ho}\ \bibnamefont {Yang}},
  \bibinfo {author} {\bibfnamefont {Ramamoorthy}\ \bibnamefont {Ramesh}}, \
  and\ \bibinfo {author} {\bibfnamefont {Yoon~H.}\ \bibnamefont {Jeong}},\
  }\bibfield  {title} {\enquote {\bibinfo {title} {Atomically flat single
  terminated oxide substrate surfaces},}\ }\href@noop {} {\bibfield  {journal}
  {\bibinfo  {journal} {Prog. Surf. Sci.}\ }\textbf {\bibinfo {volume} {92}},\
  \bibinfo {pages} {117--141} (\bibinfo {year} {2017})}\BibitemShut {NoStop}%
\bibitem [{\citenamefont {Zhang}\ \emph {et~al.}(2001)\citenamefont {Zhang},
  \citenamefont {Leinenweber}, \citenamefont {Bauer}, \citenamefont {Garvie},
  \citenamefont {McMillan},\ and\ \citenamefont
  {Wolf}}]{JACS-2001-highpressure}%
  \BibitemOpen
  \bibfield  {author} {\bibinfo {author} {\bibfnamefont {Zhihong}\ \bibnamefont
  {Zhang}}, \bibinfo {author} {\bibfnamefont {Kurt}\ \bibnamefont
  {Leinenweber}}, \bibinfo {author} {\bibfnamefont {Matt}\ \bibnamefont
  {Bauer}}, \bibinfo {author} {\bibfnamefont {Laurence A.~J.}\ \bibnamefont
  {Garvie}}, \bibinfo {author} {\bibfnamefont {Paul~F.}\ \bibnamefont
  {McMillan}}, \ and\ \bibinfo {author} {\bibfnamefont {George~H.}\
  \bibnamefont {Wolf}},\ }\bibfield  {title} {\enquote {\bibinfo {title}
  {High-pressure bulk synthesis of crystalline {C$_6$N$_9$H$_3$$\cdot$HCl}: 
  a novel {C$_3$N$_4$} graphitic derivative},}\ }\href@noop {} {\bibfield
  {journal} {\bibinfo  {journal} {J. Am. Chem. Soc.}\ }\textbf {\bibinfo
  {volume} {123}},\ \bibinfo {pages} {7788--7796} (\bibinfo {year}
  {2001})}\BibitemShut {NoStop}%
\bibitem [{\citenamefont {Klein}\ \emph {et~al.}(2019)\citenamefont {Klein},
  \citenamefont {Altman}, \citenamefont {Saballos}, \citenamefont {Walsh},
  \citenamefont {Tamerius}, \citenamefont {Meng}, \citenamefont {Puggioni},
  \citenamefont {Jacobsen}, \citenamefont {Rondinelli},\ and\ \citenamefont
  {Freedman}}]{PhysRevMaterials.3.064411}%
  \BibitemOpen
  \bibfield  {author} {\bibinfo {author} {\bibfnamefont {R.~A.}\ \bibnamefont
  {Klein}}, \bibinfo {author} {\bibfnamefont {A.~B.}\ \bibnamefont {Altman}},
  \bibinfo {author} {\bibfnamefont {R.~J.}\ \bibnamefont {Saballos}}, \bibinfo
  {author} {\bibfnamefont {J.~P.~S.}\ \bibnamefont {Walsh}}, \bibinfo {author}
  {\bibfnamefont {A.~D.}\ \bibnamefont {Tamerius}}, \bibinfo {author}
  {\bibfnamefont {Y.}~\bibnamefont {Meng}}, \bibinfo {author} {\bibfnamefont
  {D.}~\bibnamefont {Puggioni}}, \bibinfo {author} {\bibfnamefont {S.~D.}\
  \bibnamefont {Jacobsen}}, \bibinfo {author} {\bibfnamefont {J.~M.}\
  \bibnamefont {Rondinelli}}, \ and\ \bibinfo {author} {\bibfnamefont {D.~E.}\
  \bibnamefont {Freedman}},\ }\bibfield  {title} {\enquote {\bibinfo {title}
  {High-pressure synthesis of the {BiVO$_3$} perovskite},}\ }\href@noop {}
  {\bibfield  {journal} {\bibinfo  {journal} {Phys. Rev. Mater.}\ }\textbf
  {\bibinfo {volume} {3}},\ \bibinfo {pages} {064411} (\bibinfo {year}
  {2019})}\BibitemShut {NoStop}%
\bibitem [{\citenamefont {Korotcenkov}(, 2017)}]{korotcenkov2017metal}%
  \BibitemOpen
  \bibfield  {author} {\bibinfo {author} {\bibfnamefont {Ghenadii}\
  \bibnamefont {Korotcenkov}},\ }\href@noop {} {\emph {\bibinfo {title} {Metal
  Oxide-based Thin Film Structures: Formation, Characterization and Application
  of Interface-based Phenomena}}}\ (\bibinfo  {publisher} {Elsevier},\ \bibinfo
  {year} {, 2017})\BibitemShut {NoStop}%
\bibitem [{\citenamefont {Aulesti}\ \emph {et~al.}(2018)\citenamefont
  {Aulesti}, \citenamefont {Cheung}, \citenamefont {Fang}, \citenamefont {He},
  \citenamefont {Yamaura}, \citenamefont {Lai}, \citenamefont {Goh},\ and\
  \citenamefont {Chen}}]{aulesti2018APL}%
  \BibitemOpen
  \bibfield  {author} {\bibinfo {author} {\bibfnamefont {Esteban I~Paredes}\
  \bibnamefont {Aulesti}}, \bibinfo {author} {\bibfnamefont {Yiu~Wing}\
  \bibnamefont {Cheung}}, \bibinfo {author} {\bibfnamefont {Yue-Wen}\
  \bibnamefont {Fang}}, \bibinfo {author} {\bibfnamefont {Jianfeng}\
  \bibnamefont {He}}, \bibinfo {author} {\bibfnamefont {Kazunari}\ \bibnamefont
  {Yamaura}}, \bibinfo {author} {\bibfnamefont {Kwing~To}\ \bibnamefont {Lai}},
  \bibinfo {author} {\bibfnamefont {Swee~K}\ \bibnamefont {Goh}}, \ and\
  \bibinfo {author} {\bibfnamefont {Hanghui}\ \bibnamefont {Chen}},\ }\bibfield
   {title} {\enquote {\bibinfo {title} {Pressure-induced enhancement of
  non-polar to polar transition temperature in metallic {LiOsO$_3$}},}\
  }\href@noop {} {\bibfield  {journal} {\bibinfo  {journal} {Appl. Phys.
  Lett.}\ }\textbf {\bibinfo {volume} {113}},\ \bibinfo {pages} {12902}
  (\bibinfo {year} {2018})}\BibitemShut {NoStop}%
\bibitem [{\citenamefont {Pentcheva}\ and\ \citenamefont
  {Pickett}(2007)}]{PhysRevLett.99.016802}%
  \BibitemOpen
  \bibfield  {author} {\bibinfo {author} {\bibfnamefont {Rossitza}\
  \bibnamefont {Pentcheva}}\ and\ \bibinfo {author} {\bibfnamefont {Warren~E.}\
  \bibnamefont {Pickett}},\ }\bibfield  {title} {\enquote {\bibinfo {title}
  {Correlation-driven charge order at the interface between a mott and a band
  insulator},}\ }\href@noop {} {\bibfield  {journal} {\bibinfo  {journal}
  {Phys. Rev. Lett.}\ }\textbf {\bibinfo {volume} {99}},\ \bibinfo {pages}
  {016802} (\bibinfo {year} {2007})}\BibitemShut {NoStop}%
\bibitem [{\citenamefont {Pavarini}\ \emph {et~al.}(2004)\citenamefont
  {Pavarini}, \citenamefont {Biermann}, \citenamefont {Poteryaev},
  \citenamefont {Lichtenstein}, \citenamefont {Georges},\ and\ \citenamefont
  {Andersen}}]{Pavarini2004}%
  \BibitemOpen
  \bibfield  {author} {\bibinfo {author} {\bibfnamefont {E.}~\bibnamefont
  {Pavarini}}, \bibinfo {author} {\bibfnamefont {S.}~\bibnamefont {Biermann}},
  \bibinfo {author} {\bibfnamefont {A.}~\bibnamefont {Poteryaev}}, \bibinfo
  {author} {\bibfnamefont {A.~I.}\ \bibnamefont {Lichtenstein}}, \bibinfo
  {author} {\bibfnamefont {A.}~\bibnamefont {Georges}}, \ and\ \bibinfo
  {author} {\bibfnamefont {O.~K.}\ \bibnamefont {Andersen}},\ }\bibfield
  {title} {\enquote {\bibinfo {title} {Mott transition and suppression of
  orbital fluctuations in orthorhombic $3{d}^{1}$ perovskites},}\ }\href@noop
  {} {\bibfield  {journal} {\bibinfo  {journal} {Phys. Rev. Lett.}\ }\textbf
  {\bibinfo {volume} {92}},\ \bibinfo {pages} {176403} (\bibinfo {year}
  {2004})}\BibitemShut {NoStop}%
\bibitem [{\citenamefont {Fazekas}(, 1999)}]{patrik1999lecture}%
  \BibitemOpen
  \bibfield  {author} {\bibinfo {author} {\bibfnamefont {Patrick}\ \bibnamefont
  {Fazekas}},\ }\href@noop {} {\emph {\bibinfo {title} {Lecture notes on
  electron correlation and magnetism}}}\ (\bibinfo  {publisher} {World
  scientific},\ \bibinfo {year} {, 1999})\BibitemShut {NoStop}%
\bibitem [{\citenamefont {Janicka}\ \emph {et~al.}(2008)\citenamefont
  {Janicka}, \citenamefont {Velev},\ and\ \citenamefont
  {Tsymbal}}]{Janicka2008}%
  \BibitemOpen
  \bibfield  {author} {\bibinfo {author} {\bibfnamefont {Karolina}\
  \bibnamefont {Janicka}}, \bibinfo {author} {\bibfnamefont {Julian~P.}\
  \bibnamefont {Velev}}, \ and\ \bibinfo {author} {\bibfnamefont {Evgeny~Y.}\
  \bibnamefont {Tsymbal}},\ }\bibfield  {title} {\enquote {\bibinfo {title}
  {Magnetism of {LaAlO$_3$}/{SrTiO$_3$} superlattices},}\ }\href@noop {}
  {\bibfield  {journal} {\bibinfo  {journal} {J. Appl. Phys.}\ }\textbf
  {\bibinfo {volume} {103}},\ \bibinfo {pages} {07B508} (\bibinfo {year}
  {2008})}\BibitemShut {NoStop}%
\bibitem [{\citenamefont {Cohen}(1992)}]{cohen1992origin}%
  \BibitemOpen
  \bibfield  {author} {\bibinfo {author} {\bibfnamefont {Ronald~E}\
  \bibnamefont {Cohen}},\ }\bibfield  {title} {\enquote {\bibinfo {title}
  {Origin of ferroelectricity in perovskite oxides},}\ }\href@noop {}
  {\bibfield  {journal} {\bibinfo  {journal} {Nature}\ }\textbf {\bibinfo
  {volume} {358}},\ \bibinfo {pages} {136} (\bibinfo {year}
  {1992})}\BibitemShut {NoStop}%
\bibitem [{\citenamefont {Ravindran}\ \emph {et~al.}(2006)\citenamefont
  {Ravindran}, \citenamefont {Vidya}, \citenamefont {Kjekshus}, \citenamefont
  {Fjellv\aa{}g},\ and\ \citenamefont {Eriksson}}]{PhysRevB.74.224412}%
  \BibitemOpen
  \bibfield  {author} {\bibinfo {author} {\bibfnamefont {P.}~\bibnamefont
  {Ravindran}}, \bibinfo {author} {\bibfnamefont {R.}~\bibnamefont {Vidya}},
  \bibinfo {author} {\bibfnamefont {A.}~\bibnamefont {Kjekshus}}, \bibinfo
  {author} {\bibfnamefont {H.}~\bibnamefont {Fjellv\aa{}g}}, \ and\ \bibinfo
  {author} {\bibfnamefont {O.}~\bibnamefont {Eriksson}},\ }\bibfield  {title}
  {\enquote {\bibinfo {title} {Theoretical investigation of magnetoelectric
  behavior in {BiFeO$_3$}},}\ }\href@noop {} {\bibfield  {journal} {\bibinfo
  {journal} {Phys. Rev. B}\ }\textbf {\bibinfo {volume} {74}},\ \bibinfo
  {pages} {224412} (\bibinfo {year} {2006})}\BibitemShut {NoStop}%
\bibitem [{\citenamefont {Luo}\ and\ \citenamefont
  {Xiang}(2016)}]{2D-FE-lonepair}%
  \BibitemOpen
  \bibfield  {author} {\bibinfo {author} {\bibfnamefont {Wei}\ \bibnamefont
  {Luo}}\ and\ \bibinfo {author} {\bibfnamefont {Hongjun}\ \bibnamefont
  {Xiang}},\ }\bibfield  {title} {\enquote {\bibinfo {title} {Two-dimensional
  phosphorus oxides as energy and information materials},}\ }\href@noop {}
  {\bibfield  {journal} {\bibinfo  {journal} {Angewandte Chemie}\ }\textbf
  {\bibinfo {volume} {128}},\ \bibinfo {pages} {8717--8722} (\bibinfo {year}
  {2016})}\BibitemShut {NoStop}%
\bibitem [{\citenamefont {Zhao}\ \emph {et~al.}(2018)\citenamefont {Zhao},
  \citenamefont {Filippetti}, \citenamefont {Escorihuela-Sayalero},
  \citenamefont {Delugas}, \citenamefont {Canadell}, \citenamefont {Bellaiche},
  \citenamefont {Fiorentini},\ and\ \citenamefont
  {\'I\~niguez}}]{ZhaoHJ-PhysRevB.97.054107}%
  \BibitemOpen
  \bibfield  {author} {\bibinfo {author} {\bibfnamefont {Hong~Jian}\
  \bibnamefont {Zhao}}, \bibinfo {author} {\bibfnamefont {Alessio}\
  \bibnamefont {Filippetti}}, \bibinfo {author} {\bibfnamefont {Carlos}\
  \bibnamefont {Escorihuela-Sayalero}}, \bibinfo {author} {\bibfnamefont
  {Pietro}\ \bibnamefont {Delugas}}, \bibinfo {author} {\bibfnamefont {Enric}\
  \bibnamefont {Canadell}}, \bibinfo {author} {\bibfnamefont {L.}~\bibnamefont
  {Bellaiche}}, \bibinfo {author} {\bibfnamefont {Vincenzo}\ \bibnamefont
  {Fiorentini}}, \ and\ \bibinfo {author} {\bibfnamefont {Jorge}\ \bibnamefont
  {\'I\~niguez}},\ }\bibfield  {title} {\enquote {\bibinfo {title}
  {Meta-screening and permanence of polar distortion in metallized
  ferroelectrics},}\ }\href@noop {} {\bibfield  {journal} {\bibinfo  {journal}
  {Phys. Rev. B}\ }\textbf {\bibinfo {volume} {97}},\ \bibinfo {pages} {054107}
  (\bibinfo {year} {2018})}\BibitemShut {NoStop}%
\bibitem [{\citenamefont {Gu}\ \emph {et~al.}(2017)\citenamefont {Gu},
  \citenamefont {Jin}, \citenamefont {Ma}, \citenamefont {Zhang}, \citenamefont
  {Gu}, \citenamefont {Ge}, \citenamefont {Wang}, \citenamefont {Wang},
  \citenamefont {Guo},\ and\ \citenamefont {Yang}}]{gu2017coexistence}%
  \BibitemOpen
  \bibfield  {author} {\bibinfo {author} {\bibfnamefont {Jun-xing}\
  \bibnamefont {Gu}}, \bibinfo {author} {\bibfnamefont {Kui-juan}\ \bibnamefont
  {Jin}}, \bibinfo {author} {\bibfnamefont {Chao}\ \bibnamefont {Ma}}, \bibinfo
  {author} {\bibfnamefont {Qing-hua}\ \bibnamefont {Zhang}}, \bibinfo {author}
  {\bibfnamefont {Lin}\ \bibnamefont {Gu}}, \bibinfo {author} {\bibfnamefont
  {Chen}\ \bibnamefont {Ge}}, \bibinfo {author} {\bibfnamefont {Jie-su}\
  \bibnamefont {Wang}}, \bibinfo {author} {\bibfnamefont {Can}\ \bibnamefont
  {Wang}}, \bibinfo {author} {\bibfnamefont {Hai-zhong}\ \bibnamefont {Guo}}, \
  and\ \bibinfo {author} {\bibfnamefont {Guo-zhen}\ \bibnamefont {Yang}},\
  }\bibfield  {title} {\enquote {\bibinfo {title} {Coexistence of polar
  distortion and metallicity in {PbTi$_{1-x}$Nb$_x$O$_3$}},}\ }\href@noop {}
  {\bibfield  {journal} {\bibinfo  {journal} {Phys. Rev. B}\ }\textbf {\bibinfo
  {volume} {96}},\ \bibinfo {pages} {165206} (\bibinfo {year}
  {2017})}\BibitemShut {NoStop}%
\bibitem [{\citenamefont {He}\ and\ \citenamefont
  {Jin}(2016)}]{PhysRevB.94.224107}%
  \BibitemOpen
  \bibfield  {author} {\bibinfo {author} {\bibfnamefont {Xu}~\bibnamefont
  {He}}\ and\ \bibinfo {author} {\bibfnamefont {Kui-juan}\ \bibnamefont
  {Jin}},\ }\bibfield  {title} {\enquote {\bibinfo {title} {Persistence of
  polar distortion with electron doping in lone-pair driven ferroelectrics},}\
  }\href@noop {} {\bibfield  {journal} {\bibinfo  {journal} {Phys. Rev. B}\
  }\textbf {\bibinfo {volume} {94}},\ \bibinfo {pages} {224107} (\bibinfo
  {year} {2016})}\BibitemShut {NoStop}%
\bibitem [{\citenamefont {Shen}\ \emph {et~al.}(2019)\citenamefont {Shen},
  \citenamefont {Fang}, \citenamefont {Tian},\ and\ \citenamefont
  {Duan}}]{2D-FTJ-SnSe}%
  \BibitemOpen
  \bibfield  {author} {\bibinfo {author} {\bibfnamefont {Xin-Wei}\ \bibnamefont
  {Shen}}, \bibinfo {author} {\bibfnamefont {Yue-Wen}\ \bibnamefont {Fang}},
  \bibinfo {author} {\bibfnamefont {Bo-Bo}\ \bibnamefont {Tian}}, \ and\
  \bibinfo {author} {\bibfnamefont {Chun-Gang}\ \bibnamefont {Duan}},\
  }\bibfield  {title} {\enquote {\bibinfo {title} {Two-dimensional
  ferroelectric tunnel junction: the case of monolayer {In:SnSe/SnSe/Sb:SnSe}
  homostructure},}\ }\href@noop {} {\bibfield  {journal} {\bibinfo  {journal}
  {ACS Appl. Electron. Mater.}\ }\textbf {\bibinfo {volume} {1}},\ \bibinfo
  {pages} {1133--1140} (\bibinfo {year} {2019})}\BibitemShut {NoStop}%
\bibitem [{\citenamefont {Baettig}\ \emph {et~al.}(2007)\citenamefont
  {Baettig}, \citenamefont {Seshadri},\ and\ \citenamefont
  {Spaldin}}]{baettig2007anti}%
  \BibitemOpen
  \bibfield  {author} {\bibinfo {author} {\bibfnamefont {Pio}\ \bibnamefont
  {Baettig}}, \bibinfo {author} {\bibfnamefont {Ram}\ \bibnamefont {Seshadri}},
  \ and\ \bibinfo {author} {\bibfnamefont {Nicola~A}\ \bibnamefont {Spaldin}},\
  }\bibfield  {title} {\enquote {\bibinfo {title} {Anti-polarity in ideal
  {BiMnO$_3$}},}\ }\href@noop {} {\bibfield  {journal} {\bibinfo  {journal} {J.
  Am. Chem. Soc.}\ }\textbf {\bibinfo {volume} {129}},\ \bibinfo {pages}
  {9854--9855} (\bibinfo {year} {2007})}\BibitemShut {NoStop}%
\bibitem [{\citenamefont {Goian}\ \emph {et~al.}(2012)\citenamefont {Goian},
  \citenamefont {Kamba}, \citenamefont {Savinov}, \citenamefont {Nuzhnyy},
  \citenamefont {Borodavka}, \citenamefont {Van{\v{e}}k},\ and\ \citenamefont
  {Belik}}]{goian2012absence}%
  \BibitemOpen
  \bibfield  {author} {\bibinfo {author} {\bibfnamefont {V}~\bibnamefont
  {Goian}}, \bibinfo {author} {\bibfnamefont {S}~\bibnamefont {Kamba}},
  \bibinfo {author} {\bibfnamefont {M}~\bibnamefont {Savinov}}, \bibinfo
  {author} {\bibfnamefont {D}~\bibnamefont {Nuzhnyy}}, \bibinfo {author}
  {\bibfnamefont {F}~\bibnamefont {Borodavka}}, \bibinfo {author}
  {\bibfnamefont {P}~\bibnamefont {Van{\v{e}}k}}, \ and\ \bibinfo {author}
  {\bibfnamefont {AA}~\bibnamefont {Belik}},\ }\bibfield  {title} {\enquote
  {\bibinfo {title} {Absence of ferroelectricity in {BiMnO$_3$} ceramics},}\
  }\href@noop {} {\bibfield  {journal} {\bibinfo  {journal} {J. Appl. Phys.}\
  }\textbf {\bibinfo {volume} {112}},\ \bibinfo {pages} {074112} (\bibinfo
  {year} {2012})}\BibitemShut {NoStop}%
\bibitem [{\citenamefont {Seshadri}\ and\ \citenamefont
  {Hill}(2001)}]{lonepair-4}%
  \BibitemOpen
  \bibfield  {author} {\bibinfo {author} {\bibfnamefont {Ram}\ \bibnamefont
  {Seshadri}}\ and\ \bibinfo {author} {\bibfnamefont {Nicola~A.}\ \bibnamefont
  {Hill}},\ }\bibfield  {title} {\enquote {\bibinfo {title} {Visualizing the
  role of {Bi} 6$s$ “lone pairs” in the off-center distortion in
  ferromagnetic {BiMnO$_3$}},}\ }\href@noop {} {\bibfield  {journal} {\bibinfo
  {journal} {Chem. Mater.}\ }\textbf {\bibinfo {volume} {13}},\ \bibinfo
  {pages} {2892--2899} (\bibinfo {year} {2001})}\BibitemShut {NoStop}%
\bibitem [{\citenamefont {Anderson}\ and\ \citenamefont
  {Blount}(1965)}]{Anderson1965}%
  \BibitemOpen
  \bibfield  {author} {\bibinfo {author} {\bibfnamefont {P.~W.}\ \bibnamefont
  {Anderson}}\ and\ \bibinfo {author} {\bibfnamefont {E.~I.}\ \bibnamefont
  {Blount}},\ }\bibfield  {title} {\enquote {\bibinfo {title} {Symmetry
  considerations on martensitic transformations: ``ferroelectric'' metals?}}\
  }\href@noop {} {\bibfield  {journal} {\bibinfo  {journal} {Phys. Rev. Lett.}\
  }\textbf {\bibinfo {volume} {14}},\ \bibinfo {pages} {217} (\bibinfo {year}
  {1965})}\BibitemShut {NoStop}%
\bibitem [{\citenamefont {Wojde\l{}}\ and\ \citenamefont
  {\'I\~niguez}(2014)}]{Wojde2013arxiv}%
  \BibitemOpen
  \bibfield  {author} {\bibinfo {author} {\bibfnamefont {Jacek~C.}\
  \bibnamefont {Wojde\l{}}}\ and\ \bibinfo {author} {\bibfnamefont {Jorge}\
  \bibnamefont {\'I\~niguez}},\ }\bibfield  {title} {\enquote {\bibinfo {title}
  {Testing simple predictors for the temperature of a structural phase
  transition},}\ }\href@noop {} {\bibfield  {journal} {\bibinfo  {journal}
  {Phys. Rev. B}\ }\textbf {\bibinfo {volume} {90}},\ \bibinfo {pages} {014105}
  (\bibinfo {year} {2014})}\BibitemShut {NoStop}%
\bibitem [{\citenamefont {Wang}\ \emph
  {et~al.}(2012{\natexlab{b}})\citenamefont {Wang}, \citenamefont {Liu},
  \citenamefont {Burton}, \citenamefont {Jaswal},\ and\ \citenamefont
  {Tsymbal}}]{PhysRevLett.109.247601}%
  \BibitemOpen
  \bibfield  {author} {\bibinfo {author} {\bibfnamefont {Yong}\ \bibnamefont
  {Wang}}, \bibinfo {author} {\bibfnamefont {Xiaohui}\ \bibnamefont {Liu}},
  \bibinfo {author} {\bibfnamefont {J.~D.}\ \bibnamefont {Burton}}, \bibinfo
  {author} {\bibfnamefont {Sitaram~S.}\ \bibnamefont {Jaswal}}, \ and\ \bibinfo
  {author} {\bibfnamefont {Evgeny~Y.}\ \bibnamefont {Tsymbal}},\ }\bibfield
  {title} {\enquote {\bibinfo {title} {Ferroelectric instability under screened
  coulomb interactions},}\ }\href@noop {} {\bibfield  {journal} {\bibinfo
  {journal} {Phys. Rev. Lett.}\ }\textbf {\bibinfo {volume} {109}},\ \bibinfo
  {pages} {247601} (\bibinfo {year} {2012}{\natexlab{b}})}\BibitemShut
  {NoStop}%
\bibitem [{\citenamefont {Xia}\ \emph {et~al.}(2019)\citenamefont {Xia},
  \citenamefont {Chen},\ and\ \citenamefont
  {Chen}}]{PhysRevMaterials.3.054405}%
  \BibitemOpen
  \bibfield  {author} {\bibinfo {author} {\bibfnamefont {Chengliang}\
  \bibnamefont {Xia}}, \bibinfo {author} {\bibfnamefont {Yue}\ \bibnamefont
  {Chen}}, \ and\ \bibinfo {author} {\bibfnamefont {Hanghui}\ \bibnamefont
  {Chen}},\ }\bibfield  {title} {\enquote {\bibinfo {title} {Coexistence of
  polar displacements and conduction in doped ferroelectrics: An ab initio
  comparative study},}\ }\href@noop {} {\bibfield  {journal} {\bibinfo
  {journal} {Phys. Rev. Mater.}\ }\textbf {\bibinfo {volume} {3}},\ \bibinfo
  {pages} {054405} (\bibinfo {year} {2019})}\BibitemShut {NoStop}%
\bibitem [{\citenamefont {Puggioni}\ \emph {et~al.}(2018)\citenamefont
  {Puggioni}, \citenamefont {Giovannetti},\ and\ \citenamefont
  {Rondinelli}}]{PuggioniJAP-2018}%
  \BibitemOpen
  \bibfield  {author} {\bibinfo {author} {\bibfnamefont {Danilo}\ \bibnamefont
  {Puggioni}}, \bibinfo {author} {\bibfnamefont {Gianluca}\ \bibnamefont
  {Giovannetti}}, \ and\ \bibinfo {author} {\bibfnamefont {James~M.}\
  \bibnamefont {Rondinelli}},\ }\bibfield  {title} {\enquote {\bibinfo {title}
  {Polar metals as electrodes to suppress the critical-thickness limit in
  ferroelectric nanocapacitors},}\ }\href@noop {} {\bibfield  {journal}
  {\bibinfo  {journal} {J. Appl. Phys}\ }\textbf {\bibinfo {volume} {124}},\
  \bibinfo {pages} {174102} (\bibinfo {year} {2018})}\BibitemShut {NoStop}%
\bibitem [{\citenamefont {Puggioni}\ \emph {et~al.}(2015)\citenamefont
  {Puggioni}, \citenamefont {Giovannetti}, \citenamefont {Capone},\ and\
  \citenamefont {Rondinelli}}]{Puggioni2015}%
  \BibitemOpen
  \bibfield  {author} {\bibinfo {author} {\bibfnamefont {Danilo}\ \bibnamefont
  {Puggioni}}, \bibinfo {author} {\bibfnamefont {Gianluca}\ \bibnamefont
  {Giovannetti}}, \bibinfo {author} {\bibfnamefont {Massimo}\ \bibnamefont
  {Capone}}, \ and\ \bibinfo {author} {\bibfnamefont {James~M.}\ \bibnamefont
  {Rondinelli}},\ }\bibfield  {title} {\enquote {\bibinfo {title} {Design of a
  mott multiferroic from a nonmagnetic polar metal},}\ }\href@noop {}
  {\bibfield  {journal} {\bibinfo  {journal} {Phys. Rev. Lett.}\ }\textbf
  {\bibinfo {volume} {115}},\ \bibinfo {pages} {087202} (\bibinfo {year}
  {2015})}\BibitemShut {NoStop}%
\bibitem [{\citenamefont {Zhang}\ \emph {et~al.}({2018})\citenamefont {Zhang},
  \citenamefont {Xie}, \citenamefont {Kim}, \citenamefont {Stern},
  \citenamefont {Wang}, \citenamefont {Zhang}, \citenamefont {Yan},
  \citenamefont {Li}, \citenamefont {Liu}, \citenamefont {Zhao}, \citenamefont
  {Chi}, \citenamefont {Gadre}, \citenamefont {Lin}, \citenamefont {Zhou},
  \citenamefont {Uher}, \citenamefont {Chen}, \citenamefont {Chu},
  \citenamefont {Xia}, \citenamefont {Wu},\ and\ \citenamefont
  {Pan}}]{NatCommPZT-hetero}%
  \BibitemOpen
  \bibfield  {author} {\bibinfo {author} {\bibfnamefont {Yi}~\bibnamefont
  {Zhang}}, \bibinfo {author} {\bibfnamefont {Lin}\ \bibnamefont {Xie}},
  \bibinfo {author} {\bibfnamefont {Jeongwoo}\ \bibnamefont {Kim}}, \bibinfo
  {author} {\bibfnamefont {Alex}\ \bibnamefont {Stern}}, \bibinfo {author}
  {\bibfnamefont {Hui}\ \bibnamefont {Wang}}, \bibinfo {author} {\bibfnamefont
  {Kui}\ \bibnamefont {Zhang}}, \bibinfo {author} {\bibfnamefont {Xingxu}\
  \bibnamefont {Yan}}, \bibinfo {author} {\bibfnamefont {Linze}\ \bibnamefont
  {Li}}, \bibinfo {author} {\bibfnamefont {Henry}\ \bibnamefont {Liu}},
  \bibinfo {author} {\bibfnamefont {Gejian}\ \bibnamefont {Zhao}}, \bibinfo
  {author} {\bibfnamefont {Hang}\ \bibnamefont {Chi}}, \bibinfo {author}
  {\bibfnamefont {Chaitanya}\ \bibnamefont {Gadre}}, \bibinfo {author}
  {\bibfnamefont {Qiyin}\ \bibnamefont {Lin}}, \bibinfo {author} {\bibfnamefont
  {Yichun}\ \bibnamefont {Zhou}}, \bibinfo {author} {\bibfnamefont {Ctirad}\
  \bibnamefont {Uher}}, \bibinfo {author} {\bibfnamefont {Tingyong}\
  \bibnamefont {Chen}}, \bibinfo {author} {\bibfnamefont {Ying-Hao}\
  \bibnamefont {Chu}}, \bibinfo {author} {\bibfnamefont {Jing}\ \bibnamefont
  {Xia}}, \bibinfo {author} {\bibfnamefont {Ruqian}\ \bibnamefont {Wu}}, \ and\
  \bibinfo {author} {\bibfnamefont {Xiaoqing}\ \bibnamefont {Pan}},\ }\bibfield
   {title} {\enquote {\bibinfo {title} {{Discovery of a magnetic conductive
  interface in {PbZr$_{0.2}$Ti$_{0.8}$O$_3$}/{SrTiO$_3$} heterostructures}},}\
  }\href@noop {} {\bibfield  {journal} {\bibinfo  {journal} {{Nat. Commun.}}\
  }\textbf {\bibinfo {volume} {9}},\ \bibinfo {pages} {685} (\bibinfo {year}
  {{2018}})}\BibitemShut {NoStop}%
\bibitem [{\citenamefont {Li}\ \emph {et~al.}(2014)\citenamefont {Li},
  \citenamefont {Gu}, \citenamefont {Wang}, \citenamefont {Chen},\ and\
  \citenamefont {Duan}}]{PhysRevB.90.054106}%
  \BibitemOpen
  \bibfield  {author} {\bibinfo {author} {\bibfnamefont {Menglei}\ \bibnamefont
  {Li}}, \bibinfo {author} {\bibfnamefont {Yijia}\ \bibnamefont {Gu}}, \bibinfo
  {author} {\bibfnamefont {Yi}~\bibnamefont {Wang}}, \bibinfo {author}
  {\bibfnamefont {Long-Qing}\ \bibnamefont {Chen}}, \ and\ \bibinfo {author}
  {\bibfnamefont {Wenhui}\ \bibnamefont {Duan}},\ }\bibfield  {title} {\enquote
  {\bibinfo {title} {First-principles study of ${180}^{\ensuremath{\circ}}$
  domain walls in {BaTiO$_3$}: Mixed {Bloch}-{N}\'eel-{Ising} character},}\
  }\href@noop {} {\bibfield  {journal} {\bibinfo  {journal} {Phys. Rev. B}\
  }\textbf {\bibinfo {volume} {90}},\ \bibinfo {pages} {054106} (\bibinfo
  {year} {2014})}\BibitemShut {NoStop}%
\bibitem [{\citenamefont {Meyer}\ and\ \citenamefont
  {Vanderbilt}(2002)}]{PhysRevB.65.104111}%
  \BibitemOpen
  \bibfield  {author} {\bibinfo {author} {\bibfnamefont {B.}~\bibnamefont
  {Meyer}}\ and\ \bibinfo {author} {\bibfnamefont {David}\ \bibnamefont
  {Vanderbilt}},\ }\bibfield  {title} {\enquote {\bibinfo {title} {Ab initio
  study of ferroelectric domain walls in {PbTiO$_{3}$}},}\ }\href@noop {}
  {\bibfield  {journal} {\bibinfo  {journal} {Phys. Rev. B}\ }\textbf {\bibinfo
  {volume} {65}},\ \bibinfo {pages} {104111} (\bibinfo {year}
  {2002})}\BibitemShut {NoStop}%
\bibitem [{\citenamefont {Henkelman}\ \emph {et~al.}(2000)\citenamefont
  {Henkelman}, \citenamefont {Uberuaga},\ and\ \citenamefont
  {Jónsson}}]{climbNEB2000}%
  \BibitemOpen
  \bibfield  {author} {\bibinfo {author} {\bibfnamefont {Graeme}\ \bibnamefont
  {Henkelman}}, \bibinfo {author} {\bibfnamefont {Blas~P.}\ \bibnamefont
  {Uberuaga}}, \ and\ \bibinfo {author} {\bibfnamefont {Hannes}\ \bibnamefont
  {Jónsson}},\ }\bibfield  {title} {\enquote {\bibinfo {title} {A climbing
  image nudged elastic band method for finding saddle points and minimum energy
  paths},}\ }\href@noop {} {\bibfield  {journal} {\bibinfo  {journal} {J. Chem.
  Phys.}\ }\textbf {\bibinfo {volume} {113}},\ \bibinfo {pages} {9901--9904}
  (\bibinfo {year} {2000})}\BibitemShut {NoStop}%
\bibitem [{\citenamefont {Laidler}\ and\ \citenamefont
  {King}(1983)}]{TST-1983}%
  \BibitemOpen
  \bibfield  {author} {\bibinfo {author} {\bibfnamefont {Keith~J.}\
  \bibnamefont {Laidler}}\ and\ \bibinfo {author} {\bibfnamefont
  {M.~Christine}\ \bibnamefont {King}},\ }\bibfield  {title} {\enquote
  {\bibinfo {title} {Development of transition-state theory},}\ }\href@noop {}
  {\bibfield  {journal} {\bibinfo  {journal} {J. Phys. Chem.}\ }\textbf
  {\bibinfo {volume} {87}},\ \bibinfo {pages} {2657--2664} (\bibinfo {year}
  {1983})}\BibitemShut {NoStop}%
\bibitem [{\citenamefont {Tadmor}\ \emph {et~al.}(2002)\citenamefont {Tadmor},
  \citenamefont {Waghmare}, \citenamefont {Smith},\ and\ \citenamefont
  {Kaxiras}}]{TADMOR20022989}%
  \BibitemOpen
  \bibfield  {author} {\bibinfo {author} {\bibfnamefont {E.B.}\ \bibnamefont
  {Tadmor}}, \bibinfo {author} {\bibfnamefont {U.V.}\ \bibnamefont {Waghmare}},
  \bibinfo {author} {\bibfnamefont {G.S.}\ \bibnamefont {Smith}}, \ and\
  \bibinfo {author} {\bibfnamefont {E.}~\bibnamefont {Kaxiras}},\ }\bibfield
  {title} {\enquote {\bibinfo {title} {Polarization switching in {PbTiO$_3$}:
  an ab initio finite element simulation},}\ }\href@noop {} {\bibfield
  {journal} {\bibinfo  {journal} {Acta Mater.}\ }\textbf {\bibinfo {volume}
  {50}},\ \bibinfo {pages} {2989--3002} (\bibinfo {year} {2002})}\BibitemShut
  {NoStop}%
\bibitem [{\citenamefont {Wang}\ and\ \citenamefont
  {Qian}(2017)}]{Wang_20172DMater}%
  \BibitemOpen
  \bibfield  {author} {\bibinfo {author} {\bibfnamefont {Hua}\ \bibnamefont
  {Wang}}\ and\ \bibinfo {author} {\bibfnamefont {Xiaofeng}\ \bibnamefont
  {Qian}},\ }\bibfield  {title} {\enquote {\bibinfo {title} {Two-dimensional
  multiferroics in monolayer group {IV} monochalcogenides},}\ }\href@noop {}
  {\bibfield  {journal} {\bibinfo  {journal} {2D Mater.}\ }\textbf {\bibinfo
  {volume} {4}},\ \bibinfo {pages} {015042} (\bibinfo {year}
  {2017})}\BibitemShut {NoStop}%
\bibitem [{\citenamefont {Li}\ \emph {et~al.}(2018)\citenamefont {Li},
  \citenamefont {Yang}, \citenamefont {Cao}, \citenamefont {Sun}, \citenamefont
  {Peng}, \citenamefont {Zhou},\ and\ \citenamefont {Zhang}}]{LiXY-JPCC2018}%
  \BibitemOpen
  \bibfield  {author} {\bibinfo {author} {\bibfnamefont {X.~Y.}\ \bibnamefont
  {Li}}, \bibinfo {author} {\bibfnamefont {Q.}~\bibnamefont {Yang}}, \bibinfo
  {author} {\bibfnamefont {J.~X.}\ \bibnamefont {Cao}}, \bibinfo {author}
  {\bibfnamefont {L.~Z.}\ \bibnamefont {Sun}}, \bibinfo {author} {\bibfnamefont
  {Q.~X.}\ \bibnamefont {Peng}}, \bibinfo {author} {\bibfnamefont {Y.~C.}\
  \bibnamefont {Zhou}}, \ and\ \bibinfo {author} {\bibfnamefont {R.~X.}\
  \bibnamefont {Zhang}},\ }\bibfield  {title} {\enquote {\bibinfo {title}
  {Domain wall motion in perovskite ferroelectrics studied by the nudged
  elastic band method},}\ }\href@noop {} {\bibfield  {journal} {\bibinfo
  {journal} {J. Phys. Chem. C}\ }\textbf {\bibinfo {volume} {122}},\ \bibinfo
  {pages} {3091--3100} (\bibinfo {year} {2018})}\BibitemShut {NoStop}%
\bibitem [{sur()}]{surmount}%
  \BibitemOpen
  \href@noop {} {}\bibinfo {note} {At room temperature, 1 ${k_\mathrm{B}T
  \approx}$ 26 meV. In transition state theory, the probability $P$ of
  overcoming the energy barrier is proportional to e$^{-\Delta
  E/k_\mathrm{B}T}$, \ie, $P$ $\propto$ e$^{-\Delta E/k_\mathrm{B}T}$. The
  energy barrier (${\Delta E}$ = 58 meV) is about twice the $k_\mathrm{B}T$,
  hence the probability of overcoming this barrier is around
  14$\%$.}\BibitemShut {Stop}%
\bibitem [{\citenamefont {Bl\"ochl}(1994)}]{Blochl1994}%
  \BibitemOpen
  \bibfield  {author} {\bibinfo {author} {\bibfnamefont {P.~E.}\ \bibnamefont
  {Bl\"ochl}},\ }\bibfield  {title} {\enquote {\bibinfo {title} {Projector
  augmented-wave method},}\ }\href@noop {} {\bibfield  {journal} {\bibinfo
  {journal} {Phys. Rev. B}\ }\textbf {\bibinfo {volume} {50}},\ \bibinfo
  {pages} {17953--17979} (\bibinfo {year} {1994})}\BibitemShut {NoStop}%
\bibitem [{\citenamefont {Kresse}\ and\ \citenamefont {Furthm$\ddot{\rm
  u}$ller}(1996{\natexlab{a}})}]{Kresse1996}%
  \BibitemOpen
  \bibfield  {author} {\bibinfo {author} {\bibfnamefont {G.}~\bibnamefont
  {Kresse}}\ and\ \bibinfo {author} {\bibfnamefont {J.}~\bibnamefont
  {Furthm$\ddot{\rm u}$ller}},\ }\bibfield  {title} {\enquote {\bibinfo {title}
  {Efficiency of ab-initio total energy calculations for metals and
  semiconductors using a plane-wave basis set},}\ }\href@noop {} {\bibfield
  {journal} {\bibinfo  {journal} {Comp. Mater. Sci.}\ }\textbf {\bibinfo
  {volume} {6}},\ \bibinfo {pages} {15--50} (\bibinfo {year}
  {1996}{\natexlab{a}})}\BibitemShut {NoStop}%
\bibitem [{\citenamefont {Kresse}\ and\ \citenamefont {Furthm$\ddot{\rm
  u}$ller}(1996{\natexlab{b}})}]{Kresse-PRB-1996}%
  \BibitemOpen
  \bibfield  {author} {\bibinfo {author} {\bibfnamefont {G.}~\bibnamefont
  {Kresse}}\ and\ \bibinfo {author} {\bibfnamefont {J.}~\bibnamefont
  {Furthm$\ddot{\rm u}$ller}},\ }\bibfield  {title} {\enquote {\bibinfo {title}
  {Efficient iterative schemes for ab initio total-energy calculations using a
  plane-wave basis set},}\ }\href@noop {} {\bibfield  {journal} {\bibinfo
  {journal} {Phys. Rev. B}\ }\textbf {\bibinfo {volume} {54}},\ \bibinfo
  {pages} {11169--11186} (\bibinfo {year} {1996}{\natexlab{b}})}\BibitemShut
  {NoStop}%
\bibitem [{\citenamefont {Perdew}\ \emph {et~al.}(2008)\citenamefont {Perdew},
  \citenamefont {Ruzsinszky}, \citenamefont {Csonka}, \citenamefont {Vydrov},
  \citenamefont {Scuseria}, \citenamefont {Constantin}, \citenamefont {Zhou},\
  and\ \citenamefont {Burke}}]{PhysRevLett.100.136406PBEsol}%
  \BibitemOpen
  \bibfield  {author} {\bibinfo {author} {\bibfnamefont {John~P.}\ \bibnamefont
  {Perdew}}, \bibinfo {author} {\bibfnamefont {Adrienn}\ \bibnamefont
  {Ruzsinszky}}, \bibinfo {author} {\bibfnamefont {G\'abor~I.}\ \bibnamefont
  {Csonka}}, \bibinfo {author} {\bibfnamefont {Oleg~A.}\ \bibnamefont
  {Vydrov}}, \bibinfo {author} {\bibfnamefont {Gustavo~E.}\ \bibnamefont
  {Scuseria}}, \bibinfo {author} {\bibfnamefont {Lucian~A.}\ \bibnamefont
  {Constantin}}, \bibinfo {author} {\bibfnamefont {Xiaolan}\ \bibnamefont
  {Zhou}}, \ and\ \bibinfo {author} {\bibfnamefont {Kieron}\ \bibnamefont
  {Burke}},\ }\bibfield  {title} {\enquote {\bibinfo {title} {Restoring the
  density-gradient expansion for exchange in solids and surfaces},}\
  }\href@noop {} {\bibfield  {journal} {\bibinfo  {journal} {Phys. Rev. Lett.}\
  }\textbf {\bibinfo {volume} {100}},\ \bibinfo {pages} {136406} (\bibinfo
  {year} {2008})}\BibitemShut {NoStop}%
\bibitem [{\citenamefont {Dudarev}\ \emph {et~al.}(1998)\citenamefont
  {Dudarev}, \citenamefont {Botton}, \citenamefont {Savrasov}, \citenamefont
  {Humphreys},\ and\ \citenamefont {Sutton}}]{Dudarev-LDAU-PRB1998}%
  \BibitemOpen
  \bibfield  {author} {\bibinfo {author} {\bibfnamefont {S.~L.}\ \bibnamefont
  {Dudarev}}, \bibinfo {author} {\bibfnamefont {G.~A.}\ \bibnamefont {Botton}},
  \bibinfo {author} {\bibfnamefont {S.~Y.}\ \bibnamefont {Savrasov}}, \bibinfo
  {author} {\bibfnamefont {C.~J.}\ \bibnamefont {Humphreys}}, \ and\ \bibinfo
  {author} {\bibfnamefont {A.~P.}\ \bibnamefont {Sutton}},\ }\bibfield  {title}
  {\enquote {\bibinfo {title} {Electron-energy-loss spectra and the structural
  stability of nickel oxide: An {LSDA+U} study},}\ }\href@noop {} {\bibfield
  {journal} {\bibinfo  {journal} {Phys. Rev. B}\ }\textbf {\bibinfo {volume}
  {57}},\ \bibinfo {pages} {1505--1509} (\bibinfo {year} {1998})}\BibitemShut
  {NoStop}%
\bibitem [{\citenamefont {Giovannetti}\ and\ \citenamefont
  {Capone}(2014)}]{Giovannetti2014}%
  \BibitemOpen
  \bibfield  {author} {\bibinfo {author} {\bibfnamefont {Gianluca}\
  \bibnamefont {Giovannetti}}\ and\ \bibinfo {author} {\bibfnamefont {Massimo}\
  \bibnamefont {Capone}},\ }\bibfield  {title} {\enquote {\bibinfo {title}
  {Dual nature of the ferroelectric and metallic state in {LiOsO$_3$}},}\
  }\href@noop {} {\bibfield  {journal} {\bibinfo  {journal} {Phys. Rev. B}\
  }\textbf {\bibinfo {volume} {90}},\ \bibinfo {pages} {195113} (\bibinfo
  {year} {2014})}\BibitemShut {NoStop}%
\bibitem [{\citenamefont {Makov}\ and\ \citenamefont
  {Payne}(1995)}]{PhysRevB.51.4014}%
  \BibitemOpen
  \bibfield  {author} {\bibinfo {author} {\bibfnamefont {G.}~\bibnamefont
  {Makov}}\ and\ \bibinfo {author} {\bibfnamefont {M.~C.}\ \bibnamefont
  {Payne}},\ }\bibfield  {title} {\enquote {\bibinfo {title} {Periodic boundary
  conditions in ab initio calculations},}\ }\href@noop {} {\bibfield  {journal}
  {\bibinfo  {journal} {Phys. Rev. B}\ }\textbf {\bibinfo {volume} {51}},\
  \bibinfo {pages} {4014--4022} (\bibinfo {year} {1995})}\BibitemShut {NoStop}%
\bibitem [{\citenamefont {Neugebauer}\ and\ \citenamefont
  {Scheffler}(1992)}]{PhysRevB.46.16067}%
  \BibitemOpen
  \bibfield  {author} {\bibinfo {author} {\bibfnamefont {J\"org}\ \bibnamefont
  {Neugebauer}}\ and\ \bibinfo {author} {\bibfnamefont {Matthias}\ \bibnamefont
  {Scheffler}},\ }\bibfield  {title} {\enquote {\bibinfo {title}
  {Adsorbate-substrate and adsorbate-adsorbate interactions of {Na} and {K}
  adlayers on {Al(111)}},}\ }\href@noop {} {\bibfield  {journal} {\bibinfo
  {journal} {Phys. Rev. B}\ }\textbf {\bibinfo {volume} {46}},\ \bibinfo
  {pages} {16067--16080} (\bibinfo {year} {1992})}\BibitemShut {NoStop}%
\bibitem [{\citenamefont {Silvi}\ and\ \citenamefont
  {Savin}(1994)}]{silvi1994Nature}%
  \BibitemOpen
  \bibfield  {author} {\bibinfo {author} {\bibfnamefont {Bernard}\ \bibnamefont
  {Silvi}}\ and\ \bibinfo {author} {\bibfnamefont {Andreas}\ \bibnamefont
  {Savin}},\ }\bibfield  {title} {\enquote {\bibinfo {title} {Classification of
  chemical bonds based on topological analysis of electron localization
  functions},}\ }\href@noop {} {\bibfield  {journal} {\bibinfo  {journal}
  {Nature}\ }\textbf {\bibinfo {volume} {371}},\ \bibinfo {pages} {683}
  (\bibinfo {year} {1994})}\BibitemShut {NoStop}%
\bibitem [{\citenamefont {Ding}\ and\ \citenamefont
  {Duan}(2012)}]{EPL-DING2012}%
  \BibitemOpen
  \bibfield  {author} {\bibinfo {author} {\bibfnamefont {Hang-Chen}\
  \bibnamefont {Ding}}\ and\ \bibinfo {author} {\bibfnamefont {Chun-Gang}\
  \bibnamefont {Duan}},\ }\bibfield  {title} {\enquote {\bibinfo {title}
  {Electric-field control of magnetic ordering in the tetragonal-like
  $\mathrm{BiFeO}_3$},}\ }\href@noop {} {\bibfield  {journal} {\bibinfo
  {journal} {EPL (Europhys. Lett.)}\ }\textbf {\bibinfo {volume} {97}},\
  \bibinfo {pages} {57007} (\bibinfo {year} {2012})}\BibitemShut {NoStop}%
\bibitem [{\citenamefont {Aroyo}\ \emph {et~al.}(2011)\citenamefont {Aroyo},
  \citenamefont {Perez-Mato}, \citenamefont {Orobengoa}, \citenamefont {Tasci},
  \citenamefont {De~La~Flor},\ and\ \citenamefont
  {Kirov}}]{aroyo2011crystallography}%
  \BibitemOpen
  \bibfield  {author} {\bibinfo {author} {\bibfnamefont {Mois~I}\ \bibnamefont
  {Aroyo}}, \bibinfo {author} {\bibfnamefont {JM}~\bibnamefont {Perez-Mato}},
  \bibinfo {author} {\bibfnamefont {D}~\bibnamefont {Orobengoa}}, \bibinfo
  {author} {\bibfnamefont {E}~\bibnamefont {Tasci}}, \bibinfo {author}
  {\bibfnamefont {G}~\bibnamefont {De~La~Flor}}, \ and\ \bibinfo {author}
  {\bibfnamefont {A}~\bibnamefont {Kirov}},\ }\bibfield  {title} {\enquote
  {\bibinfo {title} {Crystallography online: Bilbao crystallographic server},}\
  }\href@noop {} {\bibfield  {journal} {\bibinfo  {journal} {Bulg. Chem.
  Commun}\ }\textbf {\bibinfo {volume} {43}},\ \bibinfo {pages} {183--197}
  (\bibinfo {year} {2011})}\BibitemShut {NoStop}%
\bibitem [{\citenamefont {Momma}\ and\ \citenamefont
  {Izumi}(2011)}]{momma2011vesta}%
  \BibitemOpen
  \bibfield  {author} {\bibinfo {author} {\bibfnamefont {Koichi}\ \bibnamefont
  {Momma}}\ and\ \bibinfo {author} {\bibfnamefont {Fujio}\ \bibnamefont
  {Izumi}},\ }\bibfield  {title} {\enquote {\bibinfo {title} {{VESTA} 3 for
  three-dimensional visualization of crystal, volumetric and morphology
  data},}\ }\href@noop {} {\bibfield  {journal} {\bibinfo  {journal} {J. Appl.
  Crystallogr.}\ }\textbf {\bibinfo {volume} {44}},\ \bibinfo {pages}
  {1272--1276} (\bibinfo {year} {2011})}\BibitemShut {NoStop}%
\bibitem [{\citenamefont {Togo}\ and\ \citenamefont
  {Tanaka}(2015)}]{Togo-phonopy2015}%
  \BibitemOpen
  \bibfield  {author} {\bibinfo {author} {\bibfnamefont {Atsushi}\ \bibnamefont
  {Togo}}\ and\ \bibinfo {author} {\bibfnamefont {Isao}\ \bibnamefont
  {Tanaka}},\ }\bibfield  {title} {\enquote {\bibinfo {title} {First principles
  phonon calculations in materials science},}\ }\href@noop {} {\bibfield
  {journal} {\bibinfo  {journal} {Scr. Mater.}\ }\textbf {\bibinfo {volume}
  {108}},\ \bibinfo {pages} {1--5} (\bibinfo {year} {2015})}\BibitemShut
  {NoStop}%
\end{thebibliography}
%
\end{document}


\title{Supplementary Information for: \\
Design of a multifunctional polar metal via
first-principles high-throughput structure screening}

\author{Yue-Wen Fang$^{1,2}$}
\email{fyuewen@gmail.com}
\author{Hanghui Chen$^{2,3}$}
\email{hanghui.chen@nyu.edu}
\affiliation{$^1$Department of Materials Science and Engineering, Kyoto University, Kyoto, Japan \\
$^2$NYU-ECNU Institute of Physics, New York University Shanghai China \\
$^3$Department of Physics, New York University, New York 10003, USA
}

\date{\today}
\maketitle


\clearpage
\newpage

\section{L\lowercase{ow-energy structures predicted from} CALYPSO \lowercase{search}}
Supplementary Table 1. A list of predicted crystal structures including
ten lowest energy states, perovskite anti-polar state with ${Pmma}$ symmetry,
and post-perovskite anti-polar state with $Pmmm$ symmetry.
`---' represents non-perovskite and non-post-perovskite structures.
Post-perovskite structures are explicitly
shown. `Layered', 'Rock-salt', and `Columnar' refer to different cation
orderings of $A$-site ordered double perovskite structure (the naming convention
follows Ref.~\cite{B926757C}).

	\begin{longtable}{lllllllllllllllll}
		
		Phase & & Cell parameters & & Atom type  & & Wyckoff site & X & Y & Z & Ordering & $E$ (meV)  \\
	\hline
\hline
${Pmm2}$ &&
\begin{tabular}[c]{@{}l@{}}$a$ =  5.616~\AA\\$b$ =  3.035~\AA\\$c$ =  7.597~\AA\\$\alpha$ =  90.000$^{\circ}$\\$\beta$ =  90.000$^{\circ}$\\$\gamma$ =  90.000$^{\circ}$\end{tabular} &&
\begin{tabular}[c]{@{}l@{}}Ti1\\Bi1\\Pb1\\O1\\O2\\O3\\O4\end{tabular} &&
\begin{tabular}[c]{@{}l@{}}2g\\1c\\1d\\2h\\2h\\1a\\1b\end{tabular} &
        \begin{tabular}[c]{@{}l@{}}0.483\\-0.074\\0.018\\0.264\\0.724\\0.575\\0.439\end{tabular} &
\begin{tabular}[c]{@{}l@{}}0.251\\0.000\\0.500\\0.216\\0.270\\0.000\\0.500\end{tabular} &
\begin{tabular}[c]{@{}l@{}}0.000\\0.500\\0.500\\0.500\\0.500\\0.000\\0.000\end{tabular} &
	Post-perovskite   &
0  \\
	\hline
${Pmn2_1}$ &&
\begin{tabular}[c]{@{}l@{}}$a$ =  5.520~\AA\\$b$ =  5.543~\AA\\$c$ =  7.993~\AA\\$\alpha$ =  90.000$^{\circ}$\\$\beta$ =  90.000$^{\circ}$\\$\gamma$ =  90.000$^{\circ}$\end{tabular} &&
\begin{tabular}[c]{@{}l@{}}Ti1\\Bi1\\Pb1\\O1\\O2\\O3\\O4\end{tabular} &&
\begin{tabular}[c]{@{}l@{}}4b\\2a\\2a\\4b\\4b\\2a\\2b\end{tabular} &
 \begin{tabular}[c]{@{}l@{}}0.056\\0.517\\0.039\\0.279\\0.860\\0.099\\0.557\end{tabular} &
\begin{tabular}[c]{@{}l@{}}0.742\\0.785\\0.258\\0.463\\0.045\\0.750\\0.221\end{tabular} &
\begin{tabular}[c]{@{}l@{}}0.749\\0.000\\0.000\\0.736\\0.746\\0.000\\0.000\end{tabular} &
	Rock-salt   &
93.0  \\
	\hline
	${Pmm2}$ &&
\begin{tabular}[c]{@{}l@{}}$a$ =  4.077~\AA\\$b$ =  3.919~\AA\\$c$ =  7.887~\AA\\$\alpha$ =  90.000$^{\circ}$\\$\beta$ =  90.000$^{\circ}$\\$\gamma$ =  90.000$^{\circ}$\end{tabular} &&
\begin{tabular}[c]{@{}l@{}}Ti1\\Bi1\\Pb1\\O1\\O2\\O3\\O4\end{tabular} &&
\begin{tabular}[c]{@{}l@{}}2g\\1d\\1c\\2g\\2h\\1a\\1b\end{tabular} &
 \begin{tabular}[c]{@{}l@{}}0.542\\0.129\\0.079\\-0.014\\0.457\\0.480\\0.441\end{tabular} &
\begin{tabular}[c]{@{}l@{}}0.000\\0.500\\0.500\\0.000\\0.500\\0.000\\0.000\end{tabular} &
        \begin{tabular}[c]{@{}l@{}}0.245\\0.500\\0.000\\0.253\\0.259\\0.000\\0.500\end{tabular} &
	Layered   &
93.2  \\
	\hline
	${P1}$ &&
\begin{tabular}[c]{@{}l@{}}$a$ =  3.743~\AA\\$b$ =  5.943~\AA\\$c$ =  13.435~\AA\\$\alpha$ =  93.478$^{\circ}$\\$\beta$ =  97.928$^{\circ}$\\$\gamma$ =  90.068$^{\circ}$\end{tabular} &&
\begin{tabular}[c]{@{}l@{}}Ti1\\Ti2\\Ti3\\Ti4\\Bi1\\Bi2\\Pb1\\Pb2\\O1\\O2\\O3\\O4\\O5\\O6\\O7\\O8\\O9\\O10\\O11\\O12\end{tabular} &&
\begin{tabular}[c]{@{}l@{}}1a\\1a\\1a\\1a\\1a\\1a\\1a\\1a\\1a\\1a\\1a\\1a\\1a\\1a\\1a\\1a\\1a\\1a\\1a\\1a\end{tabular} &
\begin{tabular}[c]{@{}l@{}}0.312\\0.425\\0.621\\0.719\\0.014\\0.044\\0.221\\0.865\\0.136\\0.237\\0.336\\0.371\\0.451\\0.485\\0.563\\0.590\\0.682\\0.694\\0.793\\-0.086\end{tabular} &
\begin{tabular}[c]{@{}l@{}}0.738\\0.835\\0.175\\0.710\\0.244\\0.768\\0.216\\0.252\\0.189\\0.735\\0.024\\0.587\\0.161\\0.706\\0.314\\0.884\\0.450\\-0.009\\0.665\\0.883\end{tabular} &
\begin{tabular}[c]{@{}l@{}}0.155\\0.381\\0.768\\-0.033\\0.538\\0.617\\-0.030\\0.286\\0.797\\0.002\\0.200\\0.271\\0.423\\0.498\\0.649\\0.708\\0.891\\-0.085\\0.115\\0.356\end{tabular} &
	---   &
103.4  \\
	\hline
	${I4mm}$ &&
\begin{tabular}[c]{@{}l@{}}$a$ =  5.561~\AA\\$b$ =  5.561~\AA\\$c$ =  8.161~\AA\\$\alpha$ =  90.000$^{\circ}$\\$\beta$ =  90.000$^{\circ}$\\$\gamma$ =  90.000$^{\circ}$\end{tabular} &&
\begin{tabular}[c]{@{}l@{}}Ti1\\Bi1\\Pb1\\O1\\O2\end{tabular} &&
\begin{tabular}[c]{@{}l@{}}4b\\2a\\2a\\8c\\4b\end{tabular} &
\begin{tabular}[c]{@{}l@{}}0.000\\0.000\\0.000\\0.743\\0.000\end{tabular} &
\begin{tabular}[c]{@{}l@{}}0.500\\0.000\\0.000\\0.743\\0.500\end{tabular} &
\begin{tabular}[c]{@{}l@{}}0.216\\0.508\\-0.016\\0.173\\0.437\end{tabular} &
	Rock-salt   &
109.1  \\
	\hline
${Cm}$ &&
\begin{tabular}[c]{@{}l@{}}$a$ =  7.923~\AA\\$b$ =  ~7.875~\AA\\$c$ =  ~3.972\AA\\$\alpha$ =  90.000$^{\circ}$\\$\beta$ =  90.000$^{\circ}$\\$\gamma$ =  90.000$^{\circ}$\end{tabular} &&
\begin{tabular}[c]{@{}l@{}}Ti1\\Bi1\\Pb1\\O1\\O2\\O3\\O4\end{tabular} &&
\begin{tabular}[c]{@{}l@{}}4b\\2a\\2a\\4b\\4b\\2a\\2b\end{tabular} &
\begin{tabular}[c]{@{}l@{}}0.465\\0.667\\0.201\\0.224\\0.484\\0.006\\0.460\end{tabular} &
\begin{tabular}[c]{@{}l@{}}0.257\\0.000\\0.000\\0.285\\0.245\\0.000\\0.000\end{tabular} &
\begin{tabular}[c]{@{}l@{}}0.0.060\\0.622\\0.577\\0.036\\0.551\\0.051\\0.031\end{tabular} &
	Columnar   &
120.0  \\
	\hline
	${R32}$ &&
\begin{tabular}[c]{@{}l@{}}$a$ =  5.475~\AA\\$b$ =  5.475~\AA\\$c$ =  5.475~\AA\\$\alpha$ =  61.727$^{\circ}$\\$\beta$ =  61.727$^{\circ}$\\$\gamma$ =  61.727$^{\circ}$\end{tabular} &&
\begin{tabular}[c]{@{}l@{}}Ti1\\Bi1\\Pb1\\O1\\O2\end{tabular} &&
\begin{tabular}[c]{@{}l@{}}6c\\3b\\3a\\9d\\9e\end{tabular} &
\begin{tabular}[c]{@{}l@{}}0.000\\0.000\\0.000\\0.449\\0.573\end{tabular} &
\begin{tabular}[c]{@{}l@{}}0.000\\0.000\\0.000\\0.000\\0.000\end{tabular} &
\begin{tabular}[c]{@{}l@{}}0.747\\0.500\\0.000\\0.000\\0.500\end{tabular} &
	Rock-salt   &
137.3  \\
	\hline
	${P2{_1}2{_1}2}$ &&
\begin{tabular}[c]{@{}l@{}}$a$ =  5.494~\AA\\$b$ =  7.980~\AA\\$c$ =  5.498~\AA\\$\alpha$ =  90.000$^{\circ}$\\$\beta$ =  90.000$^{\circ}$\\$\gamma$ =  90.000$^{\circ}$\end{tabular} &&
\begin{tabular}[c]{@{}l@{}}Ti1\\Bi1\\Pb1\\O1\\O2\\O3\\O4\end{tabular} &&
\begin{tabular}[c]{@{}l@{}}4c\\2b\\2a\\4c\\4c\\2a\\2b\end{tabular} &
\begin{tabular}[c]{@{}l@{}}0.501\\0.000\\0.000\\0.699\\0.704\\0.000\\0.000\end{tabular} &
\begin{tabular}[c]{@{}l@{}}0.750\\0.500\\0.000\\0.239\\0.759\\0.000\\0.500\end{tabular} &
\begin{tabular}[c]{@{}l@{}}0.252\\0.233\\0.253\\0.548\\-0.048\\0.749\\0.764\end{tabular} &
	Rock-salt   &
138.3  \\
	\hline
	${I\bar{4}2m}$ &&
\begin{tabular}[c]{@{}l@{}}$a$ =  5.492~\AA\\$b$ =  5.492~\AA\\$c$ =  5.569~\AA\\$\alpha$ =  119.545$^{\circ}$\\$\beta$ =  119.545$^{\circ}$\\$\gamma$ =  90.000$^{\circ}$\end{tabular} &&
\begin{tabular}[c]{@{}l@{}}Ti1\\Bi1\\Pb1\\O1\\O2\end{tabular} &&
\begin{tabular}[c]{@{}l@{}}4d\\2b\\2a\\8i\\4c\end{tabular} &
\begin{tabular}[c]{@{}l@{}}0.000\\0.000\\0.000\\0.201\\0.000\end{tabular} &
\begin{tabular}[c]{@{}l@{}}0.500\\0.000\\0.000\\0.201\\0.500\end{tabular} &
\begin{tabular}[c]{@{}l@{}}0.250\\0.500\\0.000\\0.260\\0.000\end{tabular} &
	Rock-salt   &
139.4  \\
	\hline
	${Pnnm}$ &&
\begin{tabular}[c]{@{}l@{}}$a$ =  5.4916~\AA\\$b$ =  5.4915~\AA\\$c$ =  7.9943~\AA\\$\alpha$ =  90.000$^{\circ}$\\$\beta$ =  90.000$^{\circ}$\\$\gamma$ =  90.000$^{\circ}$\end{tabular} &&
\begin{tabular}[c]{@{}l@{}}Ba1\\Pb1\\Ti1\\O1\\O2\\O3\end{tabular} &&
\begin{tabular}[c]{@{}l@{}}2b\\2a\\4f\\8h\\2c\\2d\end{tabular} &
\begin{tabular}[c]{@{}l@{}}0.000\\0.000\\0.000\\0.7024\\0.000\\0.000\end{tabular} &
\begin{tabular}[c]{@{}l@{}}0.000\\0.000\\0.500\\0.70239\\0.500\\0.500\end{tabular} &
\begin{tabular}[c]{@{}l@{}}0.500\\0.000\\0.750\\0.761\\0.000\\0.500\end{tabular} &
	Rock-salt   &
135.0  \\
	\hline
	${Pmma}$ &&
\begin{tabular}[c]{@{}l@{}}$a$ =  3.945~\AA\\$b$ =  3.951~\AA\\$c$ =  15.859~\AA\\$\alpha$ =  90.000$^{\circ}$\\$\beta$ =  90.000$^{\circ}$\\$\gamma$ =  90.000$^{\circ}$\end{tabular} &&
\begin{tabular}[c]{@{}l@{}}Ti1\\Bi1\\Pb1\\O1\\O2\\O3\\O4\end{tabular} &&
\begin{tabular}[c]{@{}l@{}}4j\\2e\\2a\\2d\\4j\\4i\\2f\end{tabular} &
        \begin{tabular}[c]{@{}l@{}}0.500\\0.000\\0.000\\0.500\\0.500\\0.000\\0.500\end{tabular} &
\begin{tabular}[c]{@{}l@{}}0.492\\0.882\\0.000\\0.500\\0.012\\0.531\\0.553\end{tabular} &
        \begin{tabular}[c]{@{}l@{}}0.378\\0.250\\0.000\\0.000\\0.126\\0.131\\0.250\end{tabular} &
	Layered   &
158.0  \\
\hline
${Pmmm}$ &&
\begin{tabular}[c]{@{}l@{}}$a$ =  3.057~\AA\\$b$ =  7.637~\AA\\$c$ =  5.539~\AA\\$\alpha$ =  90.000$^{\circ}$\\$\beta$ =  90.000$^{\circ}$\\$\gamma$ =  90.000$^{\circ}$\end{tabular} &&
\begin{tabular}[c]{@{}l@{}}Bi1\\Pb1\\Ti1\\O1\\O2\\O3\end{tabular} &&
\begin{tabular}[c]{@{}l@{}}1b\\1f\\2n\\4v\\1c\\1g\end{tabular} &
\begin{tabular}[c]{@{}l@{}}0.500\\0.500\\0.000\\0.500\\0.000\\0.000\end{tabular} &
\begin{tabular}[c]{@{}l@{}}0.000\\0.500\\0.249\\0.242\\0.000\\0.500\end{tabular} &
\begin{tabular}[c]{@{}l@{}}0.000\\0.000\\0.500\\0.263\\0.500\\0.500\end{tabular} &
Post-perovskite  &
218.5  \\
	\hline
	\hline
	\label{tab:structureinfo}
	\end{longtable}

In our structure search, we consider all possible cation orderings in
perovskite structure, as well as non-perovskite structures such as
post-perovskite and hexagonal structures.
        
Supplementary Table~\ref{tab:structureinfo} lists twelve crystal structures of
BiPbTi$_2$O$_6$, including ten lowest energy structures predicted by
CALYPSO, perovskite anti-polar structure with $Pmma$ symmetry and
post-perovskite anti-polar structure with $Pmmm$ symmetry.  All the
energies in Supplementary Table~\ref{tab:structureinfo} are normalized to per
formula unit (\ie, 10-atom BiPbTi$_2$O$_6$). The total energy of
post-perovskite ${Pmm2}$ structure (the one with the lowest energy) is
set as the zero point.
\clearpage
\newpage

\section{S\lowercase{tructural information of available substrates}}
Supplementary Table 2. Structural information of perovskite oxide substrates KTaO$_3$ and
NdScO$_3$, calculated by DFT-PBEsol method. The experimental lattice constants,
taken from Ref.~\cite{BISWAS2017117} are also shown in the parentheses
for comparison.

	\begin{longtable}{lllllllllllllllll}

		Substrate & & Cell parameters & & Atom type  & & Wyckoff site & X & Y & Z \\
	\hline
\hline
		KTaO$_3$ &&
		\begin{tabular}[c]{@{}l@{}}$a$ =  4.001 (3.988)~\AA\\$b$ =  4.001 (3.988)~\AA\\$c$ =  4.001 (3.988)~\AA\\$\alpha$ =  90.000$^{\circ}$\\$\beta$ =  90.000$^{\circ}$\\$\gamma$ =  90.000$^{\circ}$\end{tabular} &&
\begin{tabular}[c]{@{}l@{}}K1\\Ta1\\O1\end{tabular} &&
\begin{tabular}[c]{@{}l@{}}1b\\1a\\3d\end{tabular} &
\begin{tabular}[c]{@{}l@{}}0.500\\0.000\\0.500\end{tabular} &
\begin{tabular}[c]{@{}l@{}}0.500\\0.000\\0.000\end{tabular} &
\begin{tabular}[c]{@{}l@{}}0.500\\0.000\\0.000\end{tabular}
\\
\hline
               NdScO$_3$ &&
		\begin{tabular}[c]{@{}l@{}}$a$ =  5.496 (5.57)~\AA\\$b$ =  5.752 (5.77)~\AA\\$c$ =  7.886 (7.99)~\AA\\$\alpha$ =  90.000$^{\circ}$\\$\beta$ =  90.000$^{\circ}$\\$\gamma$ =  90.000$^{\circ}$\end{tabular} &&
\begin{tabular}[c]{@{}l@{}}Nd1\\Sc1\\O1\\O2\end{tabular} &&
\begin{tabular}[c]{@{}l@{}}4c\\4b\\8d\\4c\end{tabular} &
\begin{tabular}[c]{@{}l@{}}0.250\\0.000\\0.062\\0.250\end{tabular} &
\begin{tabular}[c]{@{}l@{}}0.482\\0.500\\0.192\\0.623\end{tabular} &
\begin{tabular}[c]{@{}l@{}}0.437\\0.000\\0.194\\0.048\end{tabular} &
	\\

	\hline
	\hline
	\label{tab:substrate}
	\end{longtable}

We study two perovskite oxide substrates KTaO$_3$ and
NdScO$_3$~\cite{BISWAS2017117}.  As shown in
Supplementary Table~\ref{tab:substrate}, the DFT (PBEsol) calculated lattice
constants of KTaO$_3$ and NdScO$_3$ are in good agreement with the
experimental lattice constants (within 1$\%$ difference).  We find
that cubic perovskite KTaO$_3$ (cubic lattice constant of $\sim$ 4.00
\AA) and orthorhombic perovskite NdScO$_3$ (pseudo-cubic lattice
constant of $\sim$ 4.08 \AA) can impose tensile strain sufficiently to
stabilize perovskite $Pmm2$ BiPbTi$_2$O$_6$ in thin film form.  In our
main text, the heterostructure of \BPTO/\PTO~ is simulated to growth
on the substrate of KTaO$_3$.

\newpage
\clearpage

\section{P\lowercase{ressure and strain study including} ${P1}$ \lowercase{and} ${I4mm}$ \lowercase{structures}}

In this section, we study more crystal structures under pressure and strain.
We consider not only the three lowest energy structures (post-perovskite
$Pmm2$, perovskite $Pmn2_1$ and perovskite $Pmm2$), but also the other
two low-energy structures (a non-perovskite ${P1}$ and perovskite
${I4mm}$). Panel \textbf{a} of Supplementary Figure~\ref{fig:phase-diagram} shows
the pressure dependence and panel \textbf{b} of Supplementary Figure~\ref{fig:phase-diagram}
shows the strain dependence. The conclusion in the main text
does not change after we consider more low-energy structures.

\begin{figure}[!htbp]
	\includegraphics[angle=0,width=0.9\textwidth]{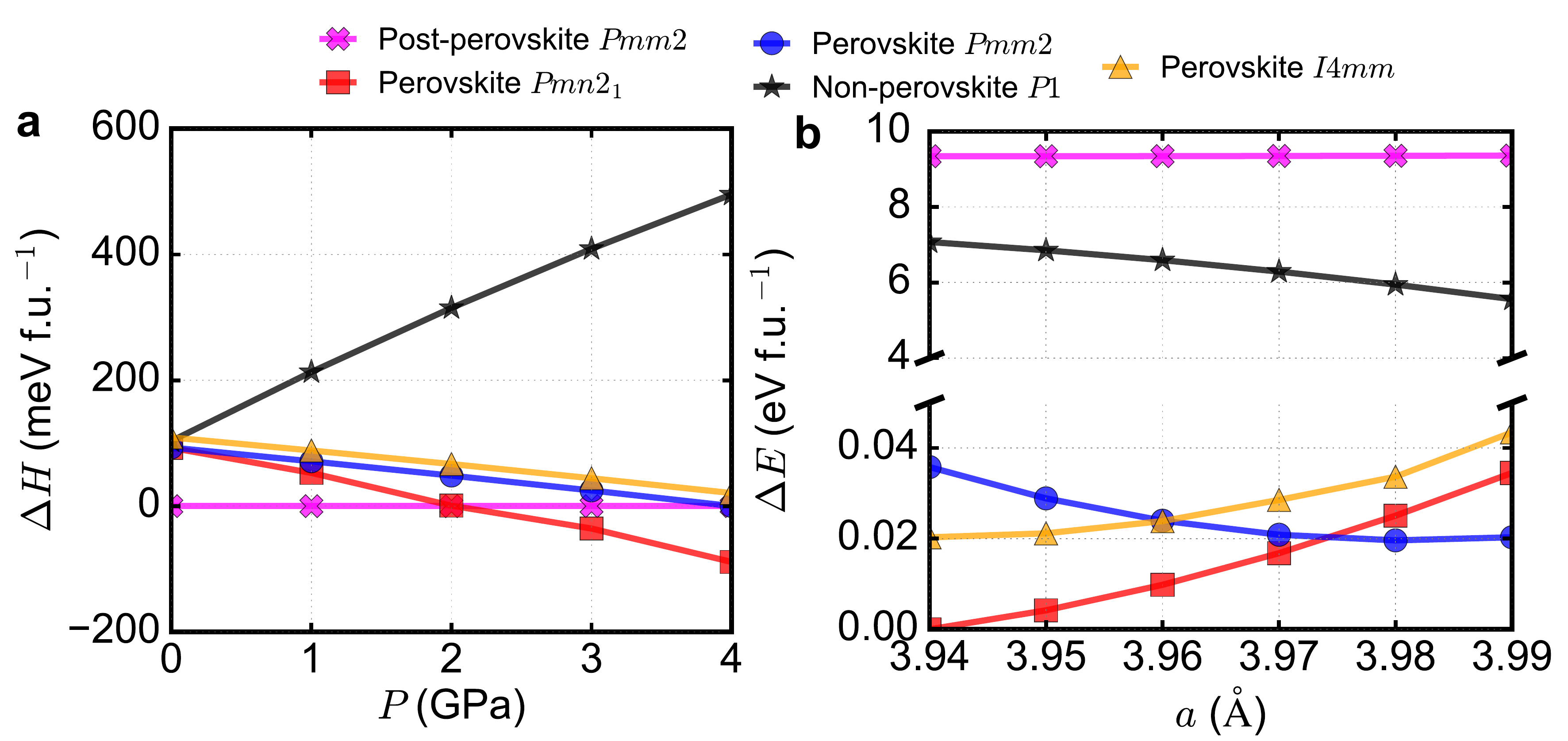}
        \caption{\label{fig:phase-diagram}
		  \textbf{Pressure and strain study
          on post-perovskite ${Pmm2}$, perovskite ${Pmn2_1}$,
          perovskite ${Pmm2}$, non-perovskite ${P1}$, and perovskite
          ${I4mm}$ structures.}
		  Panel \textbf{a} shows the
          enthalpy difference ${\Delta H}$ as a function of
          pressure. The enthalpy of post-perovskite ${Pmm2}$ structure under
          each pressure is set as zero point.  ${\Delta E}$ has the
          unit of meV f.u.$^{-1}$. Panel \textbf{b} shows the
          energy difference ${\Delta E}$ as a function of
          epitaxial strain. The energy of perovskite ${Pmn2_1}$ with
          in-plane lattice constant of 3.94 \AA~ is set as zero point.
          Note that we use a broken $y$-axis for ${\Delta E}$.
          In the top half $y$-axis, the energy runs from 4 to 10 eV f.u.$^{-1}$; in
          the bottom $y$-axis, the energy ranges from 0 to 0.04
          eV f.u.$^{-1}$.  The magenta, red, blue, black, and orange curves
          correspond to post-perovskite $Pmm2$, perovskite ${Pmn2_1}$,
          perovskite ${Pmm2}$, non-perovskite ${P1}$, and perovskite
          ${I4mm}$ structures, respectively.  }
\end{figure}

\clearpage
\newpage

\section{T\lowercase{emperature effect on the phase transitions}}
In order to investigate the temperature effect on the phase transitions between
post-perovskite $Pmm2$, perovskite ${Pmn2_1}$ and perovskite ${Pmm2}$,
we study the Helmholtz free energies of the three phases.

The Helmholtz free energy for a defect-free non-magnetic system with atomic volume $V$ at temperature $T$
can be approximated as~\cite{Bansal2016,1999Grimvall,ZHU201411}
\begin{equation}
\label{HelmholtzF}
F(V, T) = F_{\rm ph}(V, T) + F_{\rm ele}(V, T) \,,
\end{equation}
where ${F_{\rm ph}(V, T)}$ is the phonon free energy (\ie, vibrational free energy) and ${F_{\rm ele}(V, T)}$ is the thermal
electron contribution to the free energy.

The electron free energy ${F_{\rm ele}(V, T)}$
can be divided into the total energy $E_0$ at 0 K and the remaining part ${\bar{F}_{\rm ele}(V, T)}$~\cite{PhysRevB.79.134106}:
\begin{equation}
\label{ElectronicHelmholtzF}
F_{\rm ele}(V, T) = E_0 + \bar{F}_{\rm ele}(V, T) \,.
\end{equation}
$E_0$ can be calculated by standard density functional theory (DFT). In
quasi-harmonic approximation, ${\bar{F}_{\rm ele}(V, T)}$ can be
calculated by using Mermin's finite temperature formulation of
DFT~\cite{PhysRev.137.A1441,PhysRevB.79.134106}, but requires
very large supercell calculations (for BPTO, the supercells need to
contain 160$\sim$320 atoms) with a range of volumes under the
studied temperatures.

However, at elevated temperatures, when studying structural
transitions, phonon entropy plays a much more important role than
electron entropy~\cite{Bruce}.  Phonon free energy $F_{\textrm{ph}}$
can be readily calculated by using density functional perturbation
theory or frozen phonon method
~\cite{Togo-phonopy2015,PhysRevB.71.205214}.

Therefore we approximate ${F_{\rm ele}(V, T)}$ as the zero-temperature
total energy $E_0$~\cite{Togo-phonopy2015,PhysRevB.71.205214}. Furthermore
we notice that thermal expansion in solids is usually small and thus
we use the volume of the zero-temperature crystal structures. Thus,
the Helmholtz free energy in Equation~(\ref{HelmholtzF}) is approximated as
\begin{equation}
\label{HelmholtzF-simple}
F(V, T) \simeq E_0 + F_{\rm ph}(T) \,.
\end{equation}

The phonon free energy ${F_{\rm ph}(T)}$ is defined as
\begin{equation}
\label{PhononHelmholtzF}
F_{\rm ph}(T) = E_{\rm ph}(T) - TS_{\rm ph}(T) \,,
\end{equation}
where ${E_{\rm ph}(T)}$ is the phonon energy and ${S_{\rm ph}(T)}$
is the phonon entropy, both at the volume of zero-temperature crystal structure.
More specifically,
the phonon Helmholtz free energy ${F_{\rm ph}(T)}$ can be calculated from the phonon frequencies by~\cite{Togo-phonopy2015}
\begin{equation}
F_{\rm ph} = \frac{1}{2}\sum_{\textbf{q}j}\hbar\omega_{\textbf{q}j} + k_{B}T\sum_{\textbf{q}j}{\rm ln}[1-{\rm exp}(-\hbar\omega_{\textbf{q}j}/k_\mathrm{B}T)]  \,,
\end{equation}
where $\textbf{q}$, $j$, $\omega$, $T$, and $k_\mathrm{B}$ are wave vector,
band index of phonon dispersions, phonon frequency, temperature, and
Boltzmann constant, respectively. In our study, the phonon
frequencies are calculated by combining first-principles calculations
with the supercell method and finite displacement method implemented
in Phonopy~\cite{Togo-phonopy2015}.  The dimensions of supercells of
the ${Pmm2}$ post-perovskite, ${Pmm2}$ perovskite, and ${Pmn2_1}$
perovskite are 4$\times$2$\times$4 (320 atoms), 3$\times$2$\times$3
(180 atoms), and 2$\times$2$\times$2 (160 atoms) of their unit cells,
respectively. The corresponding \textbf{k}-mesh for DFT calculations
of the supercells of ${Pmm2}$ post-perovskite, ${Pmm2}$ perovskite, and
${Pmn2_1}$ perovskite are 8$\times$6$\times$6, 5$\times$3$\times$5,
and 8$\times$3$\times$7, respectively.

The obtained Helmholtz free energy ${F(T)} = E_0 +
F_{\textrm{ph}}(T)$ as a function of temperature (up to 3000 K) is
shown in Supplementary Figure~\ref{fig:free-energy}.  Under 2100 K, the
post-perovskite $Pmm2$ is the most stable phase.  There is a phase
transition between the post-perovskite $Pmm2$ and the perovskite
$Pmn2_1$ around 2100 K, hence the perovskite $Pmn2_1$ becomes the most
stable one above 2100 K.
\begin{figure}[!htbp]
	\includegraphics[angle=0,width=0.7\textwidth]{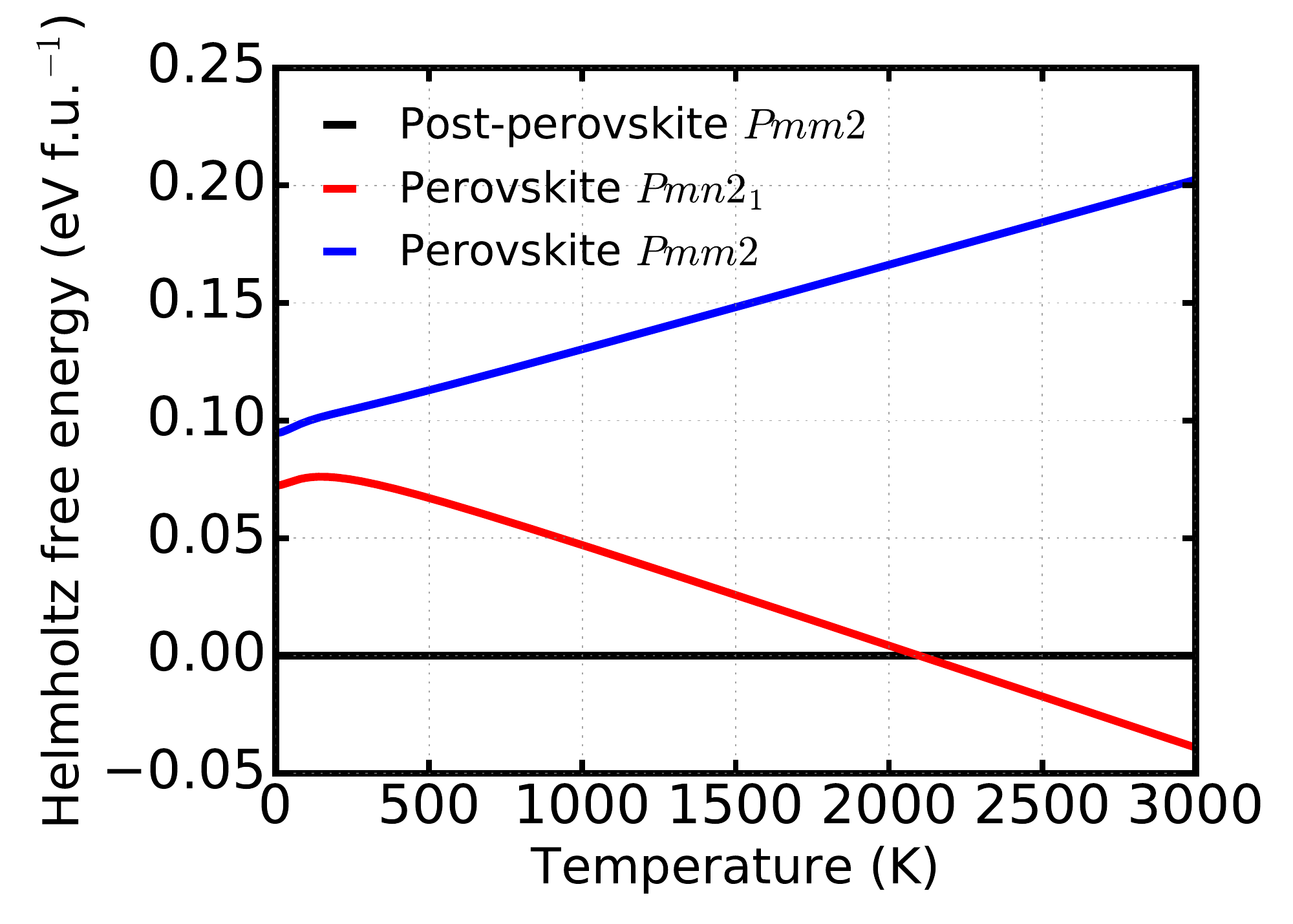}
        \caption{\label{fig:free-energy}
		\textbf{
		The approximate Helmholtz free energy
          Eq.~(\ref{PhononHelmholtzF}) of
post-perovskite $Pmm2$ (horizental black line),
perovskite $Pmm2$ (blue line) and perovskite $Pmn2_1$ (red line).}
		The Helmholtz free energy of post-perovskite $Pmm2$ is set
as the zero point.
        }
\end{figure}


\clearpage
\newpage

\section{T\lowercase{he} DOS \lowercase{of post-perovskite ${Pmm2}$ structure}}

\begin{figure}[!htbp]
	\includegraphics[angle=0,width=0.9\textwidth]{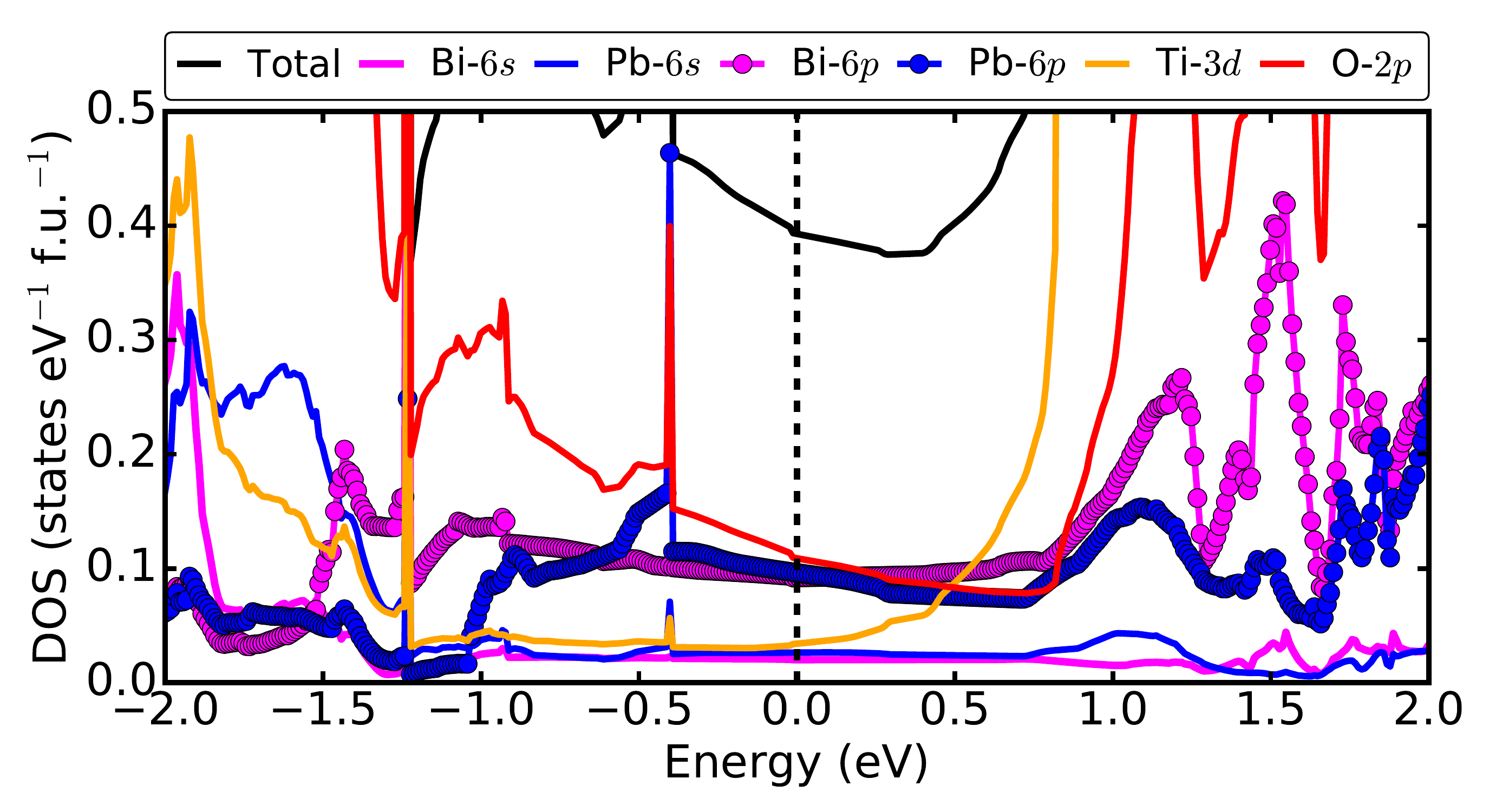}
        \caption{\label{fig:ppv-DOS}
		\textbf{The density of states of $Pmm2$ BiPbTi$_2$O$_6$.}
		The black curve is the total DOS.
  The magenta curve, magenta curve with circle markers, blue curve,
	blue curve with circle markers, orange curve and red curve
		are Bi-$6s$, Bi-6$p$, Pb-$6s$, Pb-6$p$, Ti-$3d$ and
  O-$2p$ projected DOS, respectively. The dashed line is the Fermi level.
        }
\end{figure}

Supplementary Figure~\ref{fig:ppv-DOS} shows the total density
of states (DOS) and orbital projected densities of states of
post-perovsite $Pmm2$ BiPbTi$_2$O$_6$.
The DOS at the Fermi level is mainly composed of O-2$p$,
Bi-6$p$ and Pb-6$p$ states, with very small contributions
from Ti-3$d$, Bi-6$s$ and Pb-6$s$ states.

\newpage
\clearpage

\section{P\lowercase{b}T\lowercase{i}O$_3$ \lowercase{under bi-axial strain}}

In experiment, bulk PbTiO$_3$ displays a spontaneous polarization of
about 0.75 C/m$^2$ at 295 K with $c$/$a$ ratio of 1.063 and space
group of ${P4mm}$~\cite{PhysRevLett.72.3618, PhysRevLett.95.177601}.
In this section, we use first-principles calculations (DFT-PBEsol) to
study PbTiO$_3$ under bi-axial strain. In structural optimizations,
we use an energy cutoff of 600 eV and \textbf{k} mesh of
17$\times$17$\times$17. The convergence thresholds of energy and
atomic Hellmann-Feynman forces are $10^{-9}$~eV and $10^{-4}$~eV \AA$^{-1}$,
respectively. We fix the in-plane epitaxial lattice constants ($a_x=a_y=a$) and
allow the out-of-plane lattice constant $a_z=c$ to change. All the
internal atomic coordinates are fully relaxed. We study two different
polarization orientations: if the polarization is along $z$-axis,
the state is referred to as ``out-of-plane polarization'';
if the polarization is either along $x$-axis or $y$-axis,
the state is referred to as ``in-plane polarization''.

We change the in-plane epitaxial lattice constant $a$ and study the
energy difference between the ``out-of-plane polarization'' and the
``in-plane polarization'' states as a function of $a$. The results are
shown in Supplementary Figure~\ref{fig:PbTiO3-strain}\textbf{a}.  There is a critical
lattice constant $a_c \simeq 3.97$~\AA~at which the two states are
degenerate. If the epitaxial lattice constant is less than $a_c$, the
``out-of-plane polarization'' state is more energetically favorable.
If the epitaxial lattice constant exceeds $a_c$, the ``in-plane
polarization'' state becomes more
stable. Supplementary Figure~\ref{fig:PbTiO3-strain}\textbf{b} shows the $c/a$ ratio
of both ``out-of-plane polarization'' and ``in-plane polarization''
states as a function of the in-plane epitaxial lattice constant
$a$. We note that without any strain, our calculations find that bulk
PbTiO$_3$ has lattice constants $a_x=a_y=3.89$~\AA~and
$a_z=4.17$~\AA~with its polarization pointing along $z$ axis. This
indicates that if we want to stabilize an in-plane polarization in
PbTiO$_3$ thin films, we need at least 2\% bi-axial tensile strain.

\begin{figure}[!h]
\includegraphics[angle=0,width=0.7\textwidth]{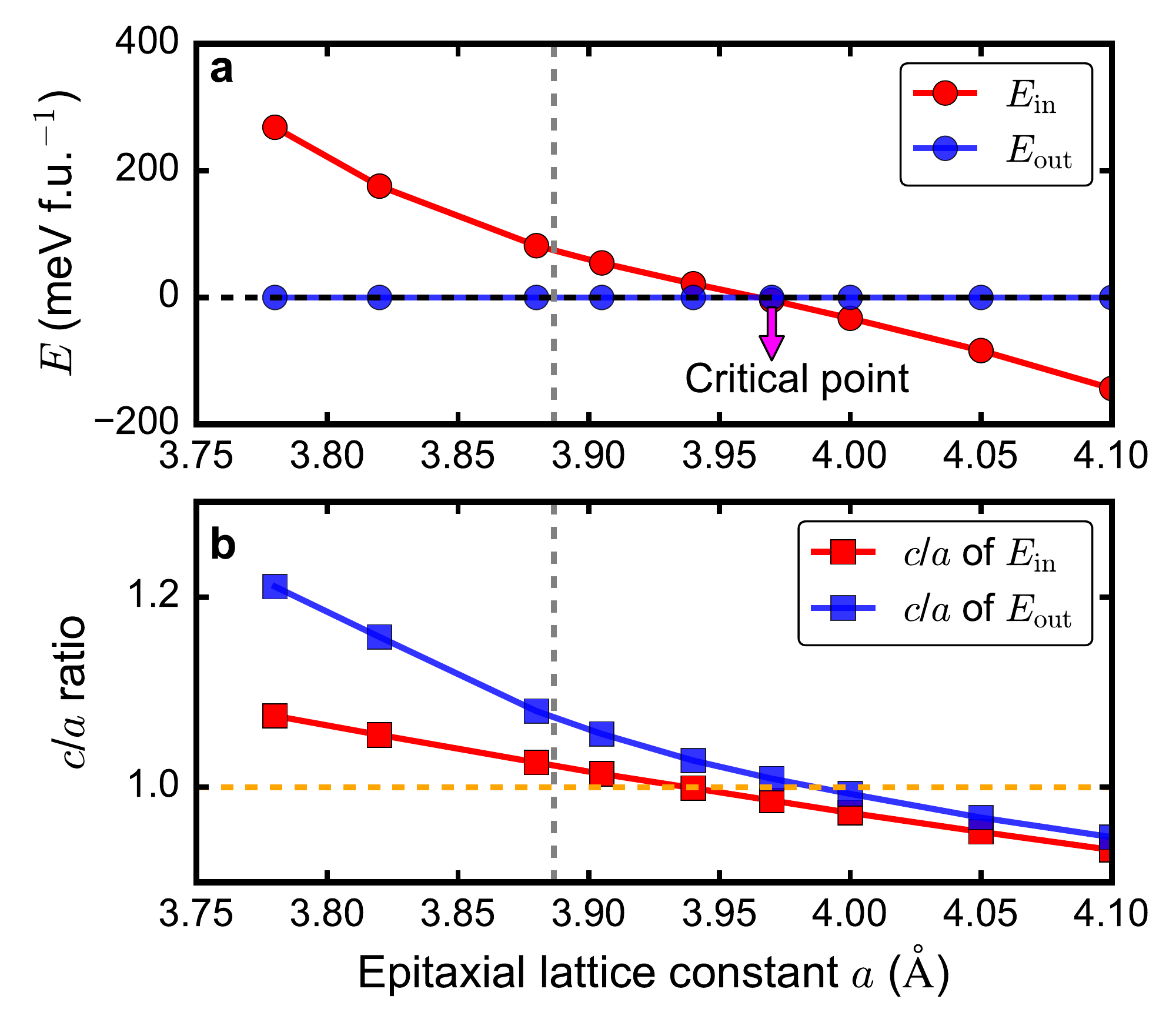}
\caption{\label{fig:PbTiO3-strain}
  \textbf{PbTiO$_3$ under bi-axial strain.}
  Panel \textbf{a}: total energies of ``in-plane polarization'' state
  ($E_{\rm in}$) and ``out-of-plane polarization'' state ($E_{\rm
    out}$) of PbTiO$_3$ under bi-axial strain. For each given
  epitaxial lattice constant $a$, $E_{\rm out}$ is chosen as the zero
  energy. At $a=3.97$~\AA, the two states are degenerate. When
  $a<3.97$~\AA, the ``out-of-plane polarization'' state is more
  energetically favorable. When $a>3.97$~\AA, the ``in-plane
  polarization'' state is more stable.  Panel \textbf{b}: the $c$/$a$
  ratio of both ``in-plane polarization'' state and ``out-of-plane
  polarization'' state as a function of epitaxial lattice constant
  $a$. In both panels, the grey dashed line refers to strain-free
  PbTiO$_3$ which has theoretical lattice constants
  $a_x=a_y=3.89$~\AA~and $a_z=4.17$~\AA.}
\end{figure}

\newpage
\clearpage

\section{D\lowercase{ensity of states of} B\lowercase{i}P\lowercase{b}T\lowercase{i}O$_6$/P\lowercase{b}T\lowercase{i}O$_3$ \lowercase{interface}}

Total density of states and
layer-resolved density of states projected onto Ti-$3d$ orbitals for
both parallel and anti-parallel states are shown in
Supplementary Figure~\ref{fig:parallel}.
The layer-resolved conduction electrons in the main text are calculated
by integrating the partial density states of Ti-$3d$ orbitals.

\begin{figure}[!h]
	\includegraphics[angle=0,width=0.99\textwidth]{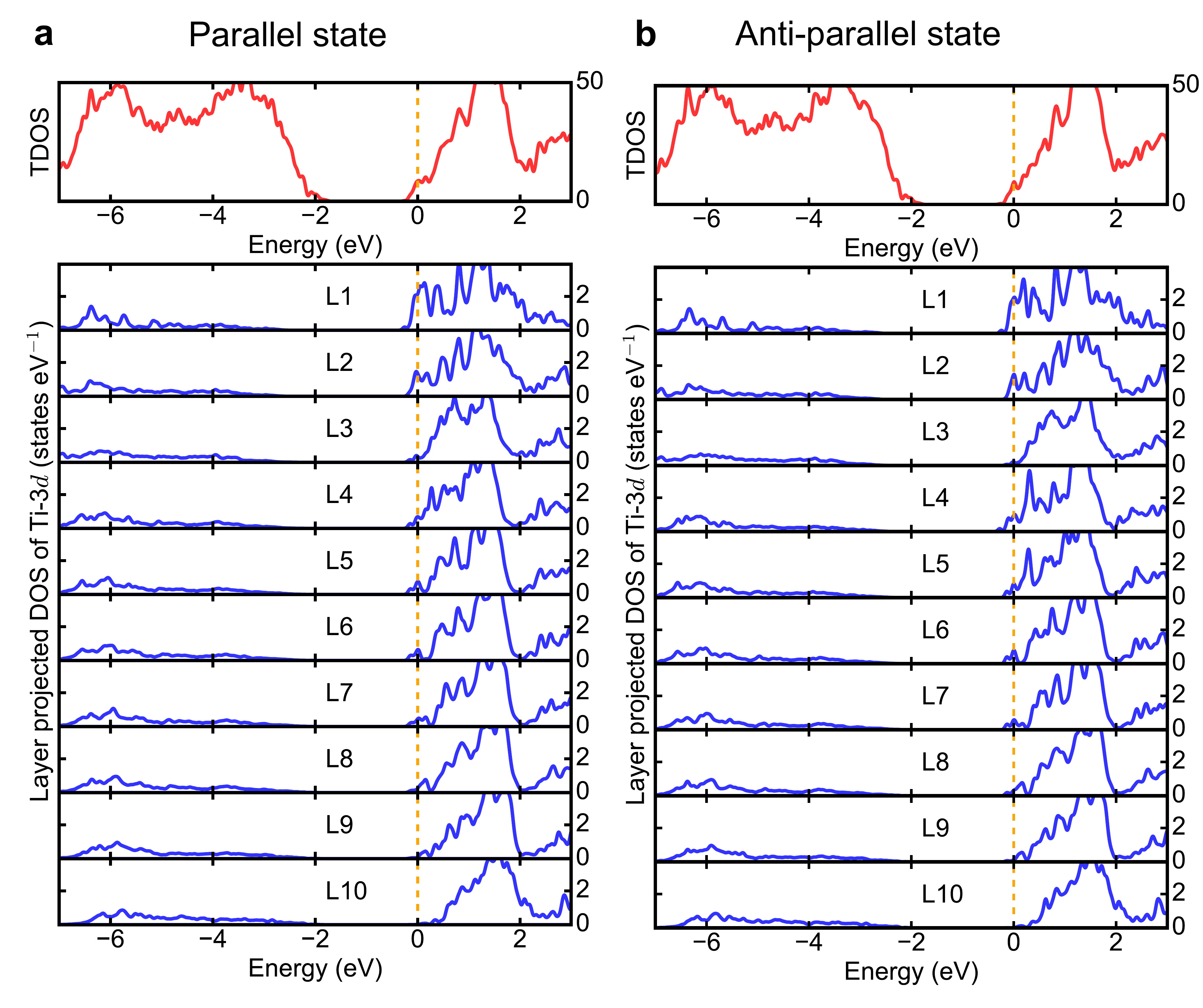}
	\caption{\label{fig:parallel}
	\textbf{Total density of states (TDOS) and
	  the layer-resolved density of states projected onto Ti-$3d$ orbitals
          (layer projected DOS).}
	Panel \textbf{a}: parallel state.
	Panel \textbf{b}: anti-parallel state.
	Red and blue curves correspond to TDOS and layer projected DOS, respectively.
	L1...L10 refer to the TiO$_2$ layers of the BiPbTi$_2$O$_6$/PbTiO$_3$
        heterostructure (the ordering is consistent with Fig. 5 in the main text).
	The orange dashed line is Fermi level, which is set to zero energy.
	}
\end{figure}

\newpage
\clearpage

\section{T\lowercase{he charge leakage in} B\lowercase{i}P\lowercase{b}T\lowercase{i}$_2$O$_6$/\lowercase{ferroelectric heterostructure}}

\begin{figure}[!t]
\includegraphics[angle=0,width=0.5\textwidth]{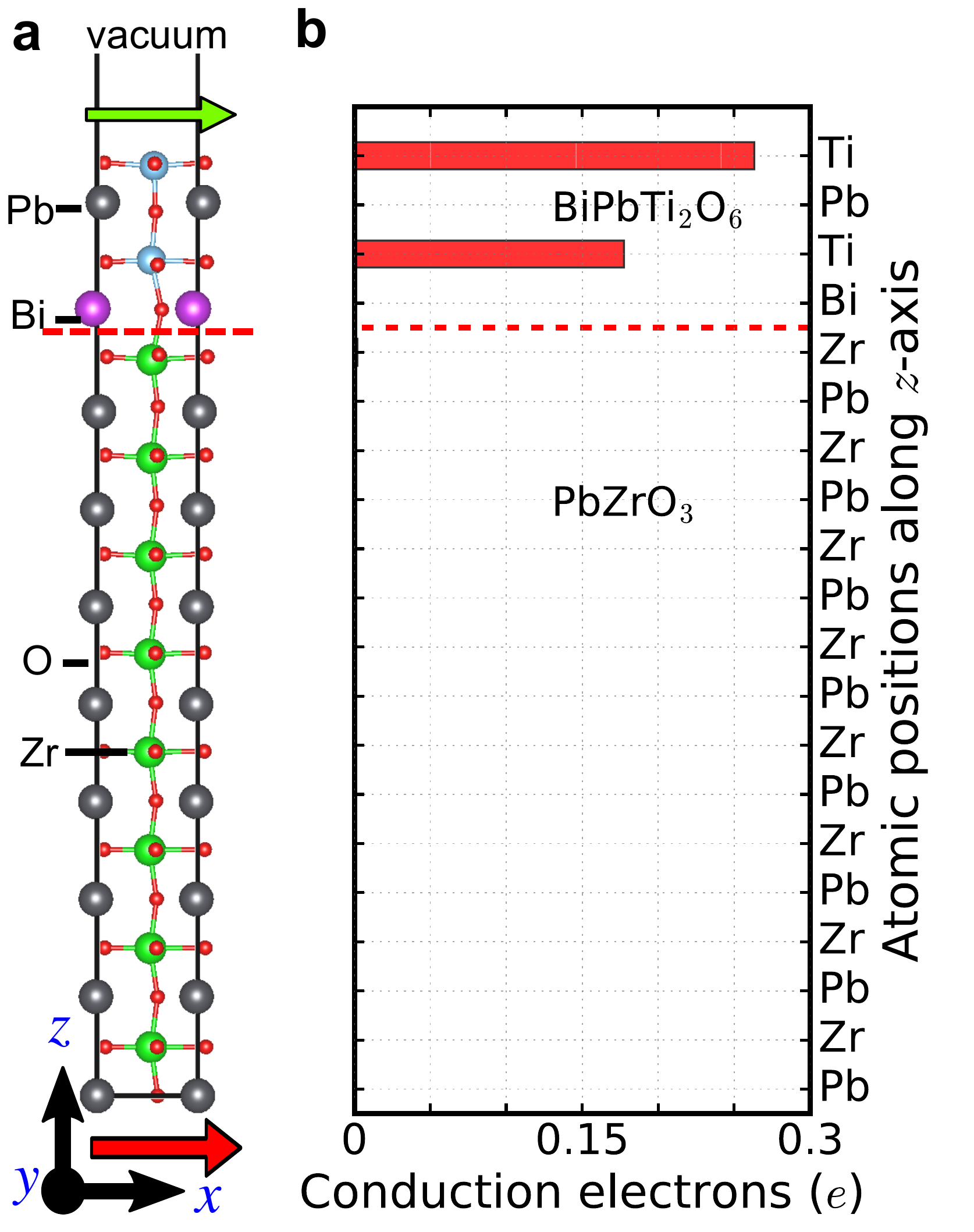}
\caption{\label{fig:PZT}
  \textbf{DFT calculation of
  \BPTO/PbZrO$_3$~heterostructure.}
\textbf{a} Atomic structure of the
  \BPTO/PbZrO$_3$~heterostructure:
  The red (green) arrow refers to PbZrO$_3$ polarization (\BPTO~polar
  displacements). \textbf{b} Layer-resolved conduction electrons on
  Ti and Zr atoms.  The red dashed lines indicate the interface between
  PbZrO$_3$~and~\BPTO.}
\end{figure}

We note that in Fig. 5 of our main text charge leakage into PbTiO$_3$
is non-negligible in the \BPTO/\PTO~heterostructure.
This charge leakage is due to ``proximate
effect'' that PbTiO$_3$ has empty Ti $d^0$ states while BiPbTi$_2$O$_6$
has 0.5$e$ in the $d$ orbitals per Ti site. Charge transfer occurs
from the Ti atoms in \BPTO~to the Ti atoms in PbTiO$_3$ thin films, a
phenomenon similar to LaTiO$_3$/SrTiO$_3$ interface~\cite{hwang2002,millis2004}.
To prevent charge leakage, we can use PbTi$_{1-x}$Zr$_x$O$_3$ (PZT) to
replace PbTiO$_3$. The mechanism is that Zr has $4d$ orbitals, whose
energy is higher than Ti $3d$ orbitals. To demonstrate that, we
replace PbTiO$_3$ by PbZrO$_3$ in our heterostructure and re-do the
calculations.  Supplementary Figure~\ref{fig:PZT} shows the conduction electrons in
Ti-$d$ and Zr-$d$ orbitals.  We find that charge leakage is completely
suppressed and all the conduction electrons are confined in Ti-$d$
orbitals in \BPTO. A direct simulation of PZT requires a much larger
supercell and a proper treatment of random alloying, which is beyond
the capability of our computation resources.  However, the physics
trend is clear: the more Ti atoms are replaced by Zr atoms, the weaker
the charge leakage. Usually PZT with $x\sim 0.2-0.5$ is widely used in
ferroelectric
heterostructures~\cite{NatCommPZT-hetero,ChoiAPL-2010,Feigl-JAP2009,WangGS-APL2001}.

\newpage
\clearpage

\section{P\lowercase{olar displacements in thin films of polar metal} B\lowercase{i}P\lowercase{b}T\lowercase{i}$_2$O$_6$}

There is ``size effects'' in ferroelectric thin
films~\cite{Li_1997JSAP}.  With the thickness of ferroelectric thin
films decreasing, depolarization fields (if not fully screened) reduce
and sometimes completely suppress the
polarization~\cite{batra1972thermodynamic,junquera2003critical}.

However, polar metals do not have such a ``size effect'' because free
electrons in metals fully screen depolarization fields in both bulk
and thin films. To support that, we perform two additional calculations.

The first calculation is a thought-experiment. We calculate a
free-standing one unit cell \BPTO~thin film. We find that down to one unit
cell, \BPTO~is still polar. The polar displacements of free-standing
one unit cell \BPTO~thin film are shown in Supplementary Figure~\ref{fig:slab}, which
are compared to bulk \BPTO. This shows that there is no ``size effect''
in thin films of polar metal \BPTO.

\begin{figure}[b]
\includegraphics[angle=0,width=0.5\textwidth]{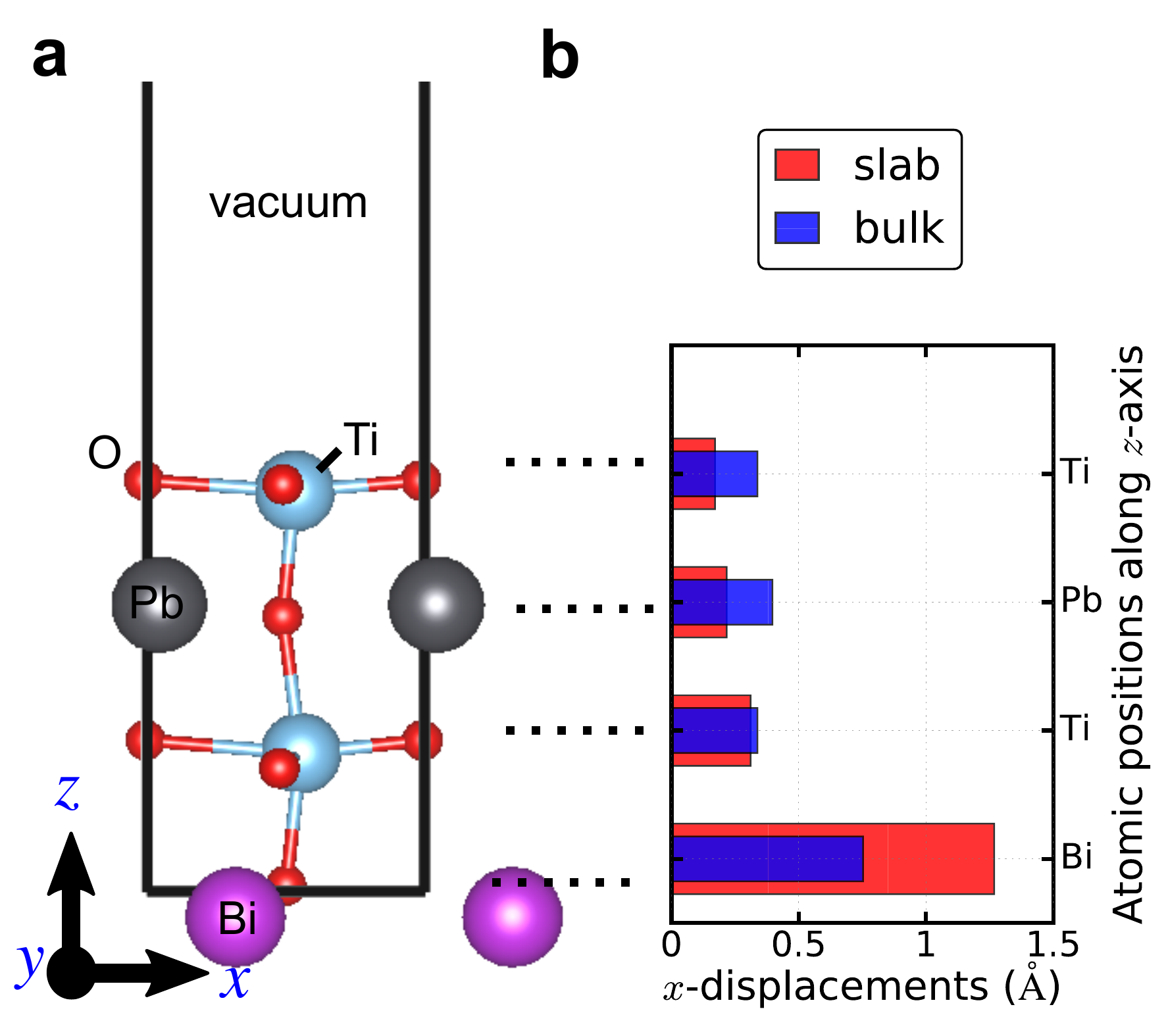}
\caption{\label{fig:slab} \textbf{The structural property of
    free-standing one unit cell \BPTO~thin film.}  \textbf{a}
  Optimized crystal structure.  \textbf{b} The comparison of
  layer-resolved polar displacements between one unit cell \BPTO~thin
  film and the bulk phase.}
\end{figure}

\begin{figure}[!t]
\includegraphics[angle=0,width=0.5\textwidth]{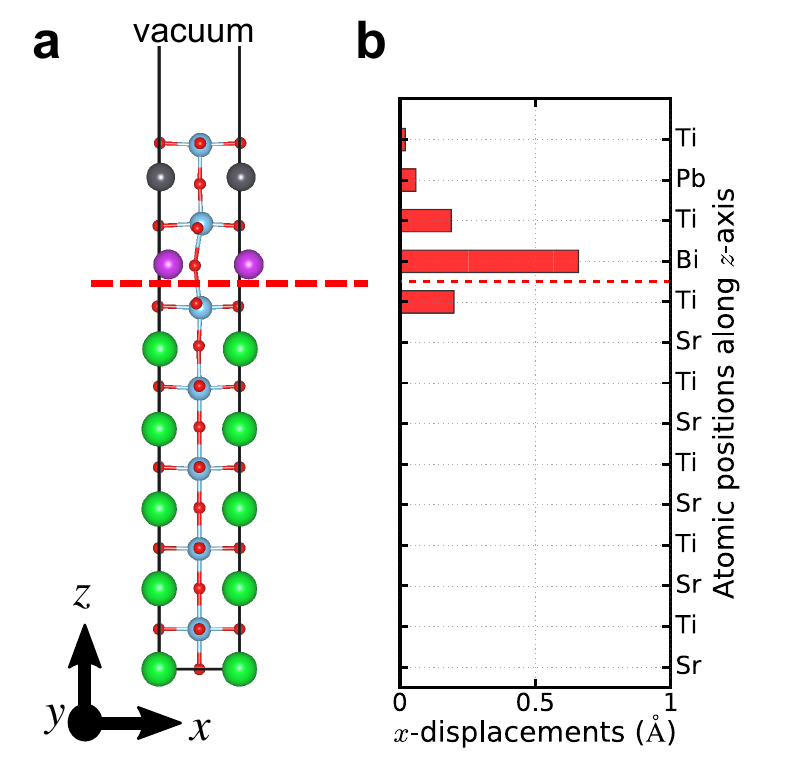}
\caption{\label{fig:STO-BPTO-disp}
\textbf{The property of \BPTO/SrTiO$_3$ interface.}
\textbf{a} The optimized structure of \BPTO/SrTiO$_3$ interface.
The vaccum layer is about 15~\AA~in our DFT model.
\textbf{b} The  polar displacements along $x$-axis in \BPTO/SrTiO$_3$ interface.
The dashed red lines indicate the interface between \BPTO~and~\STO.
	}
    \end{figure}

The second calculation is to study ``proximity effect'' in \BPTO~thin
films. Instead of a \BPTO/PbTiO$_3$ heterostructure, we calculate a
\BPTO/SrTiO$_3$ heterostructure, in which SrTiO$_3$ is paraelectric.
Supplementary Figure~\ref{fig:STO-BPTO-disp} shows the optimized structure in our DFT
calculations and the corresponding polar displacements along $x$-axis.
We find that the polar displacements in \BPTO~still exist. In
addition, the structural coupling between \BPTO~and \STO~drives the
interfacial Ti atom in \STO~to be polar (see
Supplementary Figure~\ref{fig:STO-BPTO-disp}).  This indicates that the polar
displacements in \BPTO~thin films are \textit{not} due to proximity
coupling with PbTiO$_3$.  In fact, they are so strong that they can
drive a paraelectric material (such as SrTiO$_3$) to be polar close to
the interface.

\newpage
\clearpage

\section{S\lowercase{witching of multi-layer} B\lowercase{i}P\lowercase{b}T\lowercase{i}$_2$O$_6$~\lowercase{films}}

\begin{figure}[htp]
	\includegraphics[angle=0,width=0.8\textwidth]{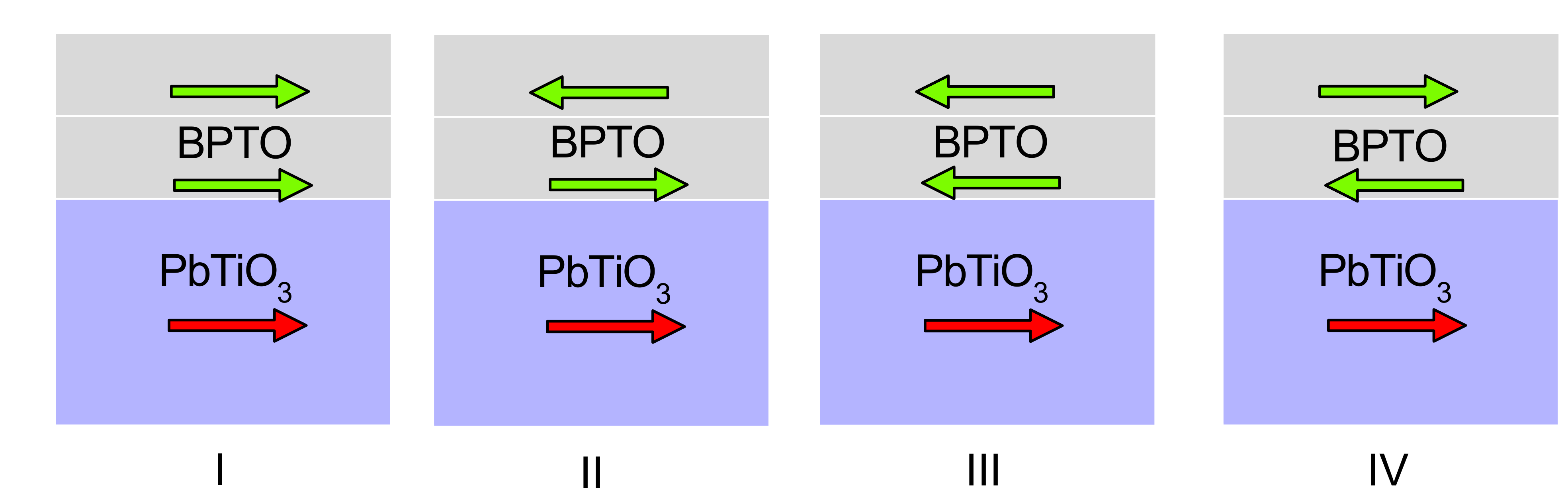}
	\caption{\label{fig:2UC-switching}
	\textbf{Switching of multi-layer \BPTO~thin films.}
	Four different configuration of the
	heterostructure composed of PbTiO$_3$ thin film and two unit cells of
	BiPbTi$_2$O$_6$. The red arrow refers to the polarization of
	PbTiO$_3$ thin film. The green arrows refer to the polar displacements of
	\textit{each} unit cell of BiPbTi$_2$O$_6$. }
\end{figure}
To support the physical picture of switching multi-layer BPTO in the main text, we perform
calculations of two-unit-cell BPTO~thin films on
PbTiO$_3$. Supplementary Figure~\ref{fig:2UC-switching} shows different configurations
of polar displacements of BPTO~and polarization of PbTiO$_3$. The red
arrow is the polarization of entire PbTiO$_3$ thin film. The green
arrow refers to the polar displacements of \textit{each} unit cell of
BPTO. Since there are two unit cells of BPTO, then there are
altogether four different configurations. Our calculations find that
their total energies are $E_{\textrm{I}} < E_{\textrm{II}} <
E_{\textrm{III}} < E_{\textrm{IV}}$.  This energy order is easy to
understand: both the interface and bulk BPTO~prefers to have a
parallel coupling between polar displacements and polarization. In
configuration I, both the polar displacements of BPTO~between the two
unit cells and the polarization of PbTiO$_3$ are parallel, which leads
to the lowest total energy. In configuration IV, the polar
displacements of BPTO~between the two unit cells are \textit{anti}parallel.
The polar displacements of the bottom layer BPTO~is also
\textit{anti}parallel to the polarization of PbTiO$_3$.
The two antiparallel couplings combined result in the highest
total energy. The switching process is as follows: we start from
configuration III in which the polarization of PbTiO$_3$ is switched
by an electric field. The interfacial coupling drives the bottom unit
cell of BPTO~to switch its polar displacements (\ie, configuration
II). Then the bulk coupling in BPTO~drives the top unit cell of
BPTO~to switch its polar displacements (\ie, configuration I).
This entire process is favored by thermodynamics because the total
energy monotonically decreases from configuration III to II and
finally to I.

\newpage
\clearpage

\section{U\lowercase{niform strain in epitaxial} B\lowercase{i}P\lowercase{b}T\lowercase{i}$_2$O$_6$~\lowercase{thin film~from experimental perspective}}

\begin{figure}[h]
	\includegraphics[angle=0,width=0.4\textwidth]{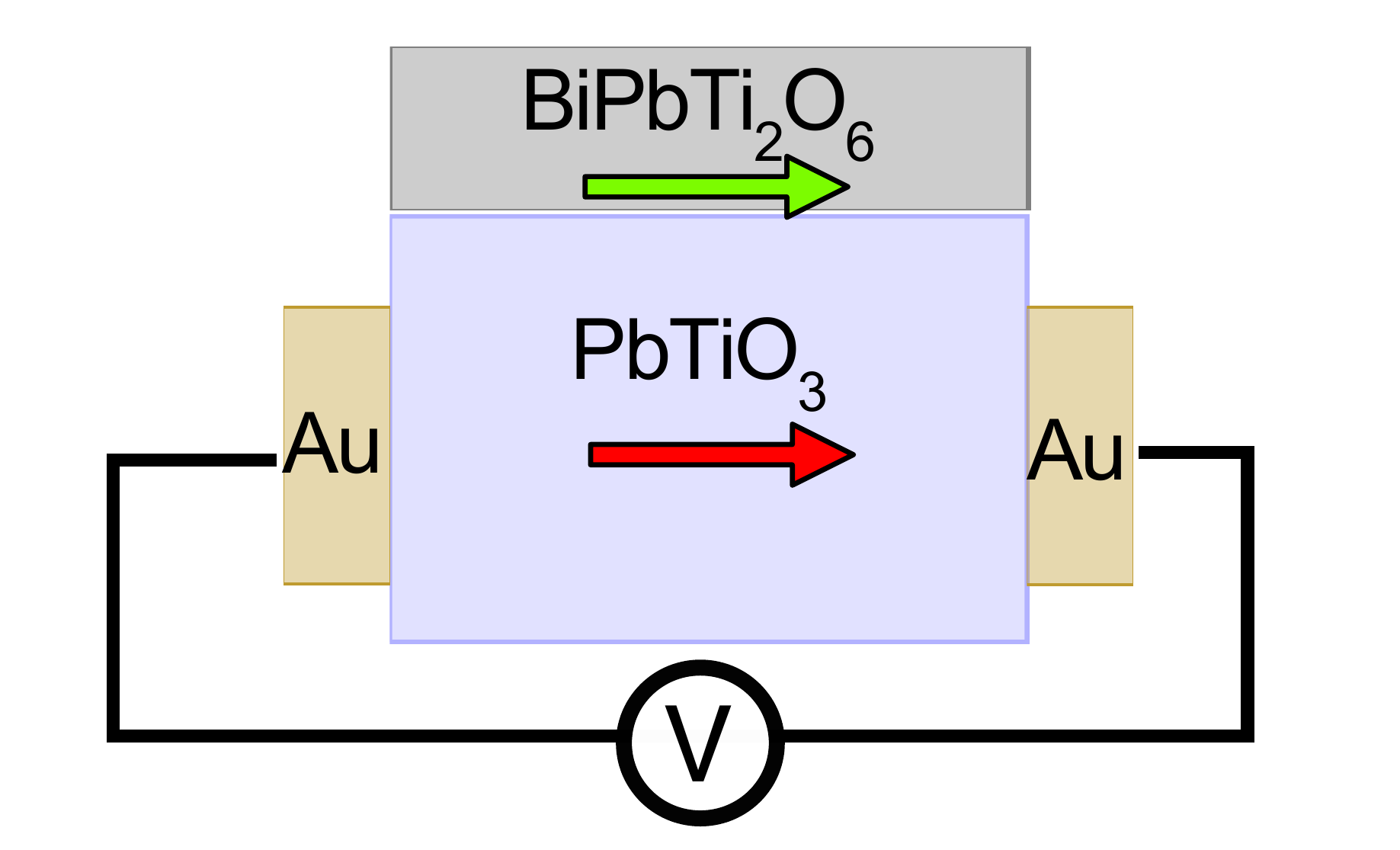}
	\caption{\label{fig:model-device}
	\textbf{A model device based on
	\BPTO/PbTiO$_3$.}
	Au only works as electrode.
	}
\end{figure}
In our DFT calculations, the BPTO/PbTiO$_3$ heterostructure is
simulated by constraining its in-plane lattice constant to that of
KTaO$_3$ substrate ($a$ = 4.00 \AA, see
Supplementary Table~\ref{tab:substrate}). The entire heterostructure is under
\textit{uniform} tensile strain. Under this strain, the $Pmm2$ perovskite phase
is indeed stabilized (see Fig. 4 in the main text).  We find the
metastable (even unstable and unusual polymorphs) phase can appear via
applying uniform epitaxial strains to thin
films~\cite{trampert1997direct,XuYaobin-ACS2019,PhysRevB.85.024113,Sando-APRev2016}.
Such epitaxial stabilization can be understood by the theory of free
energy minimization, in which the energy of coherent and semicoherent
interfaces is much lower than that of noncoherent
ones~\cite{GorbenkoChemMater2002}. Therefore, the formation of
low-energy interfaces and the minimization of overall free energy of
the system due to the contribution of volume strain energy usually
give rise to the experimentally observed metastable and even unstable
structures.

Specifically for oxide heterostructures, experimentalists find that the critical
thickness below which the entire thin film is under \textit{uniform}
strain is typically about 10 nm and sometimes can exceed 100
nm~\cite{Wang2013}.  In our case, the thickness of BPTO/PbTiO$_3$
heterostructure (in Fig. 4 in the main text) is only 4 nm, which is
far below the critical thickness for uniform strain.
Furthermore, in our BPTO/PbTiO$_3$ heterostructure, PbTiO$_3$ has a
polarization parallel to the interface. This means that the
depolarization field in PbTiO$_3$ thin film can be fully screened by
the two electrodes (see the ``toy device'' in
Supplementary Figure~\ref{fig:model-device}). Therefore the thickness of PbTiO$_3$
films can be further reduced without suppressing its polarization.
Based on the above reasons, our BPTO/PbTiO$_3$ heterostructure is
anticipated to be uniformly strained and thus can be stabilized on
KTaO$_3$ or NdScO$_3$ substrate.

\newpage
\clearpage
